\documentclass[english]{article}
\usepackage[T1]{fontenc} 
\usepackage[latin9]{inputenc}
\usepackage{geometry}
\usepackage{booktabs}
\usepackage{amsmath}
\usepackage{amsthm}
\usepackage{graphicx}
\usepackage{mathtools}
\usepackage{esint}
\usepackage{amsfonts}
\usepackage{amsbsy}
\usepackage{comment}

\makeatletter

\providecommand{\keywords}[1]
{
  \small	
  \textbf{\textit{Keywords---}} #1
}

\providecommand{\tabularnewline}{\\}

\usepackage{color}
\theoremstyle{thmstyletwo}\def\d{{\, \rm d}}

\theoremstyle{thmstylethree}%
\@ifundefined{showcaptionsetup}{}{%
 \PassOptionsToPackage{caption=false}{subfig}}
\usepackage{subfig}
\makeatother

\usepackage{babel}
\begin{document}
\title{A Physics-Informed Data-Driven Algorithm
for Ensemble Forecast of Complex Turbulent Systems }
\author{Nan Chen\textsuperscript{a} and Di Qi\textsuperscript{b}}
\date{\textsuperscript{a }Department of Mathematics, University of Wisconsin-Madison,
480 Lincoln Drive, Madison, 53706, WI, USA\\\textsuperscript{b }Department of Mathematics, Purdue University,
150 North University Street, West Lafayette, 47907, IN, USA}
\maketitle

\begin{abstract}
A new ensemble forecast algorithm, named as the physics-informed data-driven algorithm with conditional Gaussian statistics (PIDD-CG), is developed to predict the time evolution of the probability density functions (PDFs) of complex turbulent systems with  partial observations. The PIDD-CG algorithm integrates a unique multiscale statistical closure model with an extremely efficient nonlinear data assimilation scheme to represent the PDF as a mixture of conditional statistics, which overcomes the curse of dimensionality for high-dimensional systems. The multiscale features in the time evolution of each conditional statistics ensemble member  effectively captured by an appropriate combination of physics-informed analytic formulae and recurrent neural networks. An information metric is adopted as the loss function for the latter to more accurately calibrate the key turbulent signals with strong fluctuations. The proposed algorithm succeeds in forecasting both the transient and statistical equilibrium non-Gaussian PDFs of strongly turbulent systems with intermittency, regime switching and extreme events.

\end{abstract}


\section{Introduction}\label{sec1}
Complex turbulent systems are ubiquitous in many fields, such as geophysics, climate science, neural science, engineering, and plasma physics \cite{strogatz2018nonlinear, sheard2009principles, ghil2012topics}. These systems contain rich nonlinear dynamics and statistical features, including multiscale structures, intermittency, extreme events, regime switching, and strong non-Gaussian probability density functions (PDFs) \cite{lucarini2016extremes, franzke2015stochastic, wilcox1988multiscale, majda2016introduction, tao2009multiscale}. Predicting the future states of complex turbulent systems is a central challenge in contemporary science with large societal impacts. Due to the turbulent nature of such systems, the \emph{trajectory forecast} based on a single realization of the model state quickly loses track of the truth. Alternatively, the \emph{ensemble forecast}, which adopts a probabilistic characterization of the model states utilizing a Monte Carlo (MC) type approach, is the predominant strategy in predicting complex turbulent systems in practice \cite{palmer2019ecmwf, toth1997ensemble, leutbecher2008ensemble}. In the ensemble forecast, different ensemble members are sampled from an initial conditions and are subject to different random forcing, accounting for the uncertainty in the initialization and model errors, respectively. The ensemble forecast aims at providing an indication of the PDF of possible future states by tracking the evolution of the group of ensemble members.

Despite the simplicity of the general framework, there exists a major computational challenge in applying the traditional ensemble forecast method to realistic scenarios. In fact, as the dimension of the system becomes large, an exponential increase of the ensemble size is needed to maintain the accuracy of the forecast PDF, which is known as the curse of dimensionality \cite{cherkassky2007learning}. However, since the computational cost of each single model realization also shoots up significantly as the dimension of the system increases, only a small ensemble size is affordable in practical situations, such as climate and weather forecast \cite{gneiting2005weather, evans2013optimally}. As a result, although the traditional ensemble forecast method can ideally provide a reasonably accurate characterization of the mean state, the lack of a sufficient number of samples makes it extremely difficult to accurately forecast the intrinsic uncertainty of the system. Especially, the direct ensemble method often fails to capture the non-Gaussian joint PDF in high dimensions, and thus leads to large errors in characterizing many key turbulent phenomena, such as regime switching, intermittency and extreme events \cite{majda2016introduction, farazmand2018extreme}. A similar issue occurs at the initialization stage of the forecast. Since only the time series of a subset of the state variables can be observed in most of the practical problems (known as partial observations), ensemble data assimilation is often required for the state estimation of the unobserved variables \cite{evensen2009data, anderson2012localization}. Yet, the error in quantifying the initial uncertainty due to the lack of a sufficient number of samples can be rapidly amplified in the subsequent forecast, leading to large biases in predicting even short-term transient features.

In this paper, a new physics-informed data-driven conditional Gaussian (PIDD-CG) algorithm is developed that aims at efficiently and accurately forecasting the key non-Gaussian PDF for a wide class of high-dimensional complex systems.   The PIDD-CG algorithm starts with a phase space decomposition by projecting the model states into a low-dimensional subspace containing the observed state variables and a remaining multiscale high-dimensional subspace. A systematic multiscale data-driven closure approximation is developed in the low-dimensional subspace, with which  a small number of samples is sufficient to characterize the associated uncertainty propagations. The PIDD-CG algorithm then exploits an effective physics-based decomposition of the PDF in the high-dimensional subspace into a conditional Gaussian mixture and integrates the evolution equations of the conditional statistics associated with each mixture component to obtain the forecast PDF \cite{chen2017beating}.
A continuous data assimilation scheme is used to determine the characteristics in each conditional Gaussian component that is associated with one realization in the low-dimensional observed state \cite{chen2018conditional}. This captures the correlation between the two nonlinearly coupled subspaces. 
There are several remarkable advantages of the PIDD-CG algorithm. First, by creating the mixture distribution the algorithm does not suffer from the curse of dimensionality with respect to the number of mixture components \cite{chen2018rigorous}. In fact, fundamentally different from the purely data-driven approaches that often require a large number of samples,  the development of the mixture distribution in the PIDD-CG algorithm uses only the same small number of samples in characterizing the associated low-dimensional subspace and is sufficient to represent the high-dimensional full PDF thanks to the conditional mixture. Second, the governing equations of the time-evolution of the conditional statistics have closed analytic formulae \cite{liptser2013statistics}, which further reduce the computational cost and avoid the direct ensemble approximation of obtaining such statistical moments that are commonly required in applying ensemble simulations.

Yet, despite the analytically solvable properties, the computational cost of solving the governing equations of the full conditional statistics can still be demanding due to the existence of a large number of complicated nonlinear terms involving a wide spectrum of multiscale fluctuating variables, especially those associated with unresolved scales. Therefore, the PIDD-CG algorithm proposes a balanced physics-informed and data-driven construction of these governing equations, aiming at explicitly preserving the crucial dynamical structure while using data-driven approaches to effectively forecast the complicated unresolved details with a much lower computational cost. Specifically, the PIDD-CG algorithm approximates the complicated nonlinear feedbacks in these governing equations by a recurrent neural network (RNN) \cite{zaremba2014recurrent}. Since the neural network aims at predicting the statistics, a simple but effective information metric is adopted as the loss function to train the RNN \cite{qi2020using, kleeman2011information}, which significantly outweighs the traditional loss functions that are based on minimizing the path-wise errors.
Finally, in light of the evolution equations of the conditional statistics, the PIDD-CG algorithm naturally provides a systematic way of developing statistical reduced order models \cite{majda2018strategies, qi2021machine}, in which the feedback from the unresolved-scale variables is approximated by the RNNs. This further enhances the forecast efficiency for complex systems with very large dimensions in practice when the primary interest lies in the statistical forecast of certain large-scale modes.

In the rest of the paper, the PIDD-CG algorithm is illustrated based on a prototype model in geophysical turbulence: the topographic barotropic model, which displays many representative turbulent features, including extreme events and switching regimes \cite{majda2006nonlinear}. The method for a general group of nonlinear systems is described in \emph{Methods} Section \ref{sec:method}. More detailed results including a coupled dyad model as a proof-of-concept and a complete analysis of the computational performance are listed in the Supplementary Information (SI).

\section{Results}\label{sec2}
\subsection{The PIDD-CG algorithm}\label{Subsec:algorithm}
\begin{figure}[h!]
\centering
\includegraphics[width=\textwidth]{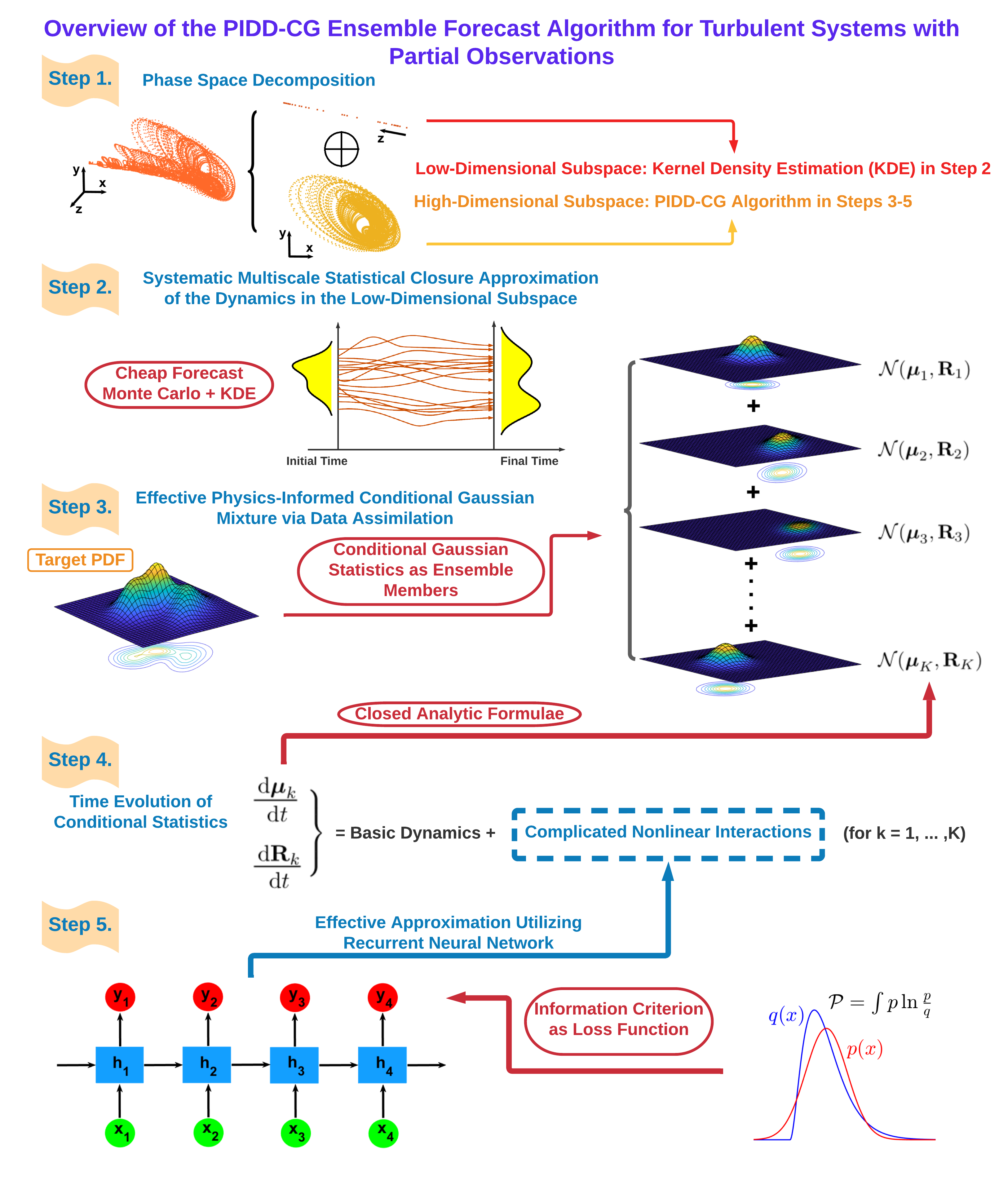}
\caption{Schematic diagram of the PIDD-CG algorithm for predicting the PDF of the high-dimensional complex turbulent systems.  }\label{Algorithm}
\end{figure}

The general framework of the PIDD-CG algorithm is schematically illustrated in Figure \ref{Algorithm}, which consists of five key steps.
\paragraph{Step 1. Phase space decomposition.} A phase space decomposition is carried out  to project the state variables into two subspaces admitting conditional Gaussian structures \cite{chen2018conditional}. A collection of the leading  state variables $\mathbf{X}$ (such as the large-scale, resolved or observed states) belongs to this relatively low-dimensional subspace. The rest of the variables $\mathbf{Y}$, which are multiscale, unresolved or unobserved, are contained in the remaining high-dimensional subspace.
\paragraph{Step 2. Systematic multiscale statistical closure of the large-scale dynamics.} A systematic multiscale statistical closure approximation is proposed to avoid running the original full-dimensional system when predicting the statistics of the low-dimensional variable $\mathbf{X}$ that is fully coupled with $\mathbf{Y}$.  The statistical closure of the equations of $\mathbf{X}$ depends explicitly on the conditional mean of only a few modes of $\mathbf{Y}$ conditioned on realizations of $\mathbf{X}$, which can be solved via closed analytic formulae (see \emph{Step 4}), while the residual part is effectively approximated by a RNN (see \emph{Step 5}). Such a unique way of building the closure model allows to forecast the statistics of $\mathbf{X}$ from an intrinsically low-dimensional subsystem, which requires only a small number of samples. Denote such a number by $J$. The forecast PDF of the low-dimensional variable $\mathbf{X}$ is then approximated by a kernel density estimation \cite{botev2010kernel} using Gaussian kernels.
\paragraph{Step 3. Effective physics-informed conditional Gaussian mixture via data assimilation.} Given the physical model formulation conditioned on each of the $J$ forecast trajectories of $\mathbf{X}$ from \emph{Step 2}, the conditional distribution of $\mathbf{Y}$ can be computed via a nonlinear data assimilation method \cite{chen2018conditional}. These $J$ conditional distributions are the $J$ conditional statistics ensemble members. Notably, the center and the bandwidth associated with each conditional statistics ensemble member are automatically optimized by the nonlinear data assimilation that takes into account the model physics. Because of this, the resulting mixture distribution consisting of the conditional statistics ensemble avoids the curse of dimensionality with respect to the number of mixture components \cite{chen2018rigorous}. In addition, since each sample of $\mathbf{Y}$ is conditioned on one realization of $\mathbf{X}$, the cross-correlation between $\mathbf{X}$ and $\mathbf{Y}$ is also captured when forming their joint distribution.
\paragraph{Step 4. Analytic formulae for the time evolution of the conditional statistics in smaller-scale dynamics.} One important feature of the PIDD-CG algorithm  is that the time evolution of each conditional statistics ensemble member from \emph{Step 3} can be solved via closed analytic formulae \cite{liptser2013statistics}. Thus, it avoids using the expensive MC methods for finding such data assimilation solutions and prevents the sampling errors when handling high-dimensional systems.
\paragraph{Step 5.  Data-driven modeling of the nonlinear feedbacks in conditional statistics with  information theory.} Despite the closed analytic formulae, the computational cost of running the evolution equations of the conditional statistics, especially the time evolution of the conditional covariance, can still be demanding for high-dimensional systems. To improve computational efficiency, a model reduction strategy is applied to approximate certain complicated nonlinear components in the unresolved fluctuation modes feedbacks using a recurrent neural network (RNN). Since the output variables are associated with the conditional distribution, a simple but effective information loss function \cite{qi2020using} is adopted as a natural metric to train the RNN.\medskip

The PIDD-CG algorithm also allows an efficient and accurate data assimilation scheme to obtain the conditional statistics ensemble at the forecast initialization stage (see the \emph{Methods} Section \ref{sec:method}), which facilitates the application of the algorithm to the more realistic situations with only partial observations.

\subsection{The topographic barotropic model}\label{Sec:baro_model}
The topographic barotropic flow is a prototype model in geophysics \cite{majda2006nonlinear}, which involves multiscale interactions and transport among the zonal mean flow and the fluctuations. It also contains many key features of interests in turbulence, such as emerging non-Gaussian PDFs, regime switching and extreme events.

The spectral formulation of the model with layered topography reads (see SI for the derivations):
\begin{subequations}\label{eq:topo_model_main}
\begin{align}
\frac{dU}{dt}= & \sum_{k}\hat{h}_{k}^{*}\hat{v}_{k}\:-d_{0}U+\sigma_{0}\dot{W}_{0}\label{eq:topo_model_U_main},\\
\frac{d\hat{v}_{k}}{dt}= & \left[-\gamma_{v,k}\left(U\right)+i\omega_{v,k}\left(U\right)\right]\hat{v}_{k}-l_{x}^{2}\hat{h}_{k}U\:-d_{v,k}\hat{v}_{k}+\sigma_{v,k}\dot{W}_{k},\label{eq:topo_model_v_main}\\
\frac{d\hat{T}_{k}}{dt}= & \left[-\gamma_{T,k}\left(U\right)+i\omega_{T,k}\left(U\right)\right]\hat{T}_{k}\:-d_{T,k}\hat{T}_{k}-\alpha\hat{v}_{k}.\label{eq:topo_model_T_main}
\end{align}
\end{subequations}
In \eqref{eq:topo_model_main}, the wavenumbers are given by $k\mathbf{l},k=\pm 1,\ldots, \pm K$, expanded along one characteristic direction $\mathbf{l}=(l_x,l_y)$ with $\lvert\mathbf{l}\rvert=1$. The state variable $U$ represents the large-scale zonal mean flow velocity while $\hat{v}_k$, $\hat{T}_k$ and $h_k$ are the coefficients of the $k$-th Fourier modes corresponding to the fluctuation components of the flow velocity $v$, the turbulent transport of passive tracer field $T$, and the topography $h$, respectively.  The notation $\cdot^*$ stands for the complex conjugate while $\dot{W}_{0}$ and $\dot{W}_{k}$ are independent white noise sources with strengths $\sigma_{0}$ and $\sigma_{k}$. The model parameters $\gamma_{v,k},\gamma_{T,k}$ and $\omega_{v,k},\omega_{T,k}$ represent the dispersion and dissipation effects. The details  of the parameter values are included in the SI.

The mean flow $U$ is driven by the topographic stress from $h$ combining all the feedbacks from the fluctuations $v$, while $U$ also inversely contributes to each spectral mode through the nonlinear advection and topographic effect. The coupling between the mean flow and the fluctuations is through the topographic stress.
The PIDD-CG algorithm is applied to two regimes with distinct dynamical and statistical features.  Depending on the statistical equilibrium distributions of $U$, $\hat{v}_k$ and $\hat{T}_k$ \cite{qi2017low}, these two regimes are named as:
\begin{itemize}
\item \emph{Strongly non-Gaussian regime:} The zonal mean flow $U$ is driven by strong white noise forcing while only small noises are added to the fluctuation modes $\hat{v}_{k}$.
\item \emph{Near-Gaussian regime:} The fluctuation modes $\hat{v}_{k}$ are subject to strong white noise forcing while the  noise strength in the zonal mean flow $U$ is relatively weak.
\end{itemize}

\begin{figure}[h]
    \noindent\begin{minipage}{1.00\linewidth}
        \includegraphics[width=1.0\linewidth]{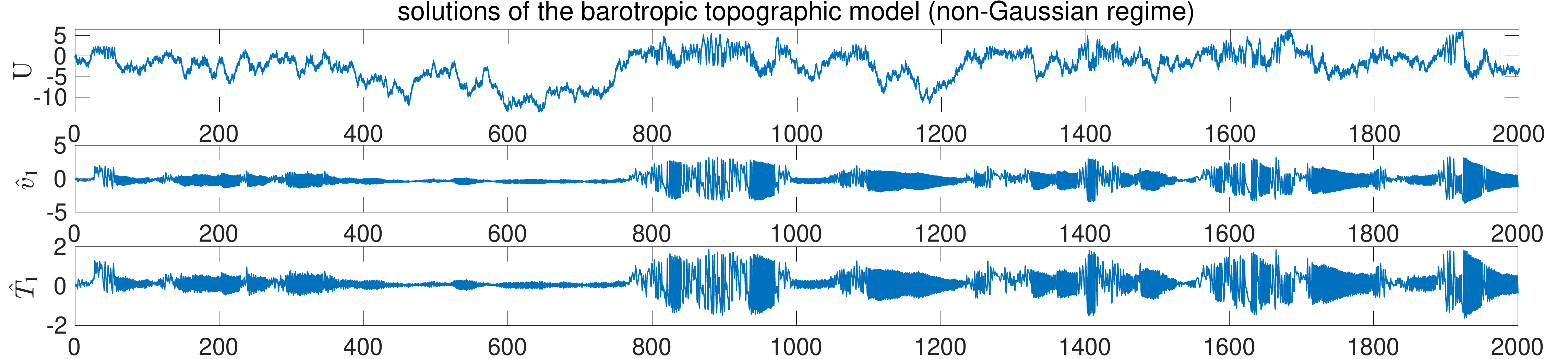}\vspace{-1.em}
        \subfloat[non-Gaussian regime]{\includegraphics[width=1.0\linewidth]{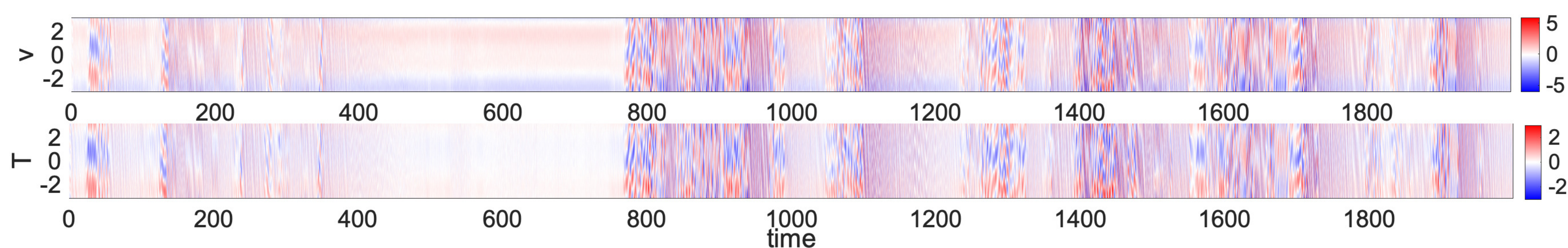}}
    \end{minipage}
    \hfill\vspace{0.5em}
    \noindent\begin{minipage}{1.00\linewidth}
        \includegraphics[width=1.0\linewidth]{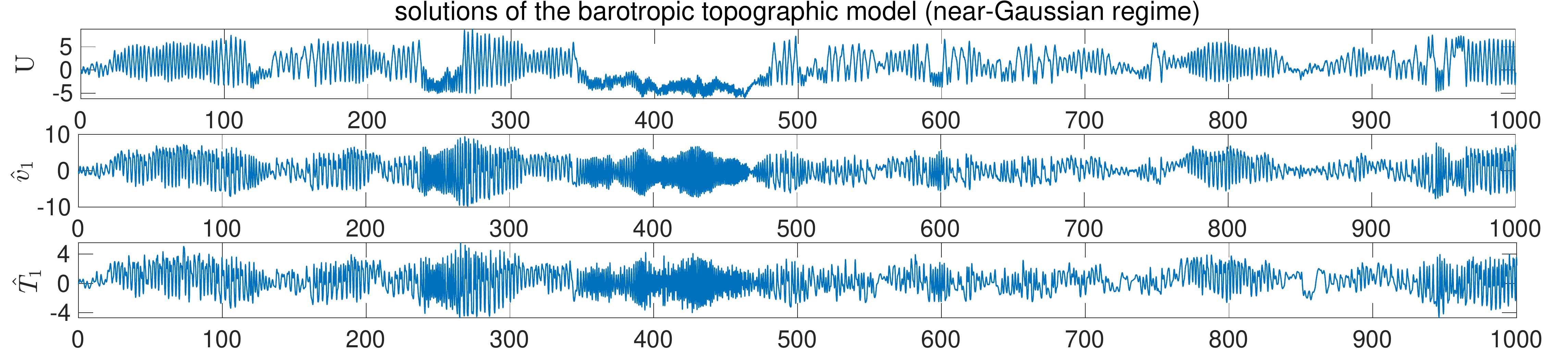}\vspace{-1.2em}
        \subfloat[near-Gaussian regime]{\includegraphics[width=1.0\linewidth]{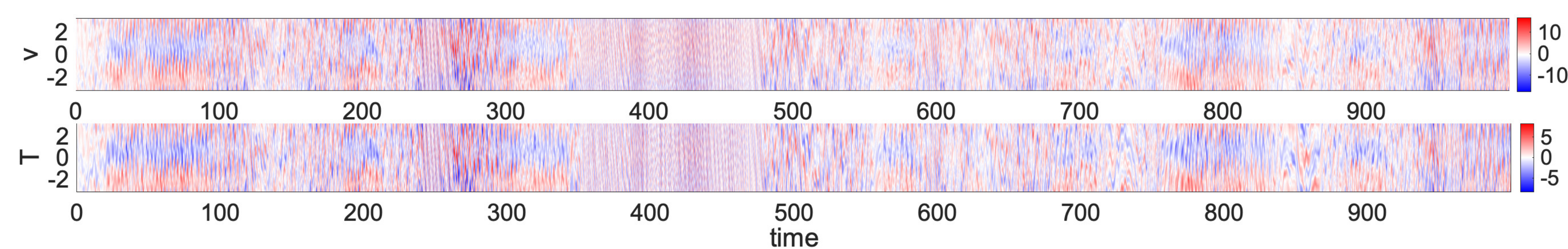}}
    \end{minipage}%
\caption{Solutions of the barotropic topographic model with in total 41 modes (i.e., $K=10$). Panel (a) shows the solution in the non-Gaussian regime while Panel (b) shows that in the near-Gaussian regime. In each panel, the zonal velocity $U$, the real parts of the first fluctuation mode $\hat{v}_1$ and $\hat{T}_1$, and the entire fields of $v$ and $T$ in physical space as a function of time are presented. }\label{fig:Trajectory-solution}\end{figure}

Figure \ref{fig:Trajectory-solution} illustrates dynamical features of both regimes.  In the non-Gaussian regime, the large white noise forcing in $U$ excites a strong competition between two alternating states: a highly intermittent flow field $v$ when the eastward jet appears ($U>0$) and a nearly steady flow when the westward jet occurs ($U<0$). The intermittent nature of the flow field triggers  non-Gaussian fat-tailed distributions of $v$ and $T$ (see e.g., the blue curves in Figure \ref{fig:PDFs}). On the other hand, in the near-Gaussian regime, strongly multiscale features emerge in the time series of all the variables, which include multiple fast scales with rapid oscillations and a slowly varying long-term tendency. In particular, there are two distinct dominant frequencies in the fast oscillations. The extremely fast oscillation appears when the zonal flow goes steadily towards the west ($U<0$) while the moderately fast one occurs when the zonal  flow becomes intermittent with an average eastward velocity ($U>0$). In contrast to the non-Gaussian regime, the near-Gaussian statistics in this regime is due to the comparable amplitude of $\hat{v}_k$ and $\hat{T}_k$ at different states.

In the following forecast tests, the zonal mean flow $U$ is assumed to be the only observed variable.

\subsection{Predicting key PDFs in the barotropic topographic model using the PIDD-CG algorithm}
In \emph{Step 1} of the PIDD-CG algorithm described in Section \ref{Subsec:algorithm}, the entire phase space is decomposed into a low-dimensional subspace that contains only the observed variable, namely the zonal mean flow $U$, and a high-dimensional subspace that includes all the fluctuation modes $\hat{v}_k$ and $\hat{T}_k$ for $k=\pm1,\ldots, \pm K$.
In \emph{Step 2}, only the explicit equations of the conditional means of the leading two complex modes of $\hat{v}_{k}$ (e.g., $k=\pm1$ and $\pm2$) are utilized to build the closure model of $U$ while all the remaining small-scale feedbacks are automatically learned by the RNN. This accounts for a strongly reduced model of 2 resolved modes compared with the full model with $K=10$.  \emph{Steps 3-5} follow directly the description in Section \ref{Subsec:algorithm} with the detailed step-by-step explanations being included in the Method Section and the SI.

\begin{figure}[h!]
    \noindent\begin{minipage}{1.00\linewidth}
        \subfloat[non-Gaussian regime]{\includegraphics[width=0.9\linewidth]{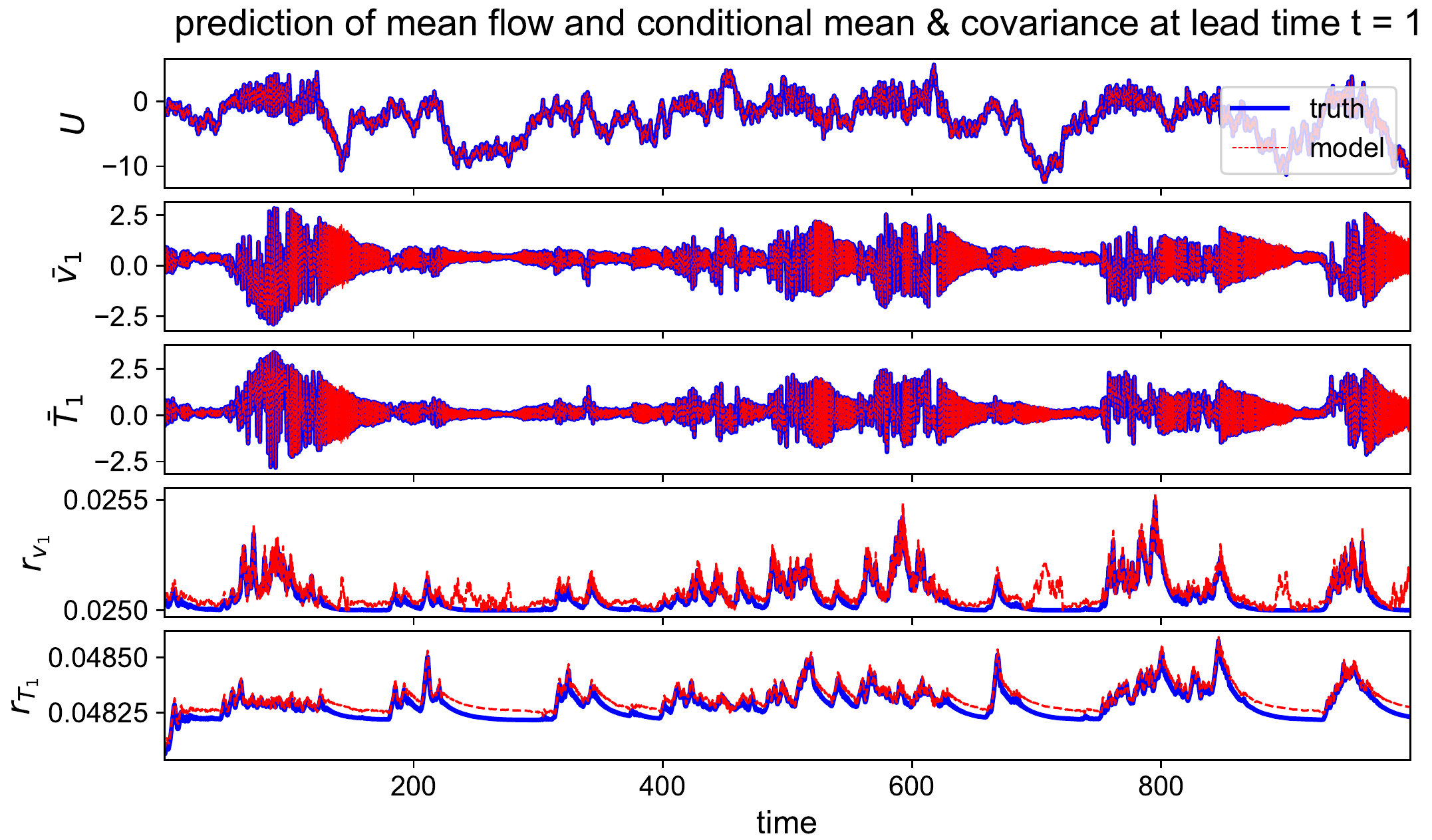}}
    \end{minipage}%
    \hfill\vspace{0.5em}
    \noindent\begin{minipage}{1.00\linewidth}
        \subfloat[near-Gaussian regime]{\includegraphics[width=0.9\linewidth]{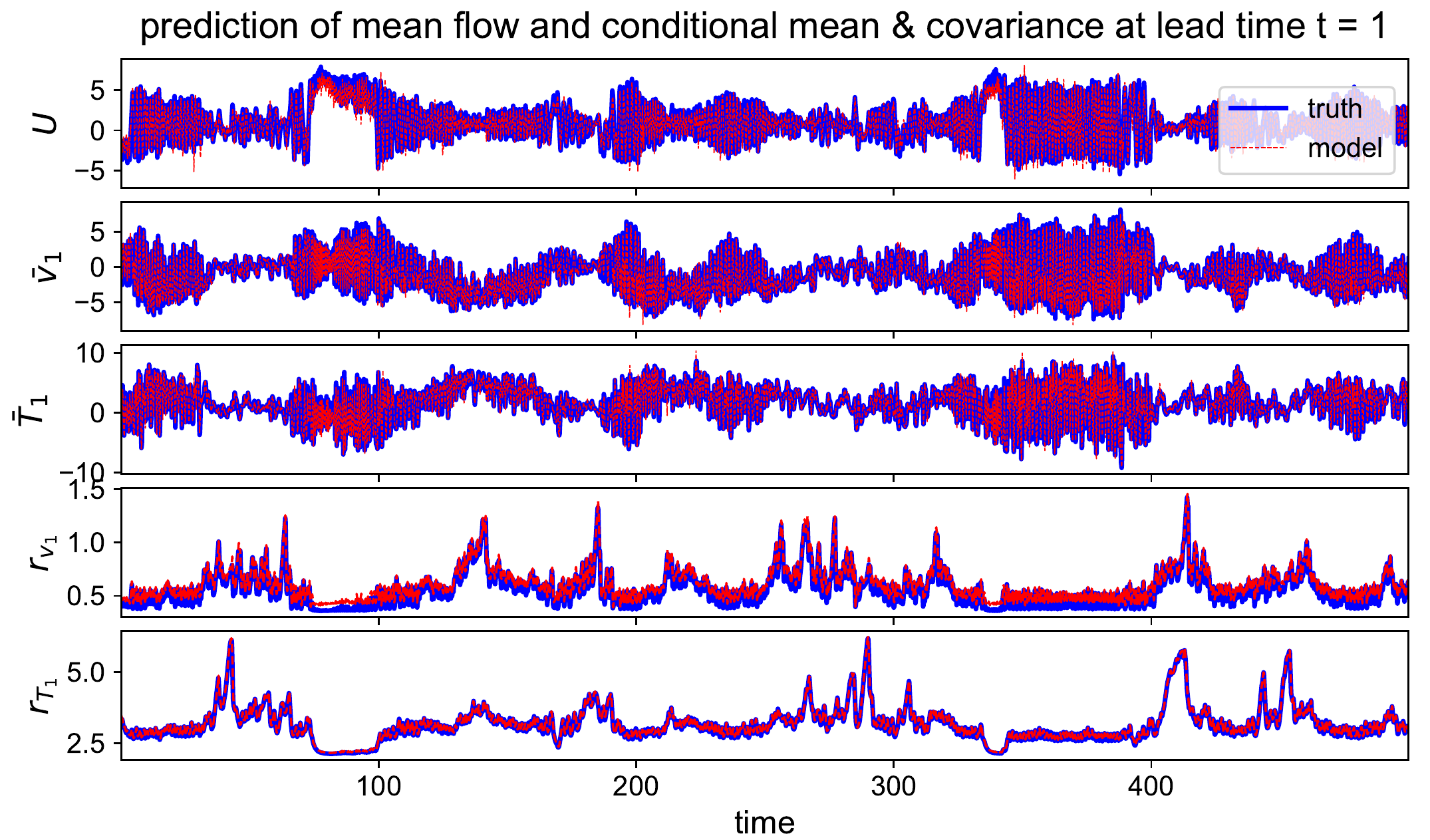}}
    \end{minipage}%
\caption{Forecast at lead time $t=1$ of the trajectories for the zonal mean flow $U$ as well as the conditional mean and conditional variance of the first fluctuation modes $\bar{v}_{1},\bar{T}_{1}$ and $r_{v_{1}},r_{T_{1}}$. Panel (a) shows the solutions in the non-Gaussian regime while Panel (b) shows those in the near-Gaussian regime. }\label{fig:Leading-time-prediction}
\end{figure}

Since the joint PDF from the PIDD-CG algorithm is given by a mixture distribution, where each mixture component is uniquely determined by $U$ and the conditional statistics of $\hat{v}_k$ and $\hat{T}_k$, it is natural to start with the study of the forecast of these quantities.
Figure \ref{fig:Leading-time-prediction} compares the forecast trajectories at the lead time $t=1$ with the truth. The lead time forecast here means each point in the forecast trajectory is the forecast value starting from $1$ unit prior to it. Such a lead time is around the decorrelation time of the first a few modes of $\hat{v}_k$ and $\hat{T}_k$ and is comparable with the time scale of the fast component of $U$ (see the SI).  
Note that the forecast at a lead around the decorrelation time of a turbulent system is extremely challenging if the entire dynamics is fully approximated by the neural network due to the quick accumulation of errors \cite{qi2020using,qi2021machine}. 
Nevertheless, basic dynamics are retained in the PIDD-CG algorithm while the neural network plays the supplementary role of only modeling the unresolved residual part. Therefore, good agreements between the truth and the forecast are achieved in the zonal mean flow $U$ as well as the conditional statistics. In particular, the multiscale structures are accurately predicted in both regimes despite the small errors in forecasting the highly oscillating small scales, which are the unpredictable part due to the turbulent nature. More importantly, both the locations and the amplitudes of the dominant intermittent features are successfully predicted by the PIDD-CG algorithm, which is a central prerequisite for accurately predicting the joint PDFs.

\begin{figure}

\begin{minipage}{1.00\textwidth}
\begin{minipage}{0.49\linewidth}
\includegraphics[width=1\linewidth]{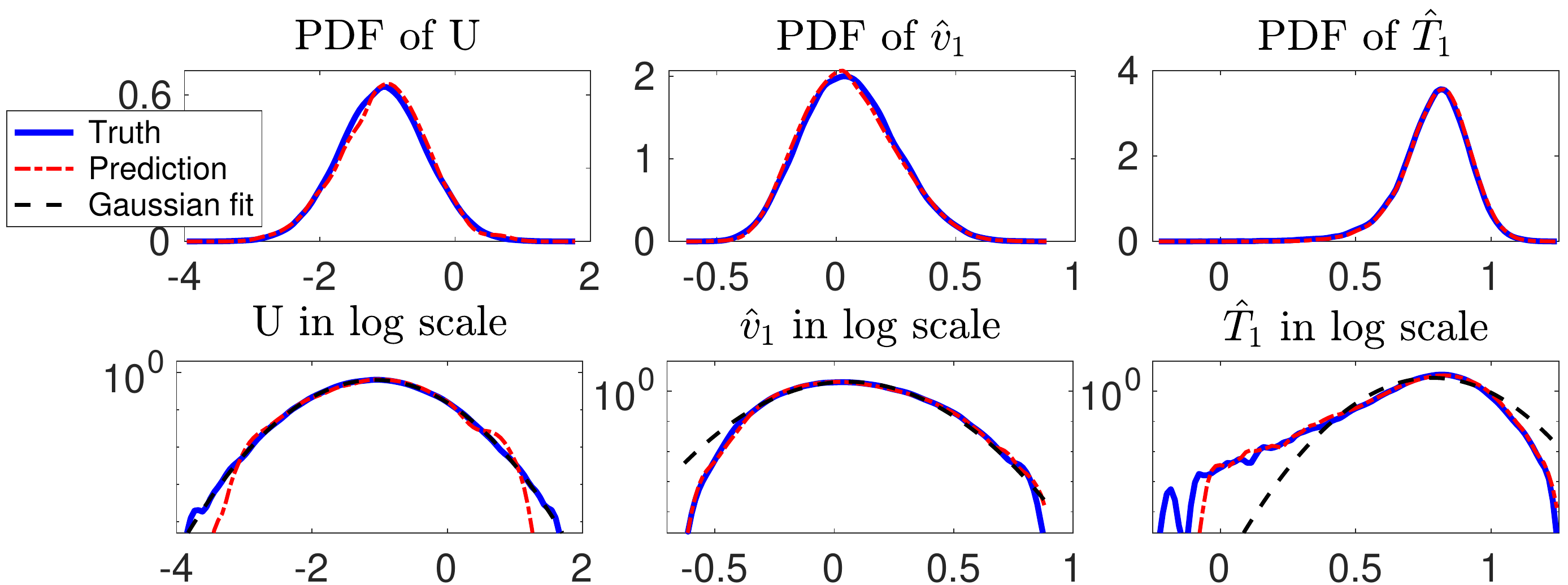}\\
        \hspace*{3em}\subfloat[non-Gaussian regime, \\ transient state at lead $t=1$]{\hspace*{-3em}\includegraphics[width=1\linewidth]{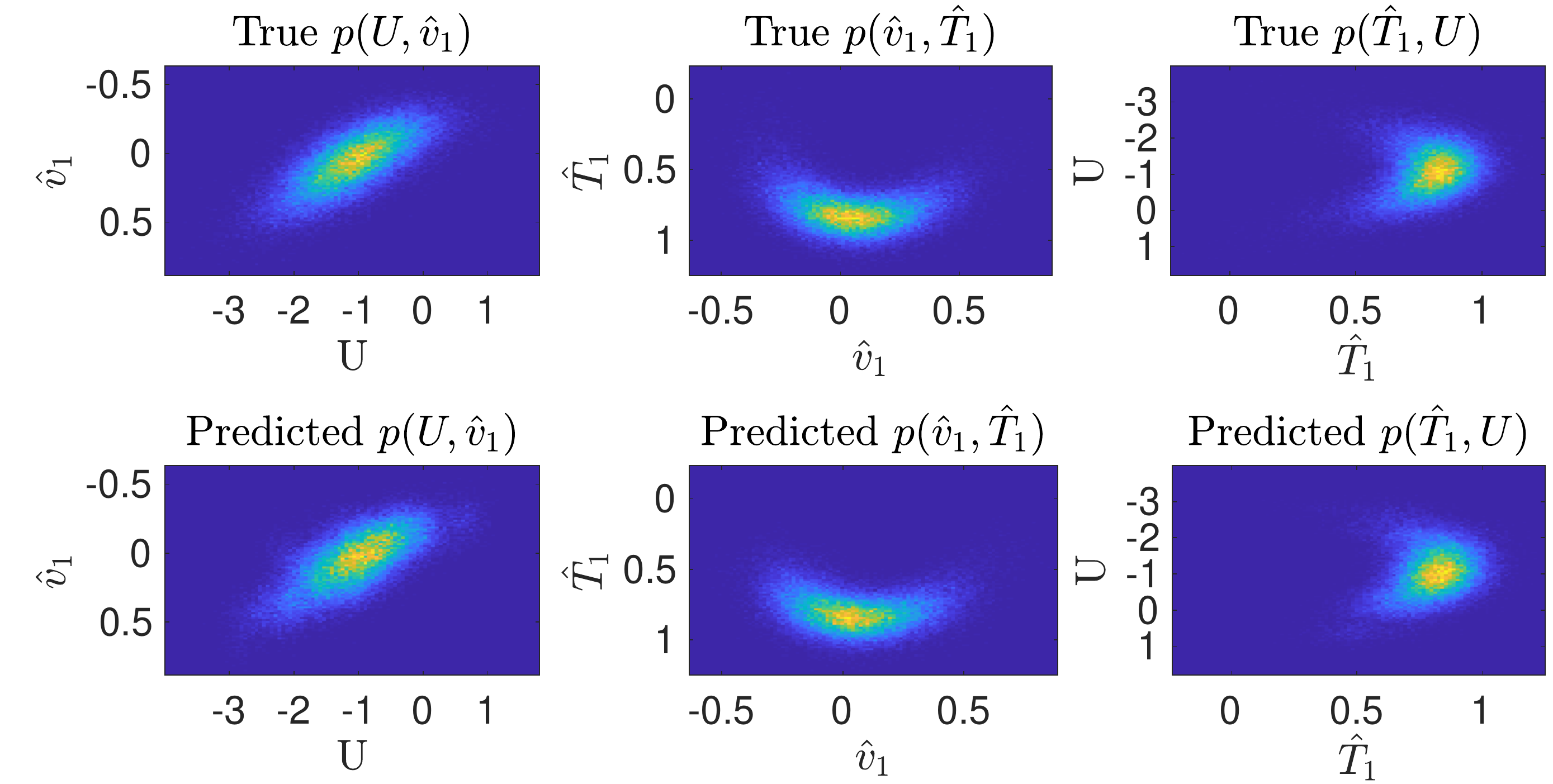}}
    \end{minipage}%
    \begin{minipage}{0.49\linewidth}
    \includegraphics[width=1\linewidth]{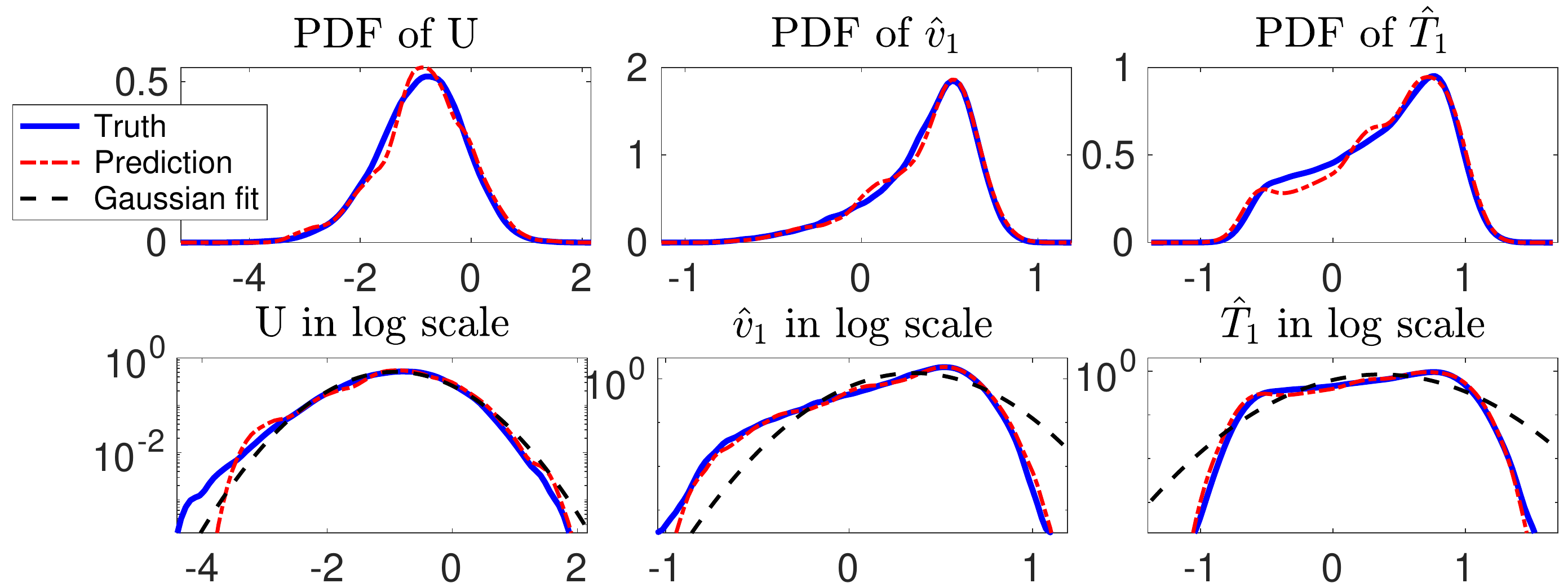}\\
        \hspace*{3em}\subfloat[non-Gaussian regime, \\equilibrium state at lead $t=2$]{\hspace*{-3em}\includegraphics[width=1\linewidth]{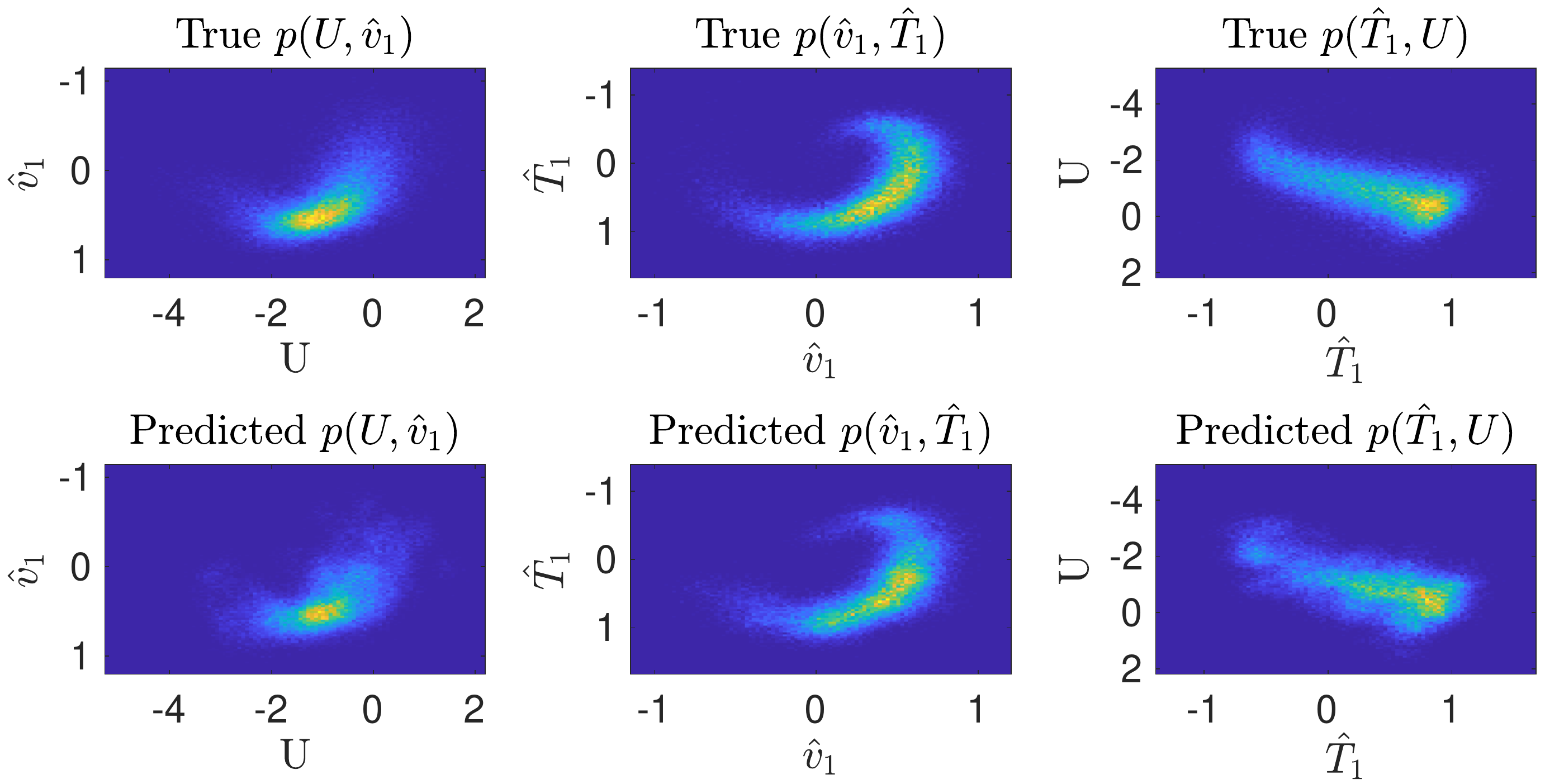}}
    \end{minipage}%
\end{minipage}
\begin{minipage}{1.00\textwidth}
\begin{minipage}{0.49\linewidth}
\includegraphics[width=1\linewidth]{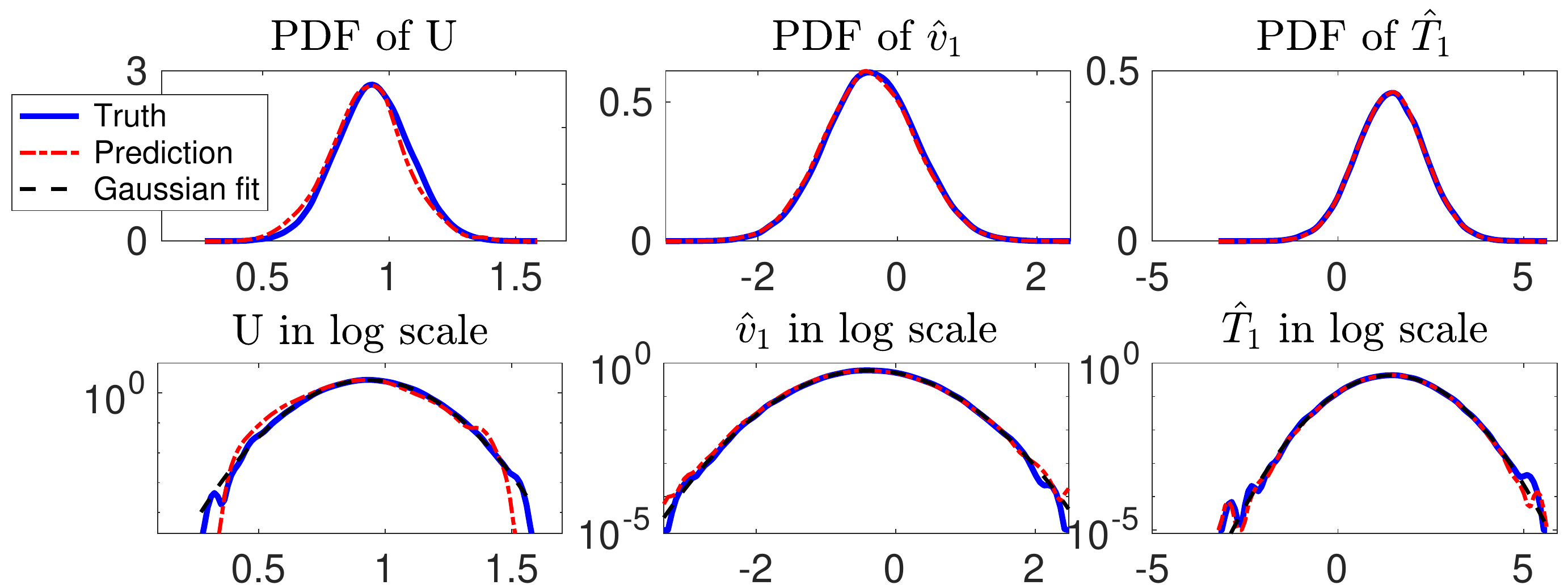}\\
        \hspace*{3em}\subfloat[near-Gaussian regime, \\ transient state at lead $t=0.2$]{\hspace*{-3em}\includegraphics[width=1\linewidth]{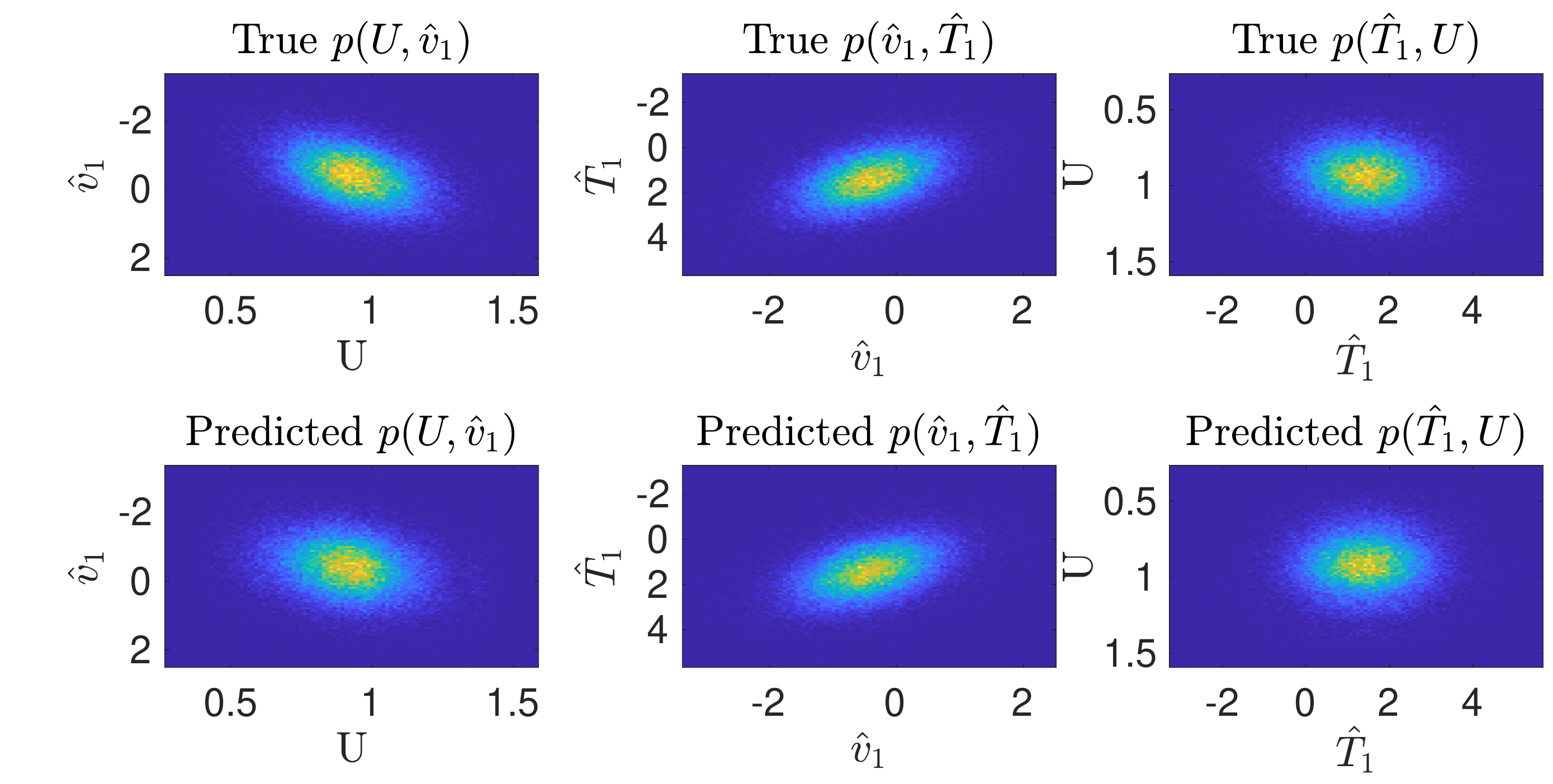}}
    \end{minipage}%
    \begin{minipage}{0.49\linewidth}
    \includegraphics[width=1\linewidth]{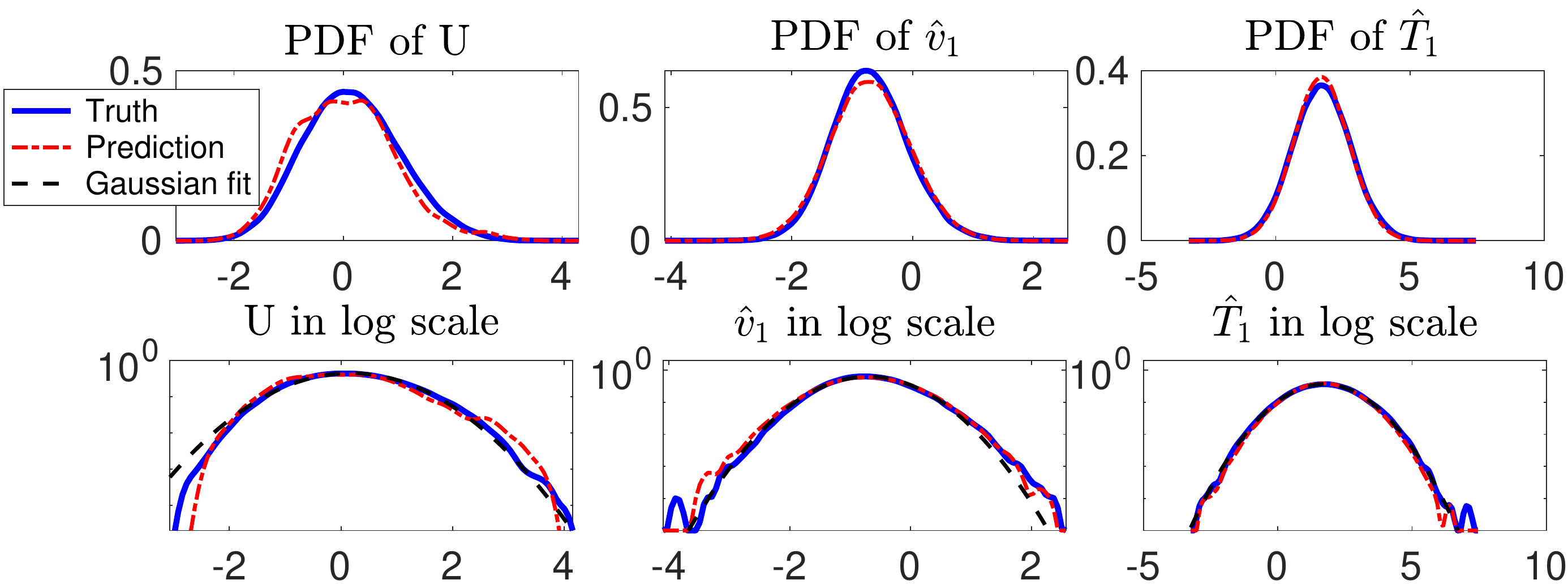}\\
        \hspace*{3em}\subfloat[near-Gaussian regime, \\equilibrium state at lead $t=1$]{\hspace*{-3em}\includegraphics[width=1\linewidth]{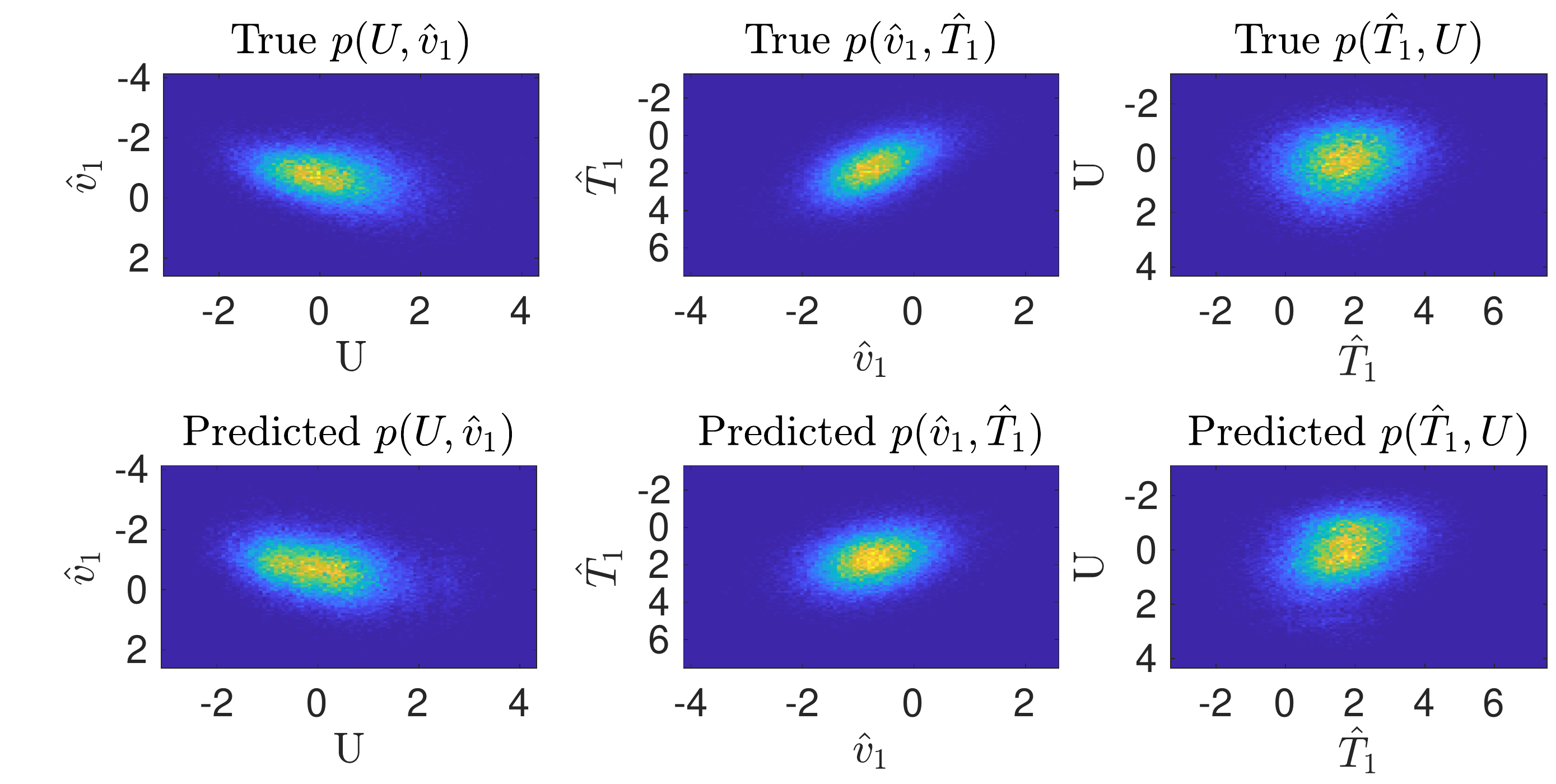}}
    \end{minipage}%
\end{minipage}
\caption{Comparison of the truth and the forecast PDFs of the zonal mean flow $U$ and the first fluctuation modes  $\hat{v}_{1},\hat{T}_{1}$. The truth is generated from a direct Monte-Carlo simulation with $J_\mathrm{MC}=50000$ ensembles, while the prediction from the PIDD-CG algorithm only uses $J=100$ ensembles. The prediction at both a transient state and a nearly statistical equilibrium state are compared in both regimes. To better demonstrate the non-Gaussianity of each variable, the comparison of the one-dimensional PDFs in the logarithm scale is also included. Note that the ranges of $x$-axes for the same variable are different with each other in two panels, representing distinct features in the transient and the nearly statistical equilibrium states.} \label{fig:PDFs}
\end{figure}

Figure \ref{fig:PDFs} shows the main prediction results for the true PDFs and the forecast ones. The forecast starts from one specific observed initial value of $U$ while the initial values of $\hat{v}_k$ and $\hat{T}_k$ are recovered from the efficient and accurate data assimilation algorithm, which is described in the Method Section \ref{sec:method}. In each panel, the one-dimensional PDFs of the zonal mean flow $U$ and the leading fluctuation modes $\hat{v}_{1}$ and $\hat{T}_{1}$ as well as the two-dimensional joint PDFs between these variables are presented. A complete comparison of the second fluctuation modes and other joint distributions at several different lead times can be found in the SI. Here, the true PDFs are generated from a direct MC simulation of the topographic barotropic flow model \eqref{eq:topo_model_main}. The MC simulation contains $J_\mathrm{MC}=50000$ samples to fully characterize the non-Gaussian statistics, which is computationally very expensive. In contrast, the PIDD-CG algorithm exploits a much smaller ensemble size with only $J=100$ samples.

For the purpose of illustrating the skill of the PIDD-CG algorithm for predicting the entire time evolution of the system, Figure \ref{fig:PDFs} includes the predicted PDFs at both a short-term transient phase and the nearly final saturated statistical equilibrium state. The non-Gaussian regime has a longer mixing time with the statistical equilibrium being reached at around the lead time $t=2$, while the solution in the near-Gaussian regime mixes faster and reaches the statistical equilibrium state within $t=1$. The PIDD-CG algorithm succeeds in capturing the transient PDFs as well as the final equilibrium state in both regimes containing distinct statistics. Particularly, the highly  skewed and fat-tailed PDFs in the non-Gaussian regime are accurately reproduced by the PIDD-CG algorithm.  Table \ref{tab:Info-error-topo} includes a quantitative assessment of predicting the one-dimensional PDFs. It shows the relative error, quantified by an information measurement (the relative entropy) \cite{kleeman2011information, majda2006nonlinear}, at different forecast lead times before the system reaches the statistical equilibrium. The forecast error remains in a negligible level at the order of $O(10^{-3})$ in most cases, and the biggest error occurs at the longest lead time, which is nevertheless at most of order $O(10^{-2})$. Among different variables, the predicted zonal mean flow $U$ shows a slightly larger error than the fluctuation modes, which is due to the relatively more severe approximation in $U$. In fact, only the conditional means of the leading two fluctuation modes are explicitly included in the development of the closure model of $U$ while the combined feedback from the remaining multiple fluctuation modes is completely approximated by the RNN.  Nevertheless, the error in predicting $U$ lies in an overall low level, which justifies the strategy in the PIDD-CG algorithm that combines the conditional mean time series with the RNN in facilitating the statistical closure of $U$.

\begin{table}
\begin{tabular}{ccccccccc}
\toprule
 & \multicolumn{4}{c}{non-Gaussian regime} &  & \multicolumn{3}{c}{near-Gaussian regime}\tabularnewline
\midrule
 & $t=0.5$ & $t=1$ & $t=1.5$ & $t=2$ &  & $t=0.2$ & $t=0.5$ & $t=1$\tabularnewline
\midrule
\midrule
$U$ & $6.80\times10^{-3}$ & $4.74\times10^{-3}$ & $7.55\times10^{-3}$ & $1.76\times10^{-2}$ &  & $1.58\times10^{-2}$ & $1.68\times10^{-2}$ & $1.85\times10^{-2}$\tabularnewline
\midrule
$\hat{v}_{1}$ & $4.56\times10^{-4}$ & $1.68\times10^{-3}$ & $4.09\times10^{-3}$ & $4.93\times10^{-3}$ &  & $7.13\times10^{-4}$ & $4.76\times10^{-3}$ & $4.06\times10^{-3}$\tabularnewline
\midrule
$\hat{v}_{2}$ & $2.89\times10^{-4}$ & $4.19\times10^{-4}$ & $9.74\times10^{-4}$ & $4.48\times10^{-3}$ &  & $1.44\times10^{-3}$ & $1.27\times10^{-2}$ & $4.14\times10^{-2}$\tabularnewline
\midrule
$\hat{T}_{1}$ & $9.21\times10^{-4}$ & $4.52\times10^{-3}$ & $5.11\times10^{-3}$ & $9.58\times10^{-3}$ &  & $4.03\times10^{-4}$ & $3.28\times10^{-3}$ & $2.55\times10^{-3}$\tabularnewline
\midrule
$\hat{T}_{2}$ & $3.40\times10^{-4}$ & $4.05\times10^{-4}$ & $4.08\times10^{-4}$ & $2.91\times10^{-3}$ &  & $4.74\times10^{-4}$ & $1.41\times10^{-3}$ & $2.21\times10^{-3}$\tabularnewline
\bottomrule
\end{tabular}

\caption{Information error (relative entropy) between the truth and forecast PDF in $U, \hat{v}_1, \hat{v}_2, \hat{T}_1$ and $\hat{T}_2$ of the barotropic topographic
model at different lead time before arriving at the statistical equilibrium state.\label{tab:Info-error-topo}}

\end{table}

\section{Discussion}

\begin{figure}
\subfloat{\includegraphics[scale=0.5]{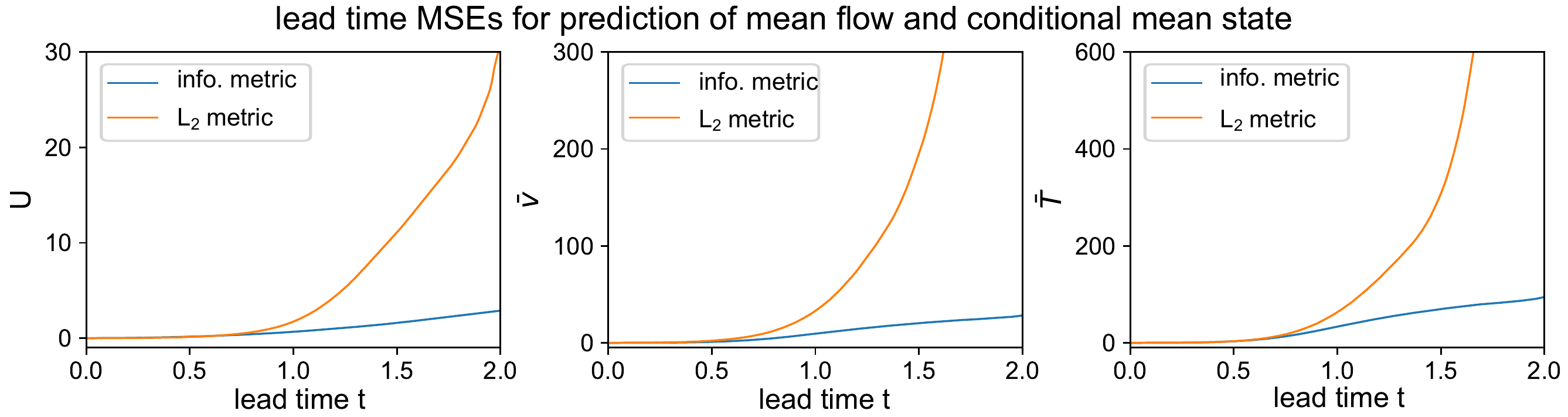}}

\subfloat{\includegraphics[scale=0.5]{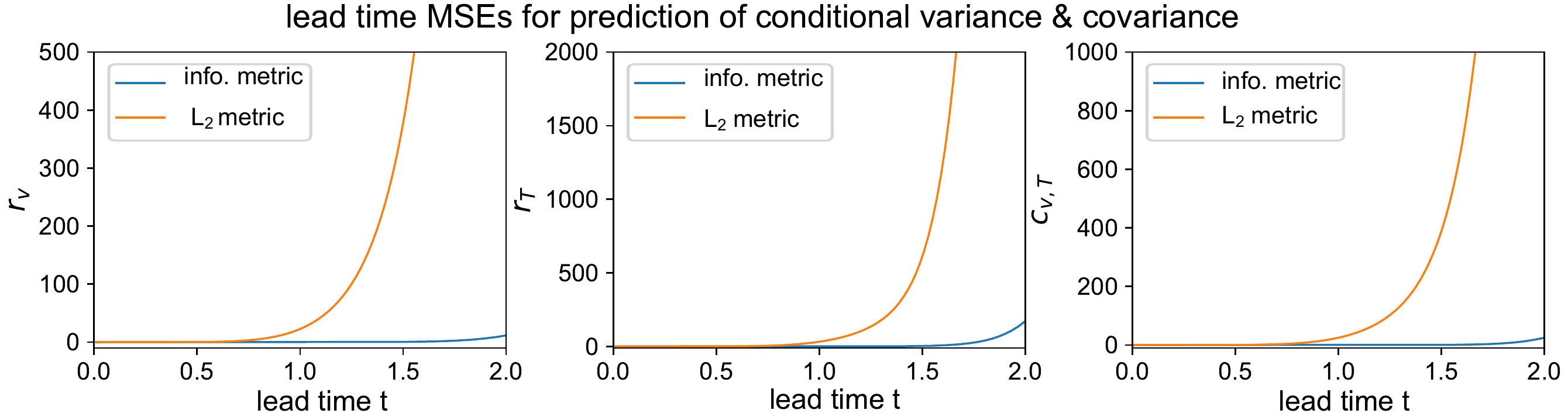}}

\caption{Prediction errors using the RNN optimized under different
loss functions in the non-Gaussian regime. The mean square errors (MSEs) in forecasting $U$, the conditional mean $\bar{v}_1, \bar{T}_1$, the conditional variance $r_{v_1}, r_{T_1}$ and the conditional cross-covariance $C_{v_1,T_1}$ under the $L_{2}$ loss and the information loss (relative entropy) are compared as a function of the lead time $t$.\label{fig:Prediction-errors-optimized}}
\end{figure}

Different from the traditional ensemble forecast methods, the PIDD-CG algorithm incorporates the evolution equations of the conditional statistics as part of the forecast scheme. A data-driven component using RNNs is incorporated into the method to facilitate an efficient estimation of the combined fluctuation feedback in the observed state and the multiscale fluctuation modes in the unobserved high-dimensional subspace. Since the learning target has become a conditional distribution, an information metric is utilized as the loss function to train the RNN, which is a natural choice to avoid unnecessary fluctuating errors in turbulent signals. Yet, a quantitative assessment of the advantage of using such an information metric compared with the traditional $L^2$ loss function is still needed. To this end, Figure \ref{fig:Prediction-errors-optimized} illustrates the forecast skill of the RNN in the non-Gaussian regime trained by  these two different losses during the optimization process. Due to the strong turbulent nature of the system, the errors accumulate in time as the model is updated recurrently. Nevertheless, the forecast error using the RNN trained with the information loss grows much slower than that using the RNN optimized via the $L^2$ loss. In fact, there is already an intrinsic barrier in the training phase if the $L^2$ loss is utilized, which indicates that the path-wise measurement is not the most appropriate choice in minimizing the statistical error. 
In particular, it is noticeable from Figure \ref{fig:Prediction-errors-optimized} that the prediction of the conditional mean using the information metric remains accurate even at a very long lead time. This is crucial for accurately recovering of the joint PDFs, as the conditional mean explicitly appears in the closure term in the $U$ equation.

Finally, it is worthwhile to point out that predicting the full spectrum of the state variables, especially in high-resolution systems, is not computational feasible. The primary interest often lies in predicting the large-scale coherent structures. Therefore, developing computationally efficient reduced order models with suitable parameterizations and forecasting the statistics of the leading a few resolved modes, which still correspond to a relatively high-dimensional PDF for direct numerical simulations, have more practical significance in more realistic applications \cite{majda2018strategies, ahmed2021closures, khodkar2018data, santos2021reduced, chekroun2020variational, berner2017stochastic, kondrashov2015data, schneider2021learning, edeling2020reducing}.
The PIDD-CG algorithm can facilitate the development of efficient statistical reduced order models as well as accelerating the associated statistical forecast. Specifically, a large portion of the state variables in $\mathbf{Y}$ can be truncated and only those that are utilized explicitly in the closure of $\mathbf{X}$ in \emph{Step 2} are preserved in the reduced order model. The contribution from the truncated modes in $\mathbf{Y}$ can be effectively approximated by RNNs, which are then added to the equations of both $\mathbf{X}$ and the resolved components of $\mathbf{Y}$ as the additional closure terms. Note that, in applying the PIDD-CG algorithm to the topographic barotropic flow model \eqref{eq:topo_model_main}, the statistical evolutions of the pairs $(\hat{v}_k, \hat{T}_k)$ and $(\hat{v}_{k'}, \hat{T}_{k'})$ with $\vert k\vert \neq \vert k'\vert $ do not explicitly influence each other but they are coupled  through the  zonal mean flow $U$ (see the SI). Therefore, the resolved subsystem consisting of these variables serves as a natural reduced order formulation for predicting the joint PDF of $U$ and $(\hat{v}_k, \hat{T}_k)$ with $k=\pm 1$ and $\pm 2$. In fact, the forecast PDF from such a 9-dimensional reduced order model is exactly the same as the associated 9-dimensional marginal distribution from the 41-dimensional model when applying the PIDD-CG algorithm to the topographic barotropic model \eqref{eq:topo_model_main}, due to the lack of explicit dependence between different $(\hat{v}_k, \hat{T}_k)$ pairs.

\section{Methods}\label{sec:method}
\subsection{General mathematical framework of the PIDD-CG  algorithm}

In this section, we present the general strategy of the PIDD-CG  algorithm step-by-step, following the schematic illustration in Figure \ref{Algorithm}. In the PIDD-CG algorithm, high computational efficiency is achieved by avoiding the direct MC sampling in the whole high-dimensional phase space of the entire system. After a proper phase space decomposition (\emph{Step 1}), the ensemble sampling is performed only inside a low-dimensional subspace, where a systematic statistical closure approximation has been applied in the leading-order states (\emph{Step 2}). On the other hand, an effective physics-informed data-driven estimation is adopted to advance the conditional statistics forecast in the remaining high-dimensional subspace (\emph{Step 3}), which allows an accurate recovery of the conditional statistics only relying on the small number of samples. In particular, the ensemble size is independent of the full dimension of the phase space \cite{chen2018rigorous}, and thus the curse of dimensionality is avoided. 
The ensemble members in the PIDD-CG algorithm consist of a conditional Gaussian mixture,  which can be tracked by closed analytic formulae for the conditional mean and covariance (\emph{Step 4}). Finally, a RNN is introduced to approximate the nonlinear unresolved-scale feedbacks in the analytic expressions of the conditional statistics as well as the statistical feedback in model closure in \emph{Step 2} and \emph{3}. In particular, since the model outputs are associated with the conditional distribution, a simple but effective information loss function is adopted to train the RNN (\emph{Step 5}). The combination of the physics-informed analytic time evolution of the statistics and the data-driven RNN closure of unresolved multiscale feedbacks accelerates the overall computational efficiency and model accuracy by a significant amount.

We start with the general formulation of turbulent dynamical systems \cite{vallis2017atmospheric, salmon1998lectures, kalnay2003atmospheric, majda2016introduction},
\begin{equation}\label{eq:abs_formu}
\frac{\mathrm{d}\mathbf{u}}{\mathrm{d}t}=\left(L+D\right)\mathbf{u}+B\left(\mathbf{u},\mathbf{u}\right)+\mathbf{F}\left(t\right)+\boldsymbol{\sigma}\left(\mathbf{u}, t\right)\dot{\mathbf{W}}\left(t\right),
\end{equation}
where the state variable $\mathbf{u}\in\mathbb{C}^{N}$ lies in a high-dimensional phase space. In \eqref{eq:abs_formu}, the first two components, $\left(L+D\right)\mathbf{u}$,
represent linear dispersion and dissipation effects, where $L^{*}=-L$ is a skew-symmetric operator; and $D$ is negative definite. The nonlinear effect is introduced through a quadratic form, $B\left(\mathbf{u},\mathbf{u}\right)$, which conserves the total energy with $\mathbf{u}\cdot B\left(\mathbf{u},\mathbf{u}\right) = 0$. Besides, the system is subject to external forcing effects that are decomposed into a deterministic component, $\mathbf{F}\left(t\right)$, and a stochastic one represented by a Gaussian random process, $\boldsymbol{\sigma}\left(\mathbf{u}, t\right)\dot{\mathbf{W}}\left(t\right)$, where $\boldsymbol{\sigma}$ measures the noise amplitude and $\dot{\mathbf{W}}$ is the white noise.

\subsubsection*{\emph{Step 1.} Phase space decomposition}
To facilitate the analytically solvable properties in the PIDD-CG algorithm, we introduce a proper phase space decomposition of general system \eqref{eq:abs_formu}. Decompose the original model state $\mathbf{u}$ into two multi-dimensional state variables, $\mathbf{X}\in\mathbb{C}^{N_1}$ and $\mathbf{Y}\in\mathbb{C}^{N_2}$, with $N_1+N_2=N$. Usually, $\mathbf{X}$ is a collection of the large-scale, resolved leading modes or observed state variables while $\mathbf{Y}$ contains the remaining relatively smaller-scale modes, including unresolved and unobserved ones. Therefore, $\mathbf{X}$ belongs to a relatively low-dimensional subspace while $\mathbf{Y}$ remains high-dimensional.
Since $\mathbf{Y}$ by design denotes the faster and smaller scale components of the system, the terms corresponding to the nonlinear self-interaction inside $\mathbf{Y}$, i.e. $B(\mathbf{Y},\mathbf{Y})$, mostly involve high frequencies and homogeneous dynamics \cite{majda2003systematic}. Thus, these terms can often be effectively parameterized either by simple stochastic noise \cite{majda2001mathematical, plant2008stochastic, mana2014toward, berner2017stochastic} or suitable closures that are nonlinear functions of $\mathbf{X}$ and conditionally linear functions of $\mathbf{Y}$ \cite{mou2022efficient}. The resulting approximate system can successfully capture the dominant dynamical and statistics features of the original one as well as reproducing very similar ensemble forecast solutions.

With such a decomposition of the model states, the following nonlinearly coupled multiscale stochastic model is reached as the approximation of the original system \eqref{eq:abs_formu}:
\begin{subequations}\label{CGNS}
\begin{align}
  \frac{\d\mathbf{X}}{\d t} &= \Big[\mathbf{A}_\mathbf{0}(\mathbf{X},t) + \mathbf{A}_\mathbf{1}(\mathbf{X},t) \mathbf{Y}(t)\Big]  + \mathbf{B}_\mathbf{1}(\mathbf{X},t)\dot{\mathbf{W}}_\mathbf{1}(t),\label{CGNS_X}\\
  \frac{\d\mathbf{Y}}{\d t} &= \Big[\mathbf{a}_\mathbf{0}(\mathbf{X},t) + \mathbf{a}_\mathbf{1}(\mathbf{X},t) \mathbf{Y}(t)\Big]  + \mathbf{b}_\mathbf{2}(\mathbf{X},t)\dot{\mathbf{W}}_\mathbf{2}(t),\label{CGNS_Y}
\end{align}
\end{subequations}
where $\mathbf{A}_0, \mathbf{a}_0, \mathbf{A}_1, \mathbf{a}_1, \mathbf{B}_1$ and $\mathbf{b}_2$ are vectors or matrices that can depend nonlinearly on the state variables $\mathbf{X}$ and time $t$  while $\dot{\mathbf{W}}_\mathbf{1}$ and $\dot{\mathbf{W}}_\mathbf{2}$ are independent white noise sources. One desirable feature of \eqref{CGNS} is that, given one realization of the time series $\mathbf{X}(s)$ for $s\in[0,t]$, the conditional distribution
\begin{equation}\label{CGNS_PDF}
    p(\mathbf{Y}(t)\vert\mathbf{X}(s),s\leq t) \sim \mathcal{N}(\boldsymbol{\mu}(t),\mathbf{R}(t))
\end{equation}
becomes Gaussian, where the conditional mean $\boldsymbol\mu$ and the conditional covariance $\mathbf{R}$ can be solved via closed analytic formulae \cite{liptser2013statistics} (details will be shown next in \emph{Step 4}).

It is worthwhile to highlight that many complex nonlinear systems already fit into the framework of \eqref{CGNS} \cite{chen2018conditional, chen2016filtering}:
\begin{itemize}
  \item Physics-constrained nonlinear stochastic models. Examples include the noisy versions of Lorenz models,  Charney-DeVore flows, and the paradigm model for topographic mean flow interactions.
  \item Stochastically coupled reaction-diffusion models in neuroscience and ecology. Examples include FitzHugh-Nagumo models and  SIR epidemic models.
  \item Multi-scale models in turbulence and geophysical flows. Example include the Boussinesq equations and rotating shallow water equation.
\end{itemize}
These examples further justify that the modeling framework \eqref{CGNS} is appropriate to characterize or approximate many nonlinear and non-Gaussian systems in various disciplines.

\subsubsection*{\emph{Step 2.} Systematic multiscale statistical closure of the large-scale dynamics}
Since the observed large-scale state variable $\mathbf{X}$ lies in a low-dimensional subspace, a small number of suitable random sample points is sufficient to effectively characterize such a low-dimensional PDF. However, as the dynamics of $\mathbf{X}$ is nonlinearly coupled with $\mathbf{Y}$, the high-dimensional system \eqref{CGNS} has to be integrated all together requiring solutions of both $\mathbf{X}$ and $\mathbf{Y}$ to obtain the ensemble forecast of $\mathbf{X}$. This is not only computationally challenging for the simulation of each single realization, but also requires a large ensemble size to accurately reconstruct the statistics of the entire system. To reduce the high computational cost, a systematic multiscale statistical closure model of $\mathbf{X}$ is thus developed, the purpose of which is to avoid running the full set of the equations of $\mathbf{Y}$ during predicting the marginal PDF of only $\mathbf{X}$.

The multiscale statistical closure here depends on the crucial feature that the dynamics of $\mathbf{Y}$ in \eqref{CGNS_X} becomes linear given one realization of the trajectory of $\mathbf{X}$. Thus the conditional mean of $\mathbf{Y}\left(t\right)$ relying on the history observation of $\mathbf{X}\left(s\right),s<t$ can be solved via closed analytic formulae (shown in \emph{Step 4}). Next, decompose $\mathbf{Y}$ as $\mathbf{Y}=(\mathbf{Y}_\mathbf{1},\mathbf{Y}_\mathbf{2})$, where $\mathbf{Y}_\mathbf{1}$ is the resolved subsscale processes and $\mathbf{Y}_\mathbf{2}$ represent the rest unresolved ones. Correspondingly, the conditional mean of $\mathbf{Y}$ can be decomposed as as $\boldsymbol{\mu} = (\boldsymbol{\mu}_\mathbf{1},\boldsymbol{\mu}_\mathbf{2})$. Accordingly, rewrite $\mathbf{A}_\mathbf{1}$ in \eqref{CGNS_X} as $\mathbf{A}_\mathbf{1} = [\mathbf{A}_{\mathbf{1,1}}, \mathbf{A}_{\mathbf{1,2}}]$. The statistical closure model of $\mathbf{X}$ in \eqref{CGNS_X} reads:
\begin{equation}\label{closure_X}
\begin{split}
\frac{\d\mathbf{X}}{\d t} &= \Big[\mathbf{A}_\mathbf{0} + \mathbf{A}_\mathbf{1}\mathbf{Y}\Big]  + \mathbf{B}_\mathbf{1}\dot{\mathbf{W}}_\mathbf{1}\\
 &= \Big[\mathbf{A}_\mathbf{0} + \mathbf{A}_\mathbf{1,1} \mathbf{Y}_\mathbf{1}+ \mathbf{A}_\mathbf{1,2} \mathbf{Y}_\mathbf{2}\Big]  + \mathbf{B}_\mathbf{1}\dot{\mathbf{W}}_\mathbf{1}\\
 &= \Big[\mathbf{A}_\mathbf{0} + \mathbf{A}_\mathbf{1,1} \boldsymbol{\mu}_\mathbf{1}\Big]  + \mathbf{B}_\mathbf{1}\dot{\mathbf{W}}_\mathbf{1}
  +\Big[\mathbf{A}_\mathbf{1,1}\big(\mathbf{Y}_\mathbf{1}-\boldsymbol{\mu}_\mathbf{1}\big)+\mathbf{A}_\mathbf{1,2} \mathbf{Y}_\mathbf{2}\Big]\\
 :&= \Big[\mathbf{A}_\mathbf{0} + \mathbf{A}_\mathbf{1,1} \boldsymbol{\mu}_\mathbf{1}\Big]  +   \mathbf{B}_\mathbf{1}\dot{\mathbf{W}}_\mathbf{1} + \mathcal{F}_\mathbf{X}.
\end{split}
\end{equation}
The contributions from the resolved and unresolved components of $\mathbf{Y}$ are separated in the above closure model \eqref{closure_X}. First, the contribution from the resolved leading modes of $\mathbf{Y}$ is modeled by the conditional mean and its stochastic deviation from the mean, namely $\mathbf{A}_{\mathbf{1},\mathbf{1}}\mathbf{Y}_\mathbf{1}=\mathbf{A}_{\mathbf{1},\mathbf{1}}\boldsymbol{\mu}_\mathbf{1}+\mathbf{A}_\mathbf{1,1}\left(\mathbf{Y}_\mathbf{1}-\boldsymbol{\mu}_\mathbf{1}\right)$. Second, the combined contribution of the unresolved fast processes $\mathbf{A}_\mathbf{1,2} \mathbf{Y}_\mathbf{2}$ is modeled together as the coupled feedback from the various multiscale fluctuations.
The conditional mean part with $\boldsymbol{\mu}_\mathbf{1}$ is explicitly modeled through the conditional dynamics of reduced order model in \emph{Step 4} and \emph{Step 5}, while the remaining components of the deviation from the mean and the fast fluctuations in a high-dimensional space are both unresolved.
In the closure approximation \eqref{closure_X}, we denote the unresolved `residual' part as $\mathcal{F}_\mathbf{X}$ by combining contributions from both the mean deviation $\mathbf{Y}_\mathbf{1}-\boldsymbol{\mu}_\mathbf{1}$ and the remaining high-dimensional unresolved fluctuations $\mathbf{Y}_\mathbf{2}$ together. Usually, these terms will include complex nonlinear coupling between multiple scales. Nevertheless, we only need their combined feedback for the prediction of the large-scale state $\mathbf{X}$.

Here, the unresolved mean feedback $\mathcal{F}_\mathbf{X}$ is effectively approximated from data by a RNN:
\begin{equation}\label{RNN1}
  \mathcal{F}_\mathbf{X}(t+1) =  \mbox{RNN}\left( \mathbf{X}(t-m:t), \boldsymbol{\mu}_\mathbf{1}(t-m:t), \mathcal{F}_\mathbf{X}(t-m:t)\right),
\end{equation}
where the input of the RNN depends on the discrete time series from a past time instant $t-m$ to the current time instant $t$ of the state variable $\mathbf{X}$, the conditional mean of the resolved leading modes $\boldsymbol\mu_\mathbf{1}$ and $\mathcal{F}_\mathbf{X}$ itself (see more details of the neural network architecture in the {SI}). Therefore, the closure model \eqref{closure_X} together with the governing equations of the conditional mean $\boldsymbol\mu_\mathbf{1}$ leads to a closed system.
An ensemble simulation of this set of equations with a small number of samples can be carried out to sufficiently characterize the low-dimensional PDF of $\mathbf{X}$ at future time instants via running the RNN \eqref{RNN1} iteratively forward starting from an initial time instant.

Finally, denote the total number of ensemble members by $J$, and the forecast value at time $t$  associated with the $j$-th ensemble member by $\mathbf{X}^{\{j\}}(t)$. Then a smoothed PDF of $\mathbf{X}(t)$ can be reached by a kernel density estimation (KDE) \cite{botev2010kernel},
\begin{equation}\label{KDE}
p(\mathbf{X}(t)) = \lim_{J\to\infty}\frac{1}{J}\sum_{j=1}^J \tilde{p}(\mathbf{X}^{\{j\}}(t)).
\end{equation}
In \eqref{KDE}, $\tilde{p}(\mathbf{X}^{\{j\}}(t))$ is the $j$-th member from the KDE that is associated with $\mathbf{X}^{\{j\}}(t)$ using the ``solve-the-equation plug-in'' algorithm \cite{botev2010kernel}, which is an appropriate KDE method for approximating non-Gaussian distributions. Note that the asymptotic expression as $J\to\infty$ is used in \eqref{KDE} for the mathematical rigor of the formula. In practice, only a finite value of $J$ is adopted as the approximation.  To facilitate the computation in \emph{Step 3}, a Gaussian kernel is used in \eqref{KDE} such that $\tilde{p}(\mathbf{X}^{\{j\}}(t))$ is a Gaussian distribution centered at $\mathbf{X}^{\{j\}}(t)$.

\subsubsection*{\emph{Step 3.} Effective physics-informed conditional Gaussian mixture  via data assimilation}
In the ensemble simulation of the large-scale dynamics in \emph{Step 2}, each ensemble member provides one trajectory of $\mathbf{X}$ up to time $t$, denoted by $\mathbf{X}^{\{j\}}(s\leq t)$, where $j=1,\ldots,J$. Given the coupled model formulation in \eqref{CGNS} and conditioned on the realization $\mathbf{X}^{\{j\}}(s\leq t)$, there is one corresponding distribution of $\mathbf{Y}$ at time instant $t$, namely $p\big(\mathbf{Y}(t)\vert \mathbf{X}^{\{j\}}(s\leq t)\big)$. Such a conditional distribution  can be viewed as the posterior distribution of the analysis state $\mathbf{Y}(t)$ from data assimilation, where $\mathbf{X}^{\{j\}}(s\leq t)$ plays the role of the observed time series.
One desirable feature in the model \eqref{CGNS} is that it becomes a conditional Gaussian distribution \eqref{CGNS_PDF}, due to the linear dynamics of $\mathbf{Y}$ conditioned on $\mathbf{X}(s\leq t)$ with Gaussian white noise \cite{chen2018conditional}. In light of these conditional Gaussian distributions, the marginal distribution of $\mathbf{Y}(t)$ is provided by a conditional Gaussian mixture,
\begin{equation}\label{Y_pdf_CGNS}
  p(\mathbf{Y}(t)) = \lim_{J\to\infty}\frac{1}{J}\sum_{j=1}^J p(\mathbf{Y}(t)\vert \mathbf{X}^{\{j\}}(s\leq t)).
\end{equation}
Combining \eqref{Y_pdf_CGNS} with \eqref{KDE} yields the formula for the joint distribution,
\begin{equation}\label{joint_pdf_CGNS}
  p(\mathbf{X}(t),\mathbf{Y}(t)) = \lim_{J\to\infty}\frac{1}{J}\sum_{j=1}^J \tilde{p}(\mathbf{X}^{\{j\}}(t))p(\mathbf{Y}(t)\vert \mathbf{X}^{\{j\}}(s\leq t)).
\end{equation}
Since the component $p(\mathbf{Y}(t)\vert \mathbf{X}^{\{j\}}(\cdot))$ with each sample $j$ is a Gaussian distribution, the overall joint distribution in \eqref{joint_pdf_CGNS} is given by a Gaussian mixture.
It has been shown in rigorous analysis \cite{chen2018rigorous} that the error bound of the joint distribution in \eqref{joint_pdf_CGNS} does not depend on the dimension of $\mathbf{Y}$. Therefore, as long as $\mathbf{X}$ stays in a relatively low dimensional subspace, it suffices to adopt a small number of samples (i.e., a small $J$) to accurately approximate the joint PDF based on the formula in \eqref{joint_pdf_CGNS}. 

Here, the physics-informed ingredient is embodied in the data assimilation that takes into account the large-scale model information to improve the conditional distribution. In fact, the center of each $p(\mathbf{Y}(t)\vert \mathbf{X}^{\{j\}}(s\leq t))$ can be very different from the actual value of $\mathbf{Y}^{\{j\}}(t)$ when it is explicitly simulated with $\mathbf{X}^{\{j\}}(t)$ from the original coupled system. This distinguishes from the KDE, where the mixture components are often centered at the simulated data points. In addition, the covariance of $p(\mathbf{Y}(t)\vert \mathbf{X}^{\{j\}}(s\leq t))$, which is the correspondence to the fixed bandwidth in the standard KDE, is determined utilizing the dynamical properties via data assimilation and can be adaptive for different $j$'s.
The optimization procedure of automatically determining the center and the bandwidth of each Gaussian mixture component in such a physics-driven method facilitates the use of only a small number of samples to accurately approximate the high-dimensional PDF.

\subsubsection*{\emph{Step 4.} Analytic formulae for time evolution of conditional statistics in smaller-scale dynamics}
To forecast the joint PDF using \eqref{joint_pdf_CGNS}, what remains to compute is the conditional Gaussian distribution $p(\mathbf{Y}(t)\vert \mathbf{X}^{\{j\}}(s\leq t))$ for the sample realizations $j=1,\ldots,J$. However, since $\mathbf{Y}$ contains all the smaller scale dynamics in a high-dimensional subspace, applying a direct ensemble method to forecast its statistics has the same the curse of dimensionality issue as simulating the original physical system \eqref{eq:abs_formu} or \eqref{CGNS}. From a different approach exploiting the important structural feature of the system \eqref{CGNS}, the dynamical equations for the  conditional mean $\boldsymbol\mu$ and the conditional covariance $\mathbf{R}$ in \eqref{CGNS_PDF} are available via the following explicit analytic formulae \cite{liptser2013statistics}
\begin{subequations}\label{CGNS_Stat}
\begin{align}
  \frac{\d \boldsymbol{\mu}}{\d t} &= (\mathbf{a}_\mathbf{0} + \mathbf{a}_\mathbf{1} \boldsymbol{\mu}) + (\mathbf{R}\mathbf{A}_\mathbf{1}^* ) (\mathbf{B}_1\mathbf{B}_1^*)^{-1} \left(\frac{\d\mathbf{X}}{\d t} - (\mathbf{A}_\mathbf{0} + \mathbf{A}_\mathbf{1}\boldsymbol{\mu})\right),\label{CGNS_Stat_Mean}\\
  \frac{\d\mathbf{R}}{\d t} &= \mathbf{a}_\mathbf{1} \mathbf{R} + \mathbf{R}\mathbf{a}_\mathbf{1}^* + \mathbf{b}_2\mathbf{b}_2^* - ( \mathbf{R}\mathbf{A}_\mathbf{1}^*)(\mathbf{B}_1\mathbf{B}_1^*)^{-1}(\mathbf{A}_\mathbf{1}\mathbf{R}),\label{CGNS_Stat_Cov}
\end{align}
\end{subequations}
with $\cdot^*$ being the complex conjugate transpose.  Once a single trajectory of $\mathbf{X}$ is given, the system \eqref{CGNS_Stat} can be solved using the standard ODE solvers, such as the Euler or the Runge-Kutta schemes. In this way, the extensive MC simulation of the entire high-dimensional system can be effectively avoided.
In addition, the analytic formulae of the moment equations in \eqref{CGNS_Stat} avoid the potential computational issues of random sampling errors and recover the true conditional statistics characterized by the deterministic solutions for the mean $\boldsymbol{\mu}$ and covariance $\mathbf{R}$.

\subsubsection*{\emph{Step 5.} Data-driven modeling of the nonlinear feedbacks in conditional statistics with information theory}
Despite the closed analytic formulae for solving the conditional statistics, the cost of running \eqref{CGNS_Stat}, especially the full spectrum of the conditional covariance $\mathbf{R}\in\mathbb{C}^{N_{2}\times N_{2}}$ in \eqref{CGNS_Stat_Cov}, remains to be computational demanding for high-dimensional systems. Therefore, an additional model reduction strategy is required to further reduce the computational cost and focus on the conditional statistics of the resolved states $\mathbf{Y_1}$. The contributions from the remaining unresolved fluctuations are modeled through a closure scheme as described in \eqref{closure_X} of \emph{Step 2}.
Usually, $\mathbf{Y_1}$ can be separately by taking the most energetic leading modes of the high-dimensional state $\mathbf{Y}$. In addition, certain localized structures \cite{whitaker2004reanalysis, anderson2012localization} can be exploited to approximate the covariance matrix with a diagonal or block diagonal structure. Thus only the entries near the diagonal line need to be computed in the algorithm. For example in our illustrative example \eqref{eq:topo_model_main} in Section \ref{Sec:baro_model}, the coupling coefficient $\mathbf{a}_\mathbf{1}$  is automatically diagonalized since it represents linear dispersion and dissipation effects in small scales as well as the noise coefficients $\mathbf{B}_\mathbf{1}$ and $\mathbf{b}_\mathbf{2}$.

To accelerate the computational efficiency, the most time consuming parts including the complicated nonlinear and possibly unresolved information in solving \eqref{CGNS_Stat} are directly learned from data. To this end, define
\begin{equation}\label{Define_FG}
\begin{split}
\mathcal{F}_\mathbf{Y} &=  \dot{\mathbf{X}} - (\mathbf{A}_\mathbf{0} + \mathbf{A}_\mathbf{1}\boldsymbol{\mu}),\\
\mathcal{G}_\mathbf{Y} &= \mathbf{R}\mathbf{A}_\mathbf{1}^*,
\end{split}
\end{equation}
where $\mathcal{F}_\mathbf{Y}$ and $\mathcal{G}_\mathbf{Y}$ are the `embedded' feedbacks to the resolved mean and covariance dynamics. The explicit expressions on the right hand sides of \eqref{Define_FG} contain the full information in $\boldsymbol{\mu}$ and $\mathbf{R}$, while the RNNs can help us learn the resolved state information without running the full spectrum of unresolved fluctuation modes. More importantly, note that the mean feedback will be the same as the unresolved term in \eqref{RNN1}, i.e., $\mathcal{F}_\mathbf{Y}=\mathcal{F}_\mathbf{X}$, when we take $\mathbf{Y}=\mathbf{Y}_\mathbf{1}$. Thus the computational cost is further reduced. As an analog to $\mathcal{F}_\mathbf{Y}$ and $\mathcal{G}_\mathbf{Y}$ in \eqref{Define_FG}, define  $\mathcal{F}_\mathbf{Y_\mathbf{1}}$ and $\mathcal{G}_\mathbf{Y_\mathbf{1}}$ as the functions constrained on the resolved subspace of $\mathbf{Y}_\mathbf{1}$. Then the associated low-order dynamics from the full formulae in \eqref{CGNS_Stat} can be rewritten as
\begin{subequations}\label{CGNS_Stat_2}
\begin{align}
  \frac{\d \boldsymbol{\mu}_\mathbf{1}}{\d t} &= (\mathbf{a}_\mathbf{0} + \mathbf{a}_\mathbf{1} \boldsymbol{\mu}_\mathbf{1}) + \mathcal{G}_\mathbf{Y_\mathbf{1}} (\mathbf{B}_1\mathbf{B}_1^*)^{-1} \mathcal{F}_\mathbf{Y_\mathbf{1}},\label{CGNS_Stat_2_Mean}\\
  \frac{\d\mathbf{R}_\mathbf{1}}{\d t} &= \mathbf{a}_\mathbf{1} \mathbf{R}_\mathbf{1} + \mathbf{R}_\mathbf{1}\mathbf{a}_\mathbf{1}^* + \mathbf{b}_2\mathbf{b}_2^* - \mathcal{G}_\mathbf{Y_\mathbf{1}}(\mathbf{B}_1\mathbf{B}_1^*)^{-1}\mathcal{G}_\mathbf{Y_\mathbf{1}}^*.\label{CGNS_Stat_2_Cov}
\end{align}
\end{subequations}
Then $\mathcal{F}_\mathbf{Y_\mathbf{1}}$ and $\mathcal{G}_\mathbf{Y_\mathbf{1}}$ are approximated by the following RNNs,
\begin{equation}\label{RNN2}
\begin{split}
  \mathcal{F}_\mathbf{Y_\mathbf{1}}(t+1) = &  \mbox{RNN}\left( \mathbf{X}(t-m:t), \boldsymbol{\mu}_\mathbf{1}(t-m:t), \mathcal{F}_\mathbf{Y_\mathbf{1}}(t-m:t)\right),\\
  \mathcal{G}_\mathbf{Y_\mathbf{1}}(t+1) = &  \mbox{RNN}\left( \mathbf{X}(t-m:t), \mathbf{R}_\mathbf{1}(t-m:t), \mathcal{G}_\mathbf{Y_\mathbf{1}}(t-m:t)\right).
\end{split}
\end{equation}

What remains is to train the RNNs. It is worthwhile to highlight that these RNNs are used to approximate certain components in the moment equations. Therefore, it is important to develop a suitable criterion as the loss function of the RNNs such that the resulting moments $\boldsymbol\mu$ and $\mathbf{R}$ or the associated PDF are forecasted as accurate as possible. Since the path-wise error is not necessarily related to the calibration of the forecast statistics, minimizing the path-wise errors in forecasting the conditional mean and conditional covariance is not the most appropriate choice of the loss function for training the RNNs. Instead, an information loss function is adopted here that specifically emphasizes the minimization of the forecast error in terms of the PDF. The information criterion used here is the so-called relative entropy or the Kullback-Leibler divergence (KL divergence) \cite{kullback1951information, kleeman2011information},
\begin{equation}\label{Relative_Entropy}
  \mathcal{P}(p^{\mathrm{ref}}(\mathbf{z}),p^f(\mathbf{z})) = \int p^{\mathrm{ref}}(\mathbf{z})\ln\frac{p^{\mathrm{ref}}(\mathbf{z})}{p^f(\mathbf{z})}d\mathbf{z},
\end{equation}
where $p^{\mathrm{ref}}(\mathbf{z})$ is the true PDF while $p^f(\mathbf{z})$ is the predicted one. 

\subsection{Implementation details of the PIDD-CG algorithm}
\subsubsection{ML training and the use of the relative entropy as cost function}
In the training process, the neural network parameters are achieved through the optimization using a proper loss function. A straightforward choice of the loss function is the standard $L^{2}$ loss, which measures the mean square error (MSE) between the truth and the predicted conditional mean or the predicted conditional covariance. That is,
\begin{equation}\label{eq:L2}
\mathcal{L}_{\mathrm{MSE}}\left(t\right)=\left\Vert \bar{u}^{\mathrm{NN}}\left(t\right)-\bar{u}^{\mathrm{ref}}\left(t\right)\right\Vert _{L^{2}}^{2}+\sum_{k}\alpha_{k}\left\Vert r_{k}^{\mathrm{NN}}\left(t\right)-r_{k}^{\mathrm{ref}}\left(t\right)\right\Vert _{L^{2}}^{2}.
\end{equation}
In \eqref{eq:L2}, $\bar{u}^{\mathrm{NN}}$ and $\bar{u}^{\mathrm{ref}}$ are the conditional mean of neural network output and the truth, respectively; and $r_k^{\mathrm{NN}}$ and $r_{k}^{\mathrm{ref}}$ are the conditional covariance entries. However, as is shown in the SI, this $L^2$-loss becomes insufficient for guiding the training convergence to the optimal critical point, especially when highly turbulent fluctuations become dominant in the solution fields.

In the regimes with stronger extreme events and many noisy small-scale fluctuations, it becomes essential to focus on the dominant solution structures and `filter out' the noises in small amplitudes in the training phase. To this end, a more balanced measurement of the training error, is introduced here. It is named as the information loss as it exploits an information metric --- the relative entropy --- as the loss function:
\begin{equation}\label{eq:KLD}
\mathcal{L}_{\mathrm{KL}}\left(\mathbf{x,y}\right)=\frac{1}{M}\sum_{j=1}^{M}L_{j,\quad}L_{j}\left(\mathbf{x}^{j},\mathbf{y}^{j}\right)=\sum_{i}\tilde{y}_{i}^{\left(j\right)}\log\frac{\tilde{y}_{i}^{\left(j\right)}}{\tilde{x}_{i}^{\left(j\right)}},
\end{equation}
where $\mathbf{x}$ is the PDF reconstructed by the predicted conditional statistics while $\mathbf{y}$ is the reconstruction of the true PDF made by the true conditional statistics. The superscript $j$ in \eqref{eq:KLD} represents the mini-batch members and the subscript $i$ goes through the dimensions of the variable. In addition, to highlight more towards the extreme events, the following two sets of positive and negative temperatures are added to rescale the data from the partition functions
\begin{equation}\label{eq:rescaling}
\tilde{x}_{i}^{+}=\frac{\exp\left(x_{i}/T_{+}\right)}{\sum_{i}\exp\left(x_{i}/T_{+}\right)},\quad\tilde{x}_{i}^{-}=\frac{\exp\left(-x_{i}/T_{-}\right)}{\sum_{i}\exp\left(-x_{i}/T_{-}\right)},
\end{equation}
where $T_{+}>0,T_{-}>0$ are the positive and negative temperatures weighting the importance of extreme events in the scaled measure.

\subsubsection{Initialization of the conditional statistics ensembles}
Assume the initial time instant for the forecast is at $t=T$.
In the traditional ensemble forecast, the initialization of the ensembles is provided by data assimilation. Consider the modeling framework in \eqref{CGNS}. Assume the time series of the large-scale variable $\mathbf{X}$ is fully observed with no additional observational error while the observation of $\mathbf{Y}$ is not available. The situation in which the observation of $\mathbf{X}$ contains observational error can be easily handled by imposing another data assimilation procedure for $\mathbf{X}$, which is however not the main scope of the current framework.
Therefore, the precise observational value of $\mathbf{X}$ at time $T$ is naturally served as the initial condition of $\mathbf{X}(T)$. This also means all the initial ensembles of $\mathbf{X}(T)$ are the same as the observed value. On the other hand, the initial ensemble of $\mathbf{Y}(T)$ is provided by sampling $J$ different samples from the conditional (or the so-called posterior) distribution \eqref{CGNS_PDF}, which is given by the data assimilation formulae in \eqref{CGNS_Stat}.

The initialization of the PIDD-CG  algorithm has two major differences compared with the traditional ensemble forecast initialization. First, each ensemble member in the PIDD-CG  algorithm is no longer a single point but instead a conditional Gaussian distribution. Second, due to the use of neural network approximations in \eqref{RNN1} and \eqref{RNN2}, each initial ensemble of $\mathcal{F}_\mathbf{X}$, $\mathcal{F}_\mathbf{Y}$ and $\mathcal{G}_\mathbf{Y}$ contains a time series that requires the past information. This is different from the traditional ensemble forecast that only exploits the state estimation at the initial time instant. The details of the initialization in the PIDD-CG  algorithm are as follows.\medskip

\noindent \emph{(a). The initialization of the state variable $\mathbf{Y}$ in \eqref{CGNS}.}\\
Since the initial ensembles of $\mathbf{X}$ are all the same, it is natural to use the same conditional statistics ensembles for the initialization of $\mathbf{Y}$ as well. In particular, all the initial ensembles of $\mathbf{Y}$ take the values where $\boldsymbol\mu$ and $\mathbf{R}$ are the posterior mean and posterior covariance computed from the direct data assimilation \eqref{CGNS_Stat}. This also makes the entire initial distribution of $\mathbf{X}$ and $\mathbf{Y}$ to be consistent as the traditional ensemble forecast method.\medskip

\noindent \emph{(b). The initialization of the functions $\mathcal{F}_\mathbf{Y}$ and $\mathcal{G}_\mathbf{Y}$ in \eqref{RNN2}.}\\
The functions $\mathbf{F}_\mathbf{Y}$ and $\mathcal{G}_\mathbf{Y}$ depend on the past information of $\boldsymbol\mu$, $\mathbf{R}$, $\mathbf{X}$ and themselves. Therefore, the data assimilation scheme \eqref{CGNS_Stat} starts from a certain time instant in the past, and then results in the time series of $\boldsymbol\mu$ and $\mathbf{R}$ from $T-m$ to the current time instant $T$. The time series of $\mathbf{X}$ from $T-m$ to $T$ is available from observations.\medskip

\noindent \emph{(c). The initialization of the functions $\mathcal{F}_\mathbf{X}$ in \eqref{RNN1}.}\\
The input of the neural network in \eqref{RNN1} requires additional path-wise information of the unobserved trajectory $\mathbf{Y}$, which are related but are not directly available using the point-wise posterior mean and posterior covariance. The trajectory of $\mathbf{Y}$ can be sampled from an infinite dimensional (or high-dimensional with temporal the discretization) joint posterior distribution, where the infinity (or high) dimensionality comes not only from the number of the state variables in $\mathbf{Y}$ but also the temporal direction. Nevertheless, the modeling framework \eqref{CGNS} allows such a high-dimensional sampling problem to be solved by integrating a backward stochastic differential equation \cite{chen2020efficient}. It is given by
 \begin{equation}\label{Sampling_Main}
  \frac{\overleftarrow{\d \mathbf{Y}}}{\d t} = \frac{\overleftarrow{\d \boldsymbol\mu_\mathbf{s}}}{\d t} - \big(\mathbf{a}_\mathbf{1} + (\mathbf{b}_2\mathbf{b}_2^*)\mathbf{R}^{-1}\big)(\mathbf{Y} - \boldsymbol\mu_\mathbf{s}) + \mathbf{b}_2\dot{\mathbf{W}}_{\mathbf{Y}}(t),
\end{equation}
where the notation $\overleftarrow{\d \cdot}/\d t$ corresponds to the negative of the usual derivative, which means that the system \eqref{Smoother_Main} is solved backward over $[0,T]$. Here $\boldsymbol\mu_\mathbf{s}(t)$ and $\mathbf{R}_\mathbf{s}(t)$ are the so-called smoother mean and smoother covariance
\begin{equation}\label{Smoother}
  p(\mathbf{Y}(t)\vert\mathbf{X}(s), s\in[0,T])\sim\mathcal{N}(\boldsymbol\mu_\mathbf{s}(t),\mathbf{R}_\mathbf{s}(t)),
\end{equation}
which are also provided with closed analytic formulae
\begin{subequations}\label{Smoother_Main}
\begin{align}
  \frac{\overleftarrow{\d \boldsymbol{\mu}_\mathbf{s}}}{\d t} &=  -\mathbf{a}_\mathbf{0} - \mathbf{a}_\mathbf{1}\boldsymbol{\mu}_\mathbf{s}  + (\mathbf{b}_2\mathbf{b}_2^*)\mathbf{R}^{-1}(\boldsymbol\mu_{\mathbf{f}} - \boldsymbol{\mu}_\mathbf{s}),\label{Smoother_Main_mu}\\
  \frac{\overleftarrow{\d \mathbf{R}_\mathbf{s}}}{\d t} &= - (\mathbf{a}_\mathbf{1} + (\mathbf{b}_2\mathbf{b}_2^*) \mathbf{R}^{-1})\mathbf{R}_\mathbf{s} - \mathbf{R}_\mathbf{s}(\mathbf{a}_\mathbf{1}^* + (\mathbf{b}_2\mathbf{b}_2^*)\mathbf{R})  + \mathbf{b}_2\mathbf{b}_2^* ,\label{Smoother_Main_R}
\end{align}
\end{subequations}
The initial condition of solving \eqref{Sampling_Main} is $(\boldsymbol\mu_\mathbf{s}(T), \mathbf{R}_\mathbf{s}(T)) = (\boldsymbol\mu(T), \mathbf{R}(T))$, which is the same as the data assimilation (filtering) estimate $(\boldsymbol\mu(T), \mathbf{R}(T))$.

\section*{Acknowledgement}

The research of N.C. is partially funded by the Office of VCRGE at UW-Madison and ONR N00014-21-1-2904. The research of D. Q. is partially supported by the start-up funds provided by Purdue University.

\bibliographystyle{plain}
\bibliography{ref}

\begin{thebibliography}{10}

\bibitem{ahmed2021closures}
Shady~E Ahmed, Suraj Pawar, Omer San, Adil Rasheed, Traian Iliescu, and Bernd~R
  Noack.
\newblock On closures for reduced order models-a spectrum of first-principle to
  machine-learned avenues.
\newblock {\em Physics of Fluids}, 33(9):091301, 2021.

\bibitem{anderson2012localization}
Jeffrey~L Anderson.
\newblock Localization and sampling error correction in ensemble {K}alman
  filter data assimilation.
\newblock {\em Monthly Weather Review}, 140(7):2359--2371, 2012.

\bibitem{berner2017stochastic}
Judith Berner, Ulrich Achatz, Lauriane Batte, Lisa Bengtsson, Alvaro de~la
  C{\'a}mara, Hannah~M Christensen, Matteo Colangeli, Danielle~RB Coleman, Daan
  Crommelin, Stamen~I Dolaptchiev, et~al.
\newblock Stochastic parameterization: Toward a new view of weather and climate
  models.
\newblock {\em Bulletin of the American Meteorological Society},
  98(3):565--588, 2017.

\bibitem{botev2010kernel}
Zdravko~I Botev, Joseph~F Grotowski, and Dirk~P Kroese.
\newblock Kernel density estimation via diffusion.
\newblock {\em The annals of Statistics}, 38(5):2916--2957, 2010.

\bibitem{chekroun2020variational}
Micka{\"e}l~D Chekroun, Honghu Liu, and James~C McWilliams.
\newblock Variational approach to closure of nonlinear dynamical systems:
  Autonomous case.
\newblock {\em Journal of Statistical Physics}, 179(5):1073--1160, 2020.

\bibitem{chen2018conditional}
Nan Chen and Andrew Majda.
\newblock Conditional gaussian systems for multiscale nonlinear stochastic
  systems: Prediction, state estimation and uncertainty quantification.
\newblock {\em Entropy}, 20(7):509, 2018.

\bibitem{chen2016filtering}
Nan Chen and Andrew~J Majda.
\newblock Filtering nonlinear turbulent dynamical systems through conditional
  {G}aussian statistics.
\newblock {\em Monthly Weather Review}, 144(12):4885--4917, 2016.

\bibitem{chen2017beating}
Nan Chen and Andrew~J Majda.
\newblock Beating the curse of dimension with accurate statistics for the
  {F}okker--{P}lanck equation in complex turbulent systems.
\newblock {\em Proceedings of the National Academy of Sciences},
  114(49):12864--12869, 2017.

\bibitem{chen2020efficient}
Nan Chen and Andrew~J Majda.
\newblock Efficient nonlinear optimal smoothing and sampling algorithms for
  complex turbulent nonlinear dynamical systems with partial observations.
\newblock {\em Journal of Computational Physics}, page 109381, 2020.

\bibitem{chen2018rigorous}
Nan Chen, Andrew~J Majda, and Xin~T Tong.
\newblock Rigorous analysis for efficient statistically accurate algorithms for
  solving {F}okker--{P}lanck equations in large dimensions.
\newblock {\em SIAM/ASA Journal on Uncertainty Quantification},
  6(3):1198--1223, 2018.

\bibitem{cherkassky2007learning}
Vladimir Cherkassky and Filip~M Mulier.
\newblock {\em Learning from data: concepts, theory, and methods}.
\newblock John Wiley \& Sons, Hoboken, New Jersy, USA, 2007.

\bibitem{edeling2020reducing}
Wouter Edeling and Daan Crommelin.
\newblock Reducing data-driven dynamical subgrid scale models by physical
  constraints.
\newblock {\em Computers \& Fluids}, 201:104470, 2020.

\bibitem{evans2013optimally}
Jason~P Evans, Fei Ji, Gab Abramowitz, and Marie Ekstr{\"o}m.
\newblock Optimally choosing small ensemble members to produce robust climate
  simulations.
\newblock {\em Environmental Research Letters}, 8(4):044050, 2013.

\bibitem{evensen2009data}
Geir Evensen.
\newblock {\em Data assimilation: the ensemble Kalman filter}.
\newblock Springer Science \& Business Media, Berlin Heidelberg, Germany, 2009.

\bibitem{farazmand2018extreme}
Mohammad Farazmand and Themistoklis Sapsis.
\newblock Extreme events: Mechanisms and prediction.
\newblock {\em Applied Mechanics Reviews}, 2018.

\bibitem{franzke2015stochastic}
Christian~LE Franzke, Terence~J O'Kane, Judith Berner, Paul~D Williams, and
  Valerio Lucarini.
\newblock Stochastic climate theory and modeling.
\newblock {\em Wiley Interdisciplinary Reviews: Climate Change}, 6(1):63--78,
  2015.

\bibitem{ghil2012topics}
Michael Ghil and Stephen Childress.
\newblock {\em Topics in geophysical fluid dynamics: atmospheric dynamics,
  dynamo theory, and climate dynamics}.
\newblock Springer Science \& Business Media, New York, USA, 2012.

\bibitem{gneiting2005weather}
Tilmann Gneiting and Adrian~E Raftery.
\newblock Weather forecasting with ensemble methods.
\newblock {\em Science}, 310(5746):248--249, 2005.

\bibitem{kalnay2003atmospheric}
Eugenia Kalnay.
\newblock {\em Atmospheric modeling, data assimilation and predictability}.
\newblock Cambridge university press, Cambridge, England, 2003.

\bibitem{khodkar2018data}
M~Amin Khodkar and Pedram Hassanzadeh.
\newblock Data-driven reduced modelling of turbulent {R}ayleigh--{B}{\'e}nard
  convection using {DMD}-enhanced fluctuation--dissipation theorem.
\newblock {\em Journal of Fluid Mechanics}, 852, 2018.

\bibitem{kleeman2011information}
Richard Kleeman.
\newblock Information theory and dynamical system predictability.
\newblock {\em Entropy}, 13(3):612--649, 2011.

\bibitem{kondrashov2015data}
Dmitri Kondrashov, Micka{\"e}l~D Chekroun, and Michael Ghil.
\newblock Data-driven non-markovian closure models.
\newblock {\em Physica D: Nonlinear Phenomena}, 297:33--55, 2015.

\bibitem{kullback1951information}
Solomon Kullback and Richard~A Leibler.
\newblock On information and sufficiency.
\newblock {\em The annals of mathematical statistics}, 22(1):79--86, 1951.

\bibitem{leutbecher2008ensemble}
Martin Leutbecher and Tim~N Palmer.
\newblock Ensemble forecasting.
\newblock {\em Journal of computational physics}, 227(7):3515--3539, 2008.

\bibitem{liptser2013statistics}
Robert~S Liptser and Albert~N Shiryaev.
\newblock {\em Statistics of random processes II: Applications}, volume~6.
\newblock Springer Science \& Business Media, Berlin Heidelberg, Germany, 2013.

\bibitem{lucarini2016extremes}
Valerio Lucarini, Davide Faranda, Jorge Miguel~Milhazes de~Freitas, Mark
  Holland, Tobias Kuna, Matthew Nicol, Mike Todd, Sandro Vaienti, et~al.
\newblock {\em Extremes and recurrence in dynamical systems}.
\newblock John Wiley \& Sons, Hoboken, New Jersey, USA, 2016.

\bibitem{majda2006nonlinear}
Andrew Majda and Xiaoming Wang.
\newblock {\em Nonlinear dynamics and statistical theories for basic
  geophysical flows}.
\newblock Cambridge University Press, Cambridge, England, 2006.

\bibitem{majda2016introduction}
Andrew~J Majda.
\newblock {\em Introduction to turbulent dynamical systems in complex systems}.
\newblock Springer, Switzerland, 2016.

\bibitem{majda2018strategies}
Andrew~J Majda and Di~Qi.
\newblock Strategies for reduced-order models for predicting the statistical
  responses and uncertainty quantification in complex turbulent dynamical
  systems.
\newblock {\em SIAM Review}, 60(3):491--549, 2018.

\bibitem{majda2001mathematical}
Andrew~J Majda, Ilya Timofeyev, and Eric Vanden~Eijnden.
\newblock A mathematical framework for stochastic climate models.
\newblock {\em Communications on Pure and Applied Mathematics}, 54(8):891--974,
  2001.

\bibitem{majda2003systematic}
Andrew~J Majda, Ilya Timofeyev, and Eric Vanden-Eijnden.
\newblock Systematic strategies for stochastic mode reduction in climate.
\newblock {\em Journal of the Atmospheric Sciences}, 60(14):1705--1722, 2003.

\bibitem{mana2014toward}
PierGianLuca~Porta Mana and Laure Zanna.
\newblock Toward a stochastic parameterization of ocean mesoscale eddies.
\newblock {\em Ocean Modelling}, 79:1--20, 2014.

\bibitem{mou2022efficient}
Changhong Mou, Nan Chen, and Traian Iliescu.
\newblock An efficient data-driven multiscale stochastic reduced order modeling
  framework for complex turbulent systems.
\newblock {\em arXiv preprint arXiv:2203.11438}, 2022.

\bibitem{palmer2019ecmwf}
Tim Palmer.
\newblock The ecmwf ensemble prediction system: Looking back (more than) 25
  years and projecting forward 25 years.
\newblock {\em Quarterly Journal of the Royal Meteorological Society},
  145:12--24, 2019.

\bibitem{plant2008stochastic}
RS~Plant and George~C Craig.
\newblock A stochastic parameterization for deep convection based on
  equilibrium statistics.
\newblock {\em Journal of the Atmospheric Sciences}, 65(1):87--105, 2008.

\bibitem{qi2021machine}
Di~Qi and John Harlim.
\newblock Machine learning-based statistical closure models for turbulent
  dynamical systems.
\newblock {\em arXiv preprint arXiv:2108.13220}, 2021.

\bibitem{qi2016low}
Di~Qi and Andrew~J Majda.
\newblock Low-dimensional reduced-order models for statistical response and
  uncertainty quantification: Two-layer baroclinic turbulence.
\newblock {\em Journal of the Atmospheric Sciences}, 73(12):4609--4639, 2016.

\bibitem{qi2017low}
Di~Qi and Andrew~J Majda.
\newblock Low-dimensional reduced-order models for statistical response and
  uncertainty quantification: Barotropic turbulence with topography.
\newblock {\em Physica D: Nonlinear Phenomena}, 343:7--27, 2017.

\bibitem{qi2020using}
Di~Qi and Andrew~J Majda.
\newblock Using machine learning to predict extreme events in complex systems.
\newblock {\em Proceedings of the National Academy of Sciences}, 117(1):52--59,
  2020.

\bibitem{salmon1998lectures}
Rick Salmon.
\newblock {\em Lectures on geophysical fluid dynamics}.
\newblock Oxford University Press, Oxford, England, 1998.

\bibitem{santos2021reduced}
Manuel Santos~Guti{\'e}rrez, Valerio Lucarini, Micka{\"e}l~D Chekroun, and
  Michael Ghil.
\newblock Reduced-order models for coupled dynamical systems: Data-driven
  methods and the koopman operator.
\newblock {\em Chaos: An Interdisciplinary Journal of Nonlinear Science},
  31(5):053116, 2021.

\bibitem{schneider2021learning}
Tapio Schneider, Andrew~M Stuart, and Jin-Long Wu.
\newblock Learning stochastic closures using ensemble {K}alman inversion.
\newblock {\em Transactions of Mathematics and Its Applications}, 5(1):tnab003,
  2021.

\bibitem{sheard2009principles}
Sarah~A Sheard and Ali Mostashari.
\newblock Principles of complex systems for systems engineering.
\newblock {\em Systems Engineering}, 12(4):295--311, 2009.

\bibitem{strogatz2018nonlinear}
Steven~H Strogatz.
\newblock {\em Nonlinear dynamics and chaos with student solutions manual: With
  applications to physics, biology, chemistry, and engineering}.
\newblock CRC press, Boca Raton, Florida, 2018.

\bibitem{tao2009multiscale}
Wei-Kuo Tao, Jiun-Dar Chern, Robert Atlas, David Randall, Marat Khairoutdinov,
  Jui-Lin Li, Duane~E Waliser, Arthur Hou, Xin Lin, Christa Peters-Lidard,
  et~al.
\newblock A multiscale modeling system: Developments, applications, and
  critical issues.
\newblock {\em Bulletin of the American Meteorological Society},
  90(4):515--534, 2009.

\bibitem{toth1997ensemble}
Zoltan Toth and Eugenia Kalnay.
\newblock Ensemble forecasting at ncep and the breeding method.
\newblock {\em Monthly Weather Review}, 125(12):3297--3319, 1997.

\bibitem{vallis2017atmospheric}
Geoffrey~K Vallis.
\newblock {\em Atmospheric and oceanic fluid dynamics}.
\newblock Cambridge University Press, Cambridge, England, 2017.

\bibitem{whitaker2004reanalysis}
Jeffrey~S Whitaker, Gilbert~P Compo, Xue Wei, and Thomas~M Hamill.
\newblock Reanalysis without radiosondes using ensemble data assimilation.
\newblock {\em Monthly Weather Review}, 132(5):1190--1200, 2004.

\bibitem{wilcox1988multiscale}
David~C Wilcox.
\newblock Multiscale model for turbulent flows.
\newblock {\em AIAA journal}, 26(11):1311--1320, 1988.

\bibitem{zaremba2014recurrent}
Wojciech Zaremba, Ilya Sutskever, and Oriol Vinyals.
\newblock Recurrent neural network regularization.
\newblock {\em arXiv preprint arXiv:1409.2329}, 2014.

\end{thebibliography}

\appendix
\renewcommand\theequation{S\arabic{equation}}
\setcounter{equation}{0}
\renewcommand\thefigure{S\arabic{figure}}
\setcounter{figure}{0}
\renewcommand\thetable{S\arabic{table}}
\setcounter{table}{0}

\section*{Supplementary Information:}
\noindent This Supplementary Information (SI) contains the numerical construction and experiments for
\begin{itemize}
  \item [A)] a coupled
dyad model as a proof-of-concept of the PIDD-CG algorithm; and
  \item [B)] a complete analysis of the computational
performance of the barotropic topographic model with multiscale coupling.
\end{itemize}

\section{The coupled dyad model}

First, we verify the effectiveness of the PIDD-CG algorithm on a prototype
two-mode dyad model as a standard proof of concept. The purpose of using this model with the simplest possible  turbulent structure 
is to illustrate the basic ideas in the algorithm construction
and the key features in the method to predict crucial statistics in
a simple and clean setup.

\subsection{Model description}

The dyad model is described by the two nonlinearly coupled states
$\left(u,v\right)$ following the dynamics
\begin{equation}
\begin{aligned}\frac{du}{dt}= & -d_{u}u+cuv+F_{u}+\sigma_{u}\dot{W}_{u},\\
\frac{dv}{dt}= & -d_{v}v-cu^{2}+F_{v}+\sigma_{v}\dot{W}_{v}.
\end{aligned}
\label{eq:dyad_model}
\end{equation}
Naturally, we can view $u$ as the `observed state' (as $\mathbf{X}$
in the main text) and $v$ the `unresolved process' ($\mathbf{Y}$
in the main text) satisfying the general conditional Gaussian framework (Eqn. (3) in the main text). Different
statistics can be generated by varying the model parameters $\left(d_{u},d_{v},c,F_{u},F_{v},\sigma_{u},\sigma_{v}\right)$.
The common parameters are taken as $\left(d_{u},d_{v},c,F_{u},F_{v}\right)=\left(0.8,0.8,1.2,1,1\right)$.
In particular, we consider two typical statistical regimes by changing
the noise forcing strength for: i) near-Gaussian regime in $u$ with
$\left(\sigma_{u},\sigma_{v}\right)=\left(3,0.2\right)$ ; and ii)
non-Gaussian regime in $u$ with $\left(\sigma_{u},\sigma_{v}\right)=\left(0.5,2\right)$. For conciseness, these two regimes are referred to as \emph{near-Gaussian $u$} and \emph{non-Gaussian $u$} in the following.
The typical trajectories of the dyad model are illustrated in Figure
\ref{eq:dyad_model}. It can be seen that different strongly turbulent features including significant skewness and kurtosis appear in the solutions of $u$ and $v$. Thus the dyad model
becomes a desirable first test for confirming the basic features
and prediction skill of the general PIDD-CG algorithm.

\begin{figure}
\subfloat{\includegraphics[scale=0.37]{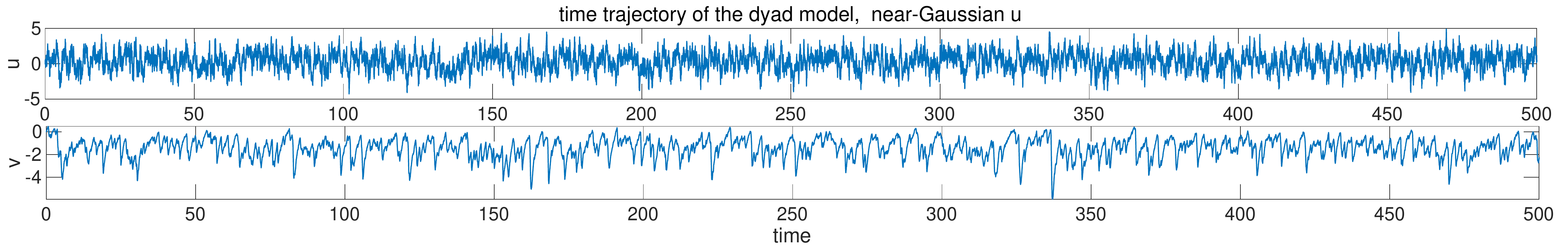}

}

\vspace{-1em}

\subfloat{\includegraphics[scale=0.37]{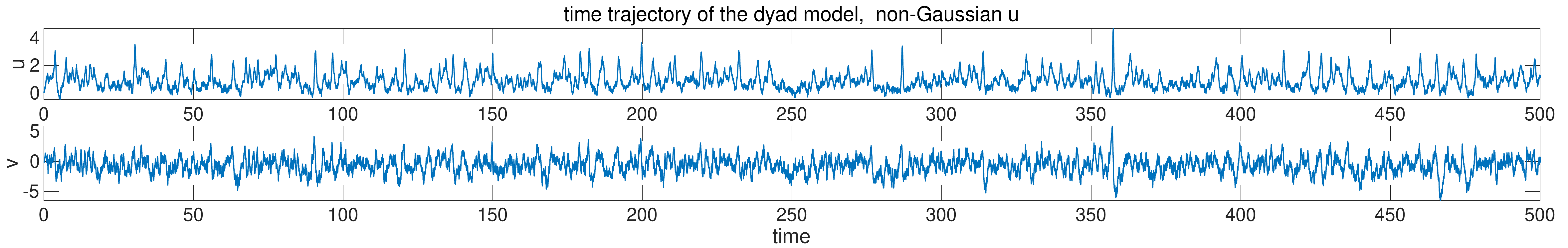}}

\caption{Trajectories of the dyad model in different statistical regimes with
near-Gaussian observed state $u$ (upper row) and non-Gaussian observed
state $u$ (lower row).\label{fig:Trajectories-dyad}}

\end{figure}
The conditional Gaussian dynamics can be proposed for the dyad model
(\ref{eq:dyad_model}) based on the general framework (3) in the main
text. Given a realization of the observed state $u$, the unresolved
state $v$ has conditional Gaussian statistics $v\sim\mathcal{N}\left(\bar{v},r_{v}\right)$
with $\bar{v}\equiv\bar{v}\left(t;u\left(\cdot\right)\right)$ and
$r_{v}\equiv r_{v}\left(t;u\left(\cdot\right)\right)$ being the conditional mean
and conditional variance dependent on the history trajectory of $u\left(s\right),s\leq t$.
Therefore, we can find the explicit dynamical equations for the conditional
statistics as
\begin{equation}
\begin{aligned}\frac{d\bar{v}}{dt}= & -d_{v}\bar{v}-cu^{2}+F_{u}+\sigma_{u}^{-2}\mathcal{F}_{u}\cdot\mathcal{G}_{v},\\
\frac{dr_{v}}{dt}= & -2d_{v}r_{v}+\sigma_{v}^{2}-\sigma_{u}^{-2}\mathcal{G}_{v}^{2}.
\end{aligned}
\label{eq:dyad_condGau}
\end{equation}
In the above formulation, we have the unresolved nonlinear coupling
terms $\mathcal{F}_{u}$ and $\mathcal{G}_{v}$ as follows
\begin{equation}
\mathcal{F}_{u}=\dot{u}+d_{u}u-f_{u}-cu\bar{v},\quad\mathcal{G}_{v}=cur_{v}.\label{eq:dyad_coupling}
\end{equation}
In the effective approximation of the nonlinear terms, we propose
to use RNNs to replace the crucial nonlinear feedback terms in the PIDD-CG algorithm
as
\begin{equation}
\begin{aligned}\mathcal{F}_{u}\left(t+1\right)= & \mathrm{RNN}\left(u\left(t-m:t\right),\bar{v}\left(t-m:t\right),\mathcal{F}_{u}\left(t-m:t\right)\right),\\
\mathcal{G}_{v}\left(t+1\right)= & \mathrm{RNN}\left(u\left(t-m:t\right),r_{v}\left(t-m:t\right),\mathcal{G}_{v}\left(t-m:t\right)\right).
\end{aligned}
\label{eq:dyad_rnn}
\end{equation}
Notice that we need to include the history of the nonlinear terms
$\mathcal{F}_{u},\mathcal{G}_{v}$ in the inputs of the RNNs.

\subsection{Forecast results}

\subsubsection{Training and lead time prediction}

\begin{figure}[h!]
\subfloat{\includegraphics[scale=0.44]{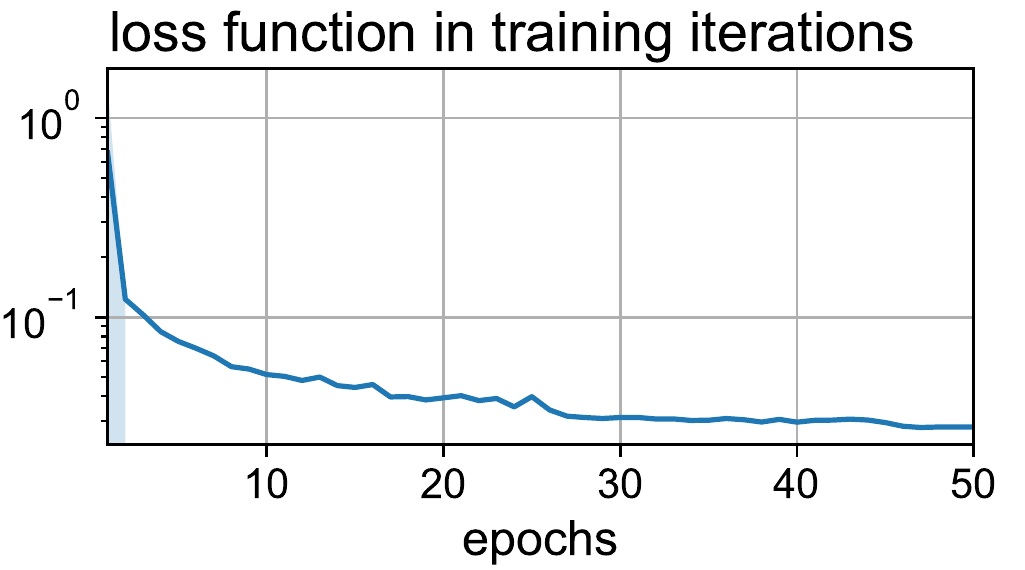}\includegraphics[scale=0.44]{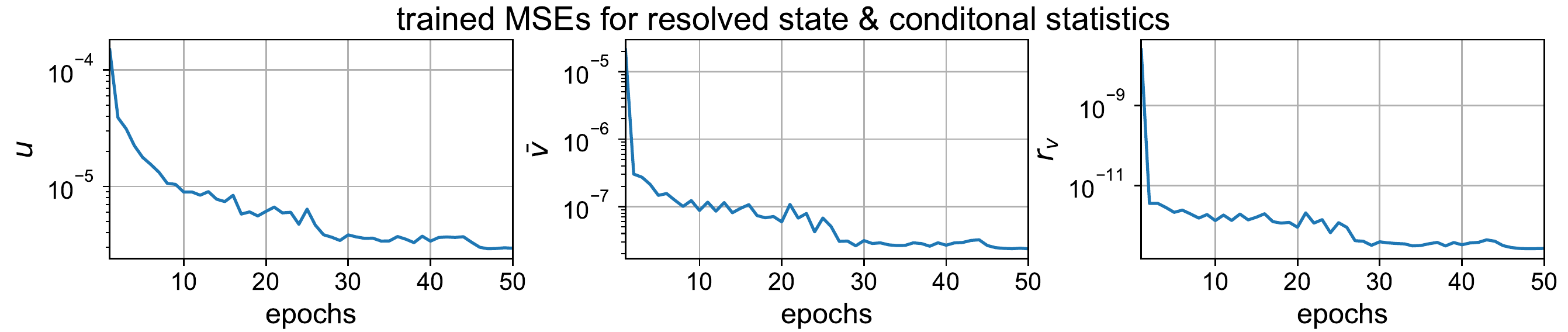}

}

\subfloat{\includegraphics[scale=0.38]{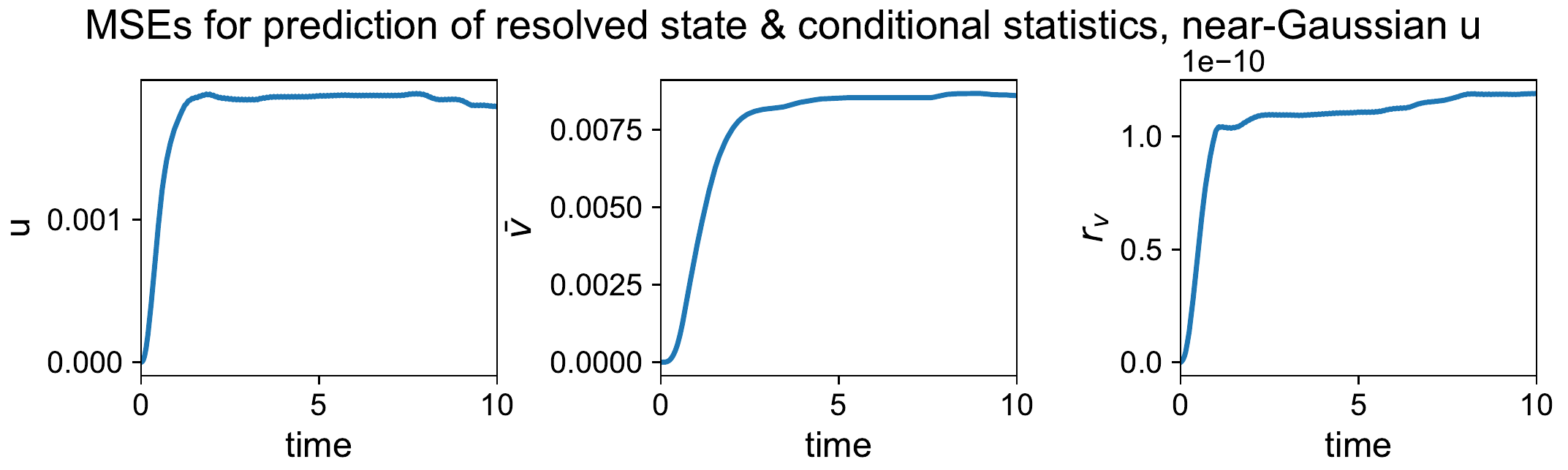}\includegraphics[scale=0.38]{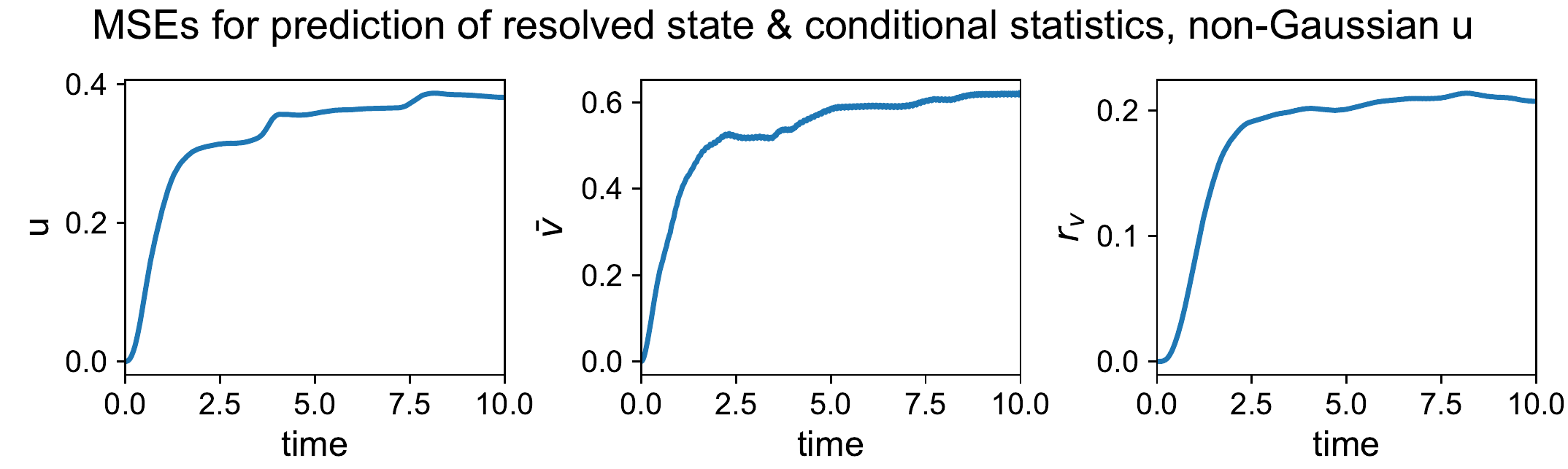}}

\caption{Training and lead-time prediction errors in the dyad model. The first
row shows the iterations of training loss and training errors in the
resolved state $u$ and conditional mean and variance in the unresolved
state $v$. The second row shows the development of lead time errors
using the trained model in the two parameter regimes.\label{fig:Training-and-lead-time-dyad}}
\end{figure}

\begin{figure}[h!]
\subfloat[near-Gaussian $u$]{\includegraphics[scale=0.4]{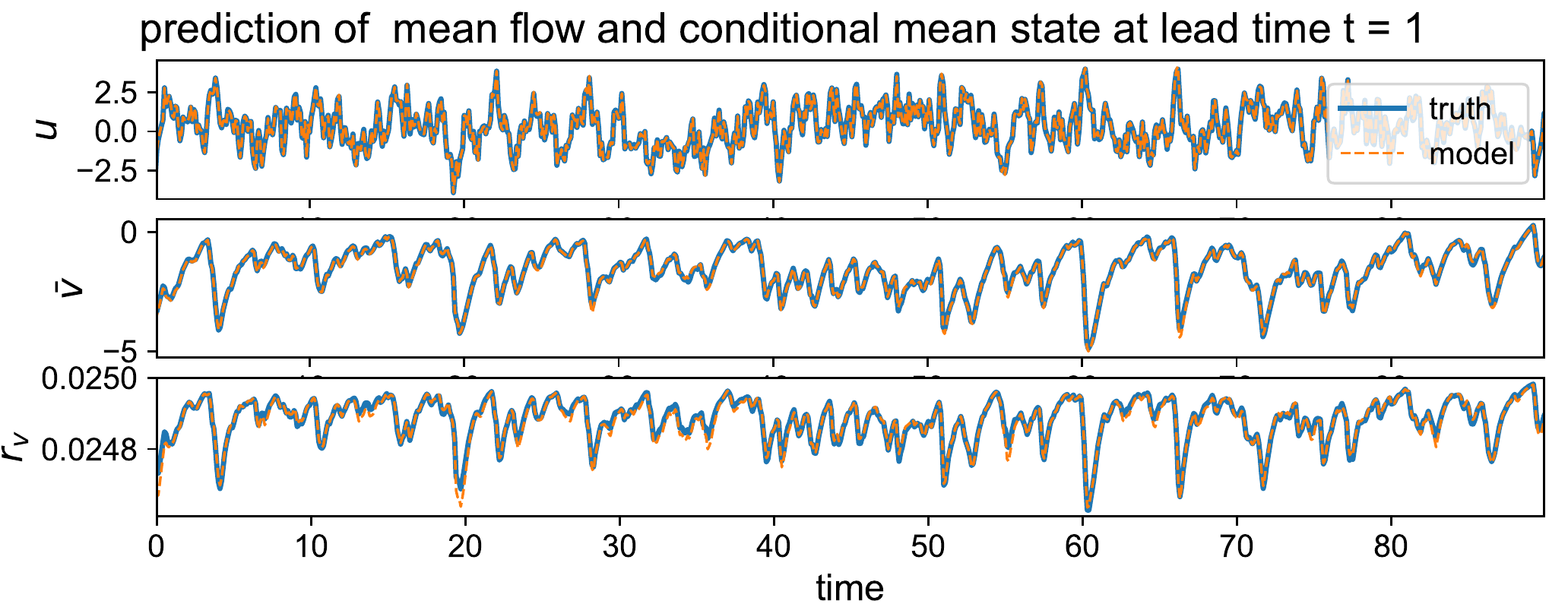}

}\subfloat[non-Gaussian $u$]{\includegraphics[scale=0.4]{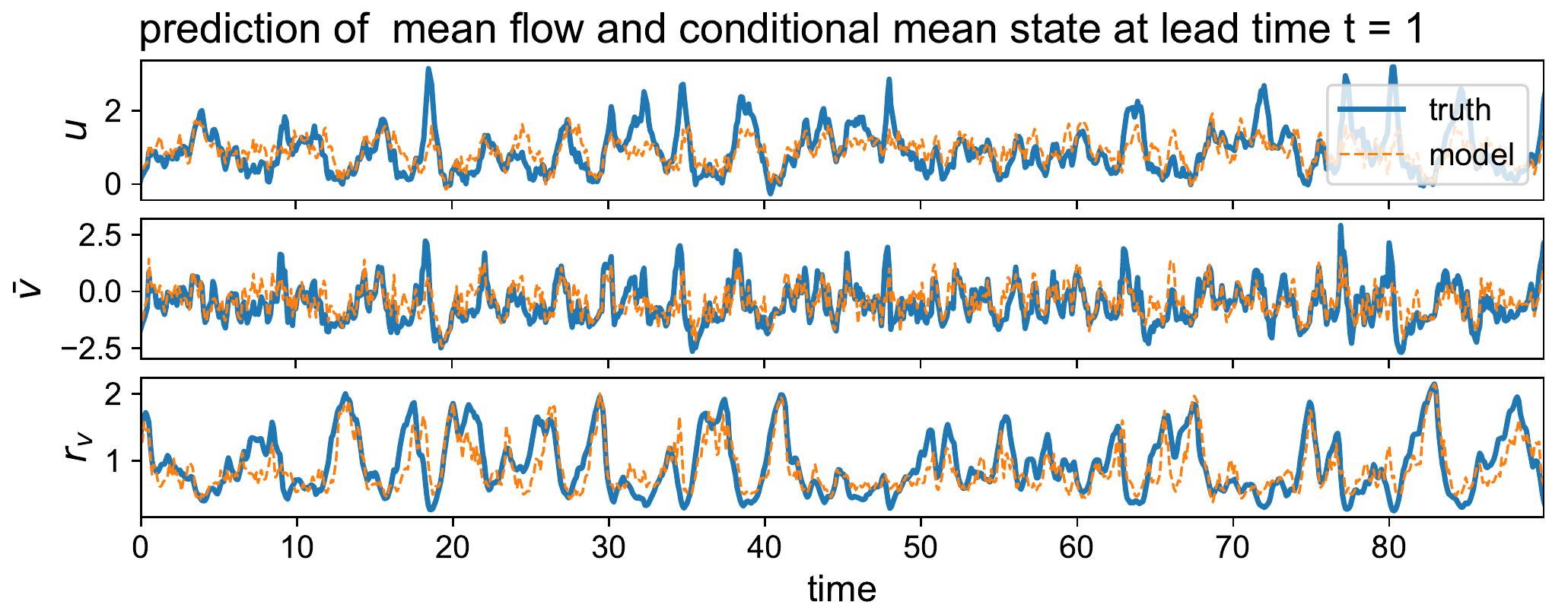}

}

\caption{Lead time prediction of the resolved state $u$ and conditional mean
and variance in the unresolved state $v$ in the two test regimes.\label{fig:Lead-time-prediction-dyad}}

\end{figure}

First, we train the RNNs proposed in (\ref{eq:dyad_rnn})
by using the standard LSTM network. The loss function is from
the information metric (16) in the main text and the model is trained
in 50 epochs. The error evolution of the loss and the mean square errors (MSEs) in the
target states are compared in the first row of Figure \ref{fig:Training-and-lead-time-dyad}.
Rapid convergence and accurate training are achieved. Then we confirm
the trajectory prediction performance in the two tested statistical
regimes. The second row of Figure \ref{fig:Training-and-lead-time-dyad}
displays the prediction errors in the states with different lead times.
The errors all saturate in small amplitudes, inferring accurate prediction
for both the resolved state $u$ and the conditional statistics for
the unresolved state $v$. The truth and model prediction of the trajectory
realizations are also compared in Figure \ref{fig:Lead-time-prediction-dyad}.
It confirms the good performance by using the PIDD-CG algorithm to recover
the true trajectory by directly learning the nonlinear dynamics from
data. Notice that the near-Gaussian regime is usually easier to predict
and can stay accurate with smaller errors for longer lead time prediction.
In the trajectory prediction of the resolved state $u$ in its non-Gaussian
regime, larger errors will emerge as the model is iterated for longer
time forecast (right panel of Figure \ref{fig:Lead-time-prediction-dyad}).

\subsubsection{Prediction of PDFs}

\begin{figure}
\subfloat[equilibrium, near-Gaussian $u$]{\includegraphics[scale=0.45]{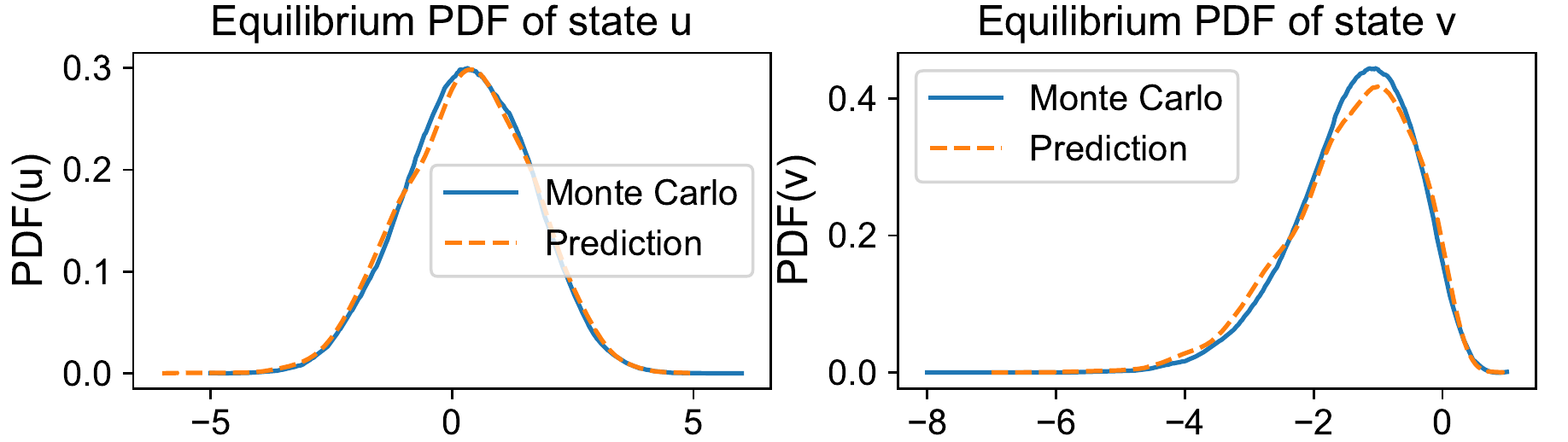}
}\subfloat[equilibrium,  non-Gaussian $u$]{\includegraphics[scale=0.45]{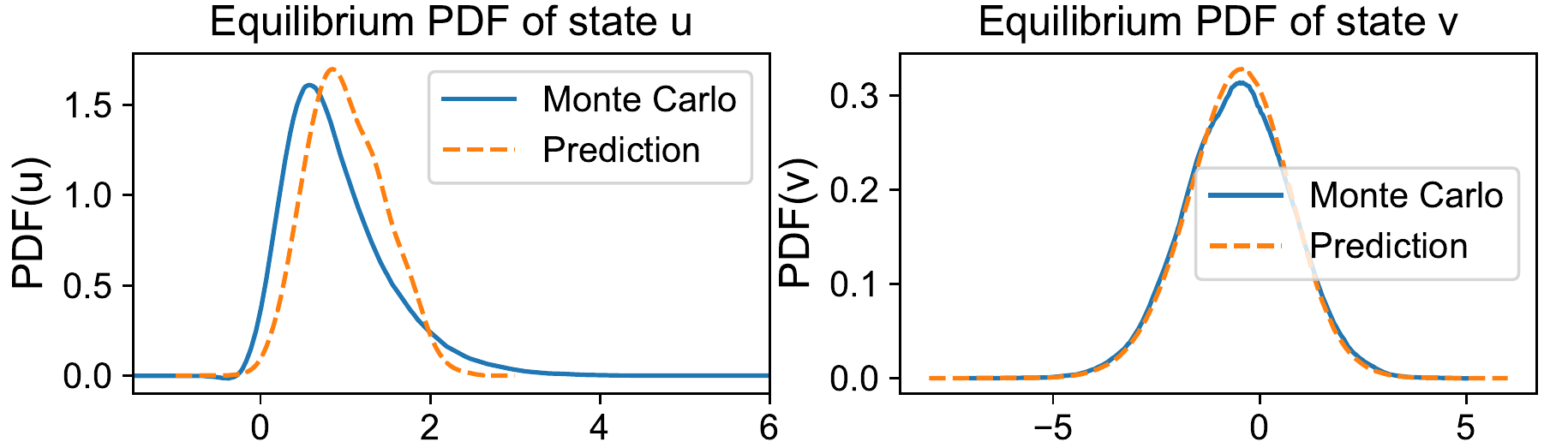}

}

\subfloat[transient,  near-Gaussian $u$]{\includegraphics[scale=0.46]{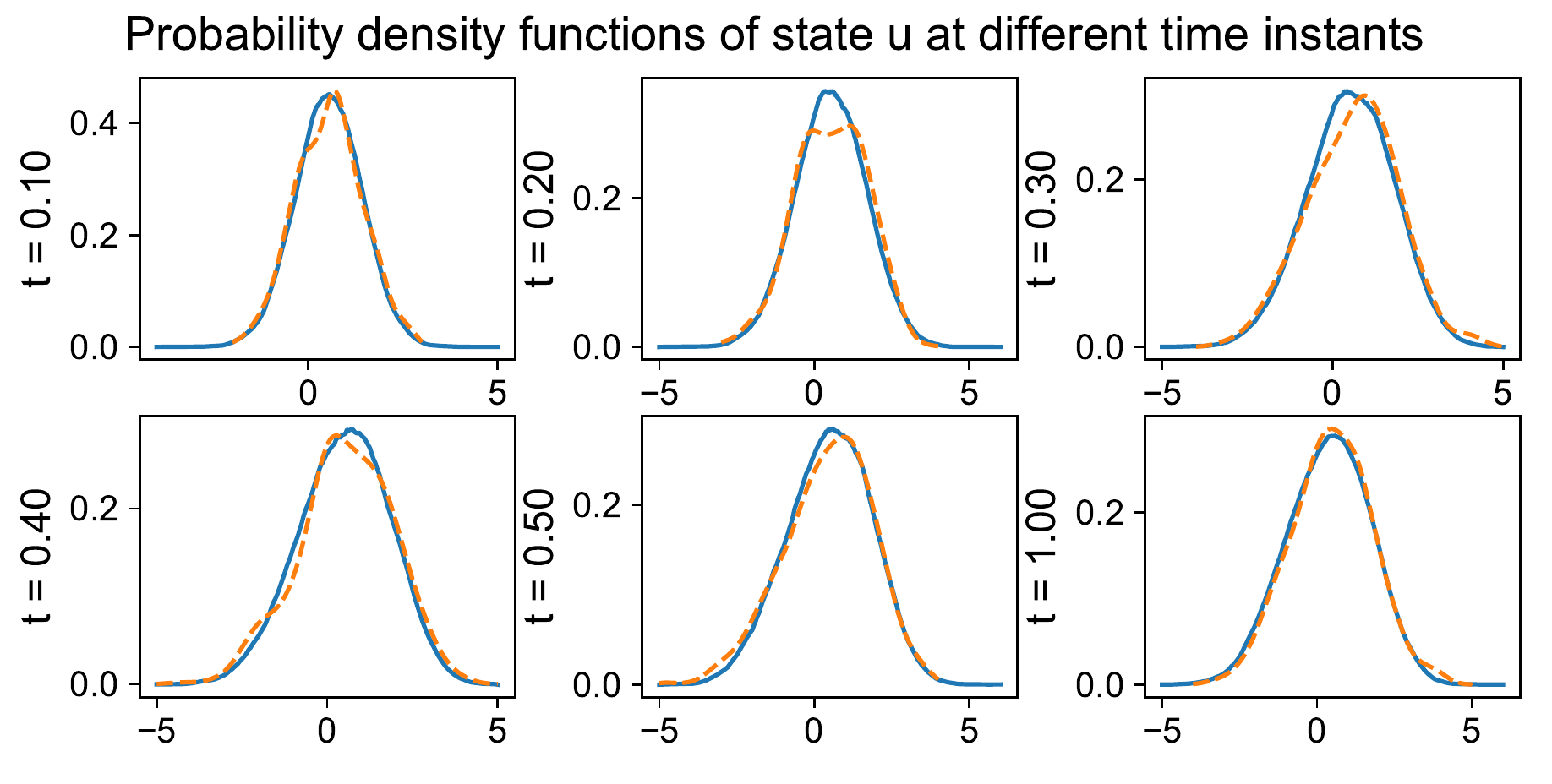}\includegraphics[scale=0.46]{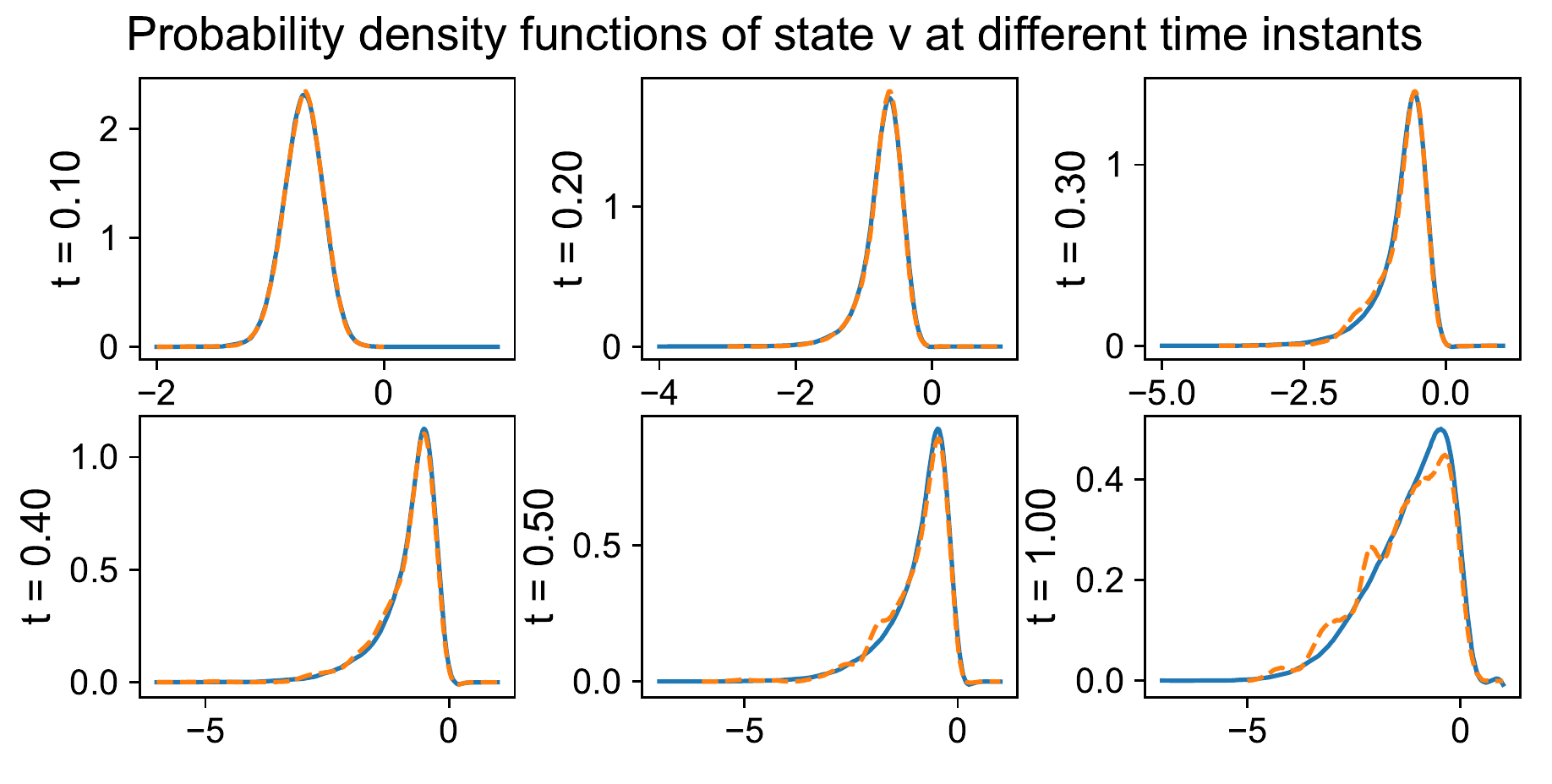}

}

\subfloat[transient, non-Gaussian $u$]{\includegraphics[scale=0.46]{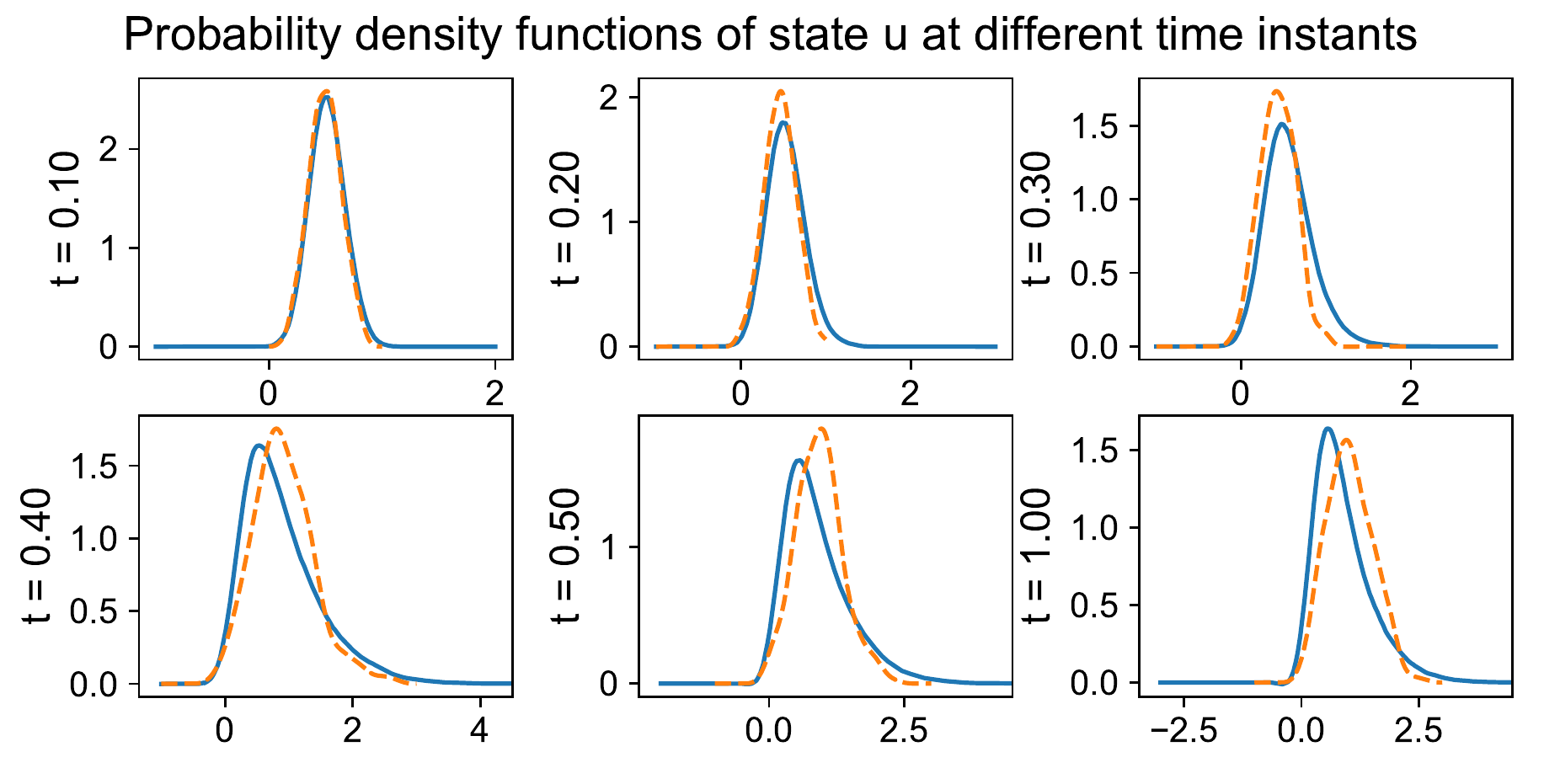}\includegraphics[scale=0.46]{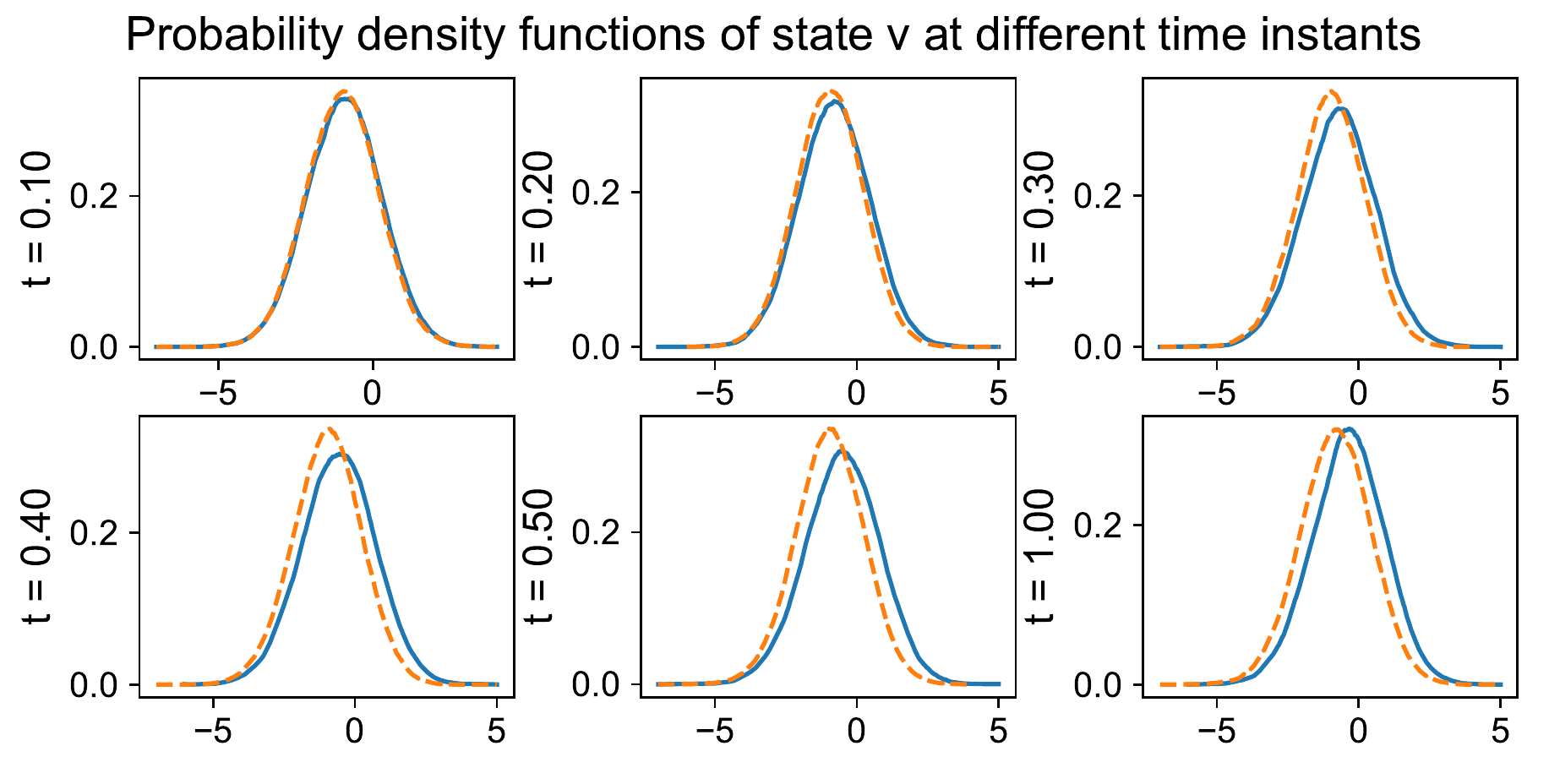}

}

\caption{Prediction of equilibrium (first row) and transient (second and third rows) PDFs of the dyad states at different time
instants before the equilibrium. The truth from Monte-Carlo samples
is shown in solid blue line, and the reduced model prediction in dashed
orange line. The two tested regimes with different statistics are
compared.\label{fig:Prediction-of-transient-dyad}}

\end{figure}

Next, we use the PIDD-CG algorithm to predict the evolution of
state PDFs from a given initial distribution. The initial distribution is taken as Gaussian and is very different from the final highly non-Gaussian equilibrium state. To show the true statistics
as the reference solution, we carry out direct Monte-Carlo simulation with a large
ensemble size $N=50000$. In the PIDD-CG algorithm, we  only take
$M=100$ samples to recover the resolved space of $u$. The first row of Figure
\ref{fig:Prediction-of-transient-dyad} shows the prediction of the final equilibrium
PDFs in the two tested regimes. As is implied from the trajectory prediction in Figure \ref{fig:Lead-time-prediction-dyad},
the near-Gaussian regime of $u$ gives accurate recovery of the PDFs.
On the other hand, the non-Gaussian regime of $u$ is more challenging to capture
the entire non-Gaussian statistics with a very small sample size. Nevertheless,
the major statistical structures are still captured using the PIDD-CG algorithm. The second and third rows of Figure \ref{fig:Prediction-of-transient-dyad} show the predicted
PDFs at several different lead time instants before the final statistical
equilibrium. 
We observe the development of skewed
PDFs in time in the transient states. Again, the efficient PIDD-CG algorithm
is able to capture the statistical development of the transient PDFs.
Especially in the non-Gaussian regime of $u$, the forecast is accurate
at the starting time then errors are gradually developed as
we try to predict the PDFs at a longer lead time due to the strong nonlinearity in the dynamics.

\section{The barotropic topographic model with multiscale coupling}

Next, we display detailed results about applying the PIDD-CG algorithm
on the multiscale barotropic topographic model discussed as the major test model in the main text.

\subsection{Model description}
\subsubsection{The starting model}
The topographic barotropic flow is a prototype model in geophysics. It is expressed as follows \cite{majda2006nonlinear},
\begin{equation}\label{barotropic}
\begin{aligned}
\frac{\partial q}{\partial t}+\nabla^{\bot}\psi\cdot\nabla q & =  \mathcal{D}\left(\Delta\right)\psi+F_{q},\\
\frac{dU}{dt}+ \fint \frac{\partial h}{\partial x}\psi^{\prime} & =  -d_{U}U+F_{U},
\end{aligned}
\end{equation}
which is defined in a two-dimensional domain $D: \mathbf{x}=(x,y)\in\left[-\pi,\pi\right]^{2}$ with double periodic boundary conditions. In \eqref{barotropic}, $\mathcal{D}\left(\Delta\right)$ is the dissipation operator, $h$ is the topographic effect, and $F_{q}$ and $F_{U}$ are external forcings. The state variable $U$ represents the large-scale zonal flow velocity while $q$ and $\psi$ are the potential vorticity and the stream function, respectively. They are related by
\begin{equation}\label{eq:vor_stream}
q=q^{\prime}+f=\nabla^{2}\psi^{\prime}+h+\beta y,\quad\psi=-U\left(t\right)y+\psi^{\prime},
\end{equation}
where the prime terms denote the fluctuations subject to the large-scale mean flow. The averaged integration in the mean dynamics $U$ in \eqref{barotropic} is defined as $\fint fd\mathbf{x}=\frac{1}{\lvert D\rvert}\int_{D}fd\mathbf{x}$, where $\lvert D\rvert$ is the total area of the domain.
The topographic barotropic flow model in \eqref{barotropic} is supplemented by a passive tracer model that characterizes the advection-diffusions
of the transport of a tracer density field $T\left(\mathbf{x},t\right)$, such that
\begin{equation}\label{eq:tracer}
\frac{\partial T}{\partial t}+\mathbf{u}\cdot\nabla T=-d_{T}T+\kappa\Delta T,
\end{equation}
where $\mathbf{u}=\nabla^{\bot}\psi$, $d_T$ is the drag term and $\kappa$ is the diffusion coefficient.  The model \eqref{barotropic}--\eqref{eq:tracer} exhibits very rich dynamical and statistical features, such as the switching behavior between blocked and unblocked zonal flow regimes, non-Gaussian distributions and extreme events.

\subsubsection{The barotropic model with layered topography}
A particularly interesting case of the above barotropic model \eqref{barotropic} is the one with layered topography, where the topography and stream function have the following expansion form
\begin{equation}
h\left(x,y\right)=\sum_{k}\hat{h}_{k}e^{ik\mathbf{l\cdot x}},\quad\psi\left(x,y,t\right)=\sum_{k}\hat{\psi}_{k}\left(t\right)e^{ik\mathbf{l\cdot x}}.
\end{equation}
The expansion is along one characteristic wavenumber direction $\mathbf{l}=(l_x,l_y)$ with $\lvert\mathbf{l}\rvert=1$. The corresponding full velocity field can be found combining the zonal mean flow and the fluctuations
\begin{equation}
\mathbf{u}=\left(U+u^{\prime},v^{\prime}\right)=\left(U-il_{y}\sum_{k}k\hat{\psi}_{k}e^{ik\mathbf{l\cdot x}},il_{x}\sum_{k}k\hat{\psi}_{k}e^{ik\mathbf{l\cdot x}}\right).\label{eq:velocity}
\end{equation}
Then nonlinear coupling term, $\nabla^{\bot}\psi\cdot\nabla q$, vanishes under the above layered topography expansion.  As a further simplification of the passive tracer model, we introduce a background mean gradient $\boldsymbol{\alpha}=\left(\alpha_{x},\alpha_{y}\right)$ on top of the tracer field fluctuations $T^{\prime}$ and a stochastic velocity field $\mathbf{u}$ is assumed such that
\begin{equation}
\begin{aligned}T\left(\mathbf{x},t\right)= & \:\boldsymbol{\alpha}\cdot\mathbf{x}+T^{\prime}\left(\mathbf{x},t\right),\\
\mathbf{u}\left(\mathbf{x},t\right)= & \left(U\left(t\right),v\left(x,t\right)\right).
\end{aligned}
\label{eq:assump}
\end{equation}
Here the zonal cross-sweep $U$ and the  fluctuations $v$ can be adopted from the topographic barotropic model solution. The same layered structure can be also assumed for the tracer fluctuation state so that we consider the tracer mean gradient $\alpha_{x}\equiv0,\alpha_{y}=\alpha$ with
\begin{equation}
T^{\prime}\left(x,y,t\right)=\sum_{k}\hat{T}_{k}\left(t\right)e^{ik\mathbf{l\cdot x}}.
\end{equation}
The resulting equation gives a simplified formulation for the turbulent
transport of passive tracer field
\begin{equation}
\frac{\partial T^{\prime}}{\partial t}+U\frac{\partial T^{\prime}}{\partial x}=-d_{T}T^{\prime}+\kappa\frac{\partial^{2}T^{\prime}}{\partial x^{2}}\:-\alpha v\left(x,t\right).\label{eq:tracer_simplified}
\end{equation}
The above equation for the tracer fluctuation field provides a judicious
simplified formulation in modeling the tracer passive transport with
many interesting statistical features such as intermittency and skewed
statistics. 

\subsubsection{The spectral formulation of the barotropic model with layered topography }
Based on the above justifications, the original equations (\ref{barotropic}) can be reformulated in the following form for each Fourier spectral mode
\begin{subequations}\label{eq:topo_model}
\begin{align}
\frac{dU}{dt}= & \sum_{k}h_{k}^{*}\hat{v}_{k}\:-d_{0}U+\sigma_{0}\dot{W}_{0}\label{eq:topo_model_U},\\
\frac{d\hat{v}_{k}}{dt}= & \left[-\gamma_{v,k}\left(U\right)+i\omega_{v,k}\left(U\right)\right]\hat{v}_{k}-l_{x}^{2}\hat{h}_{k}U\:-d_{v,k}\hat{v}_{k}+\sigma_{v,k}\dot{W}_{k},\label{eq:topo_model_v}\\
\frac{d\hat{T}_{k}}{dt}= & \left[-\gamma_{T,k}\left(U\right)+i\omega_{T,k}\left(U\right)\right]\hat{T}_{k}\:-d_{T,k}\hat{T}_{k}-\alpha\hat{v}_{k},\label{eq:topo_model_T}
\end{align}
\end{subequations}
where $k$ is the wavenumber with $\lvert k \rvert=1,\ldots, K$ and model parameters
\begin{equation}
\begin{gathered}
\gamma_{T,k}=d_{T}+\kappa k^{2},\qquad\omega_{T,k}=-k\left(U+u\right)\\
\gamma_{v,k}=0, \qquad\omega_{v,k}=l_x(k^{-1}\beta-kU),\qquad\sigma_{v,k}=-ik^{-1}\sigma_{k}.
\end{gathered}
\end{equation}
Note that additional dependent parameters $\gamma_{v,k},\gamma_{T,k}$ are introduced in \eqref{eq:topo_model} as extra parameterization for the unresolved multiscale interactions between fluctuation modes.

Notably, given one realization of $U$, the processes $\hat{v}_{k}$ and $\hat{T}_{k}$ in \eqref{eq:topo_model} become conditionally linear and Gaussian. Thus, the system automatically fits into the general modeling framework (3) proposed in the main text.

\subsubsection{Step-by-step illustration of predicting the barotropic topographic model using the PIDD-CG algorithm}
\paragraph{Step 1. Phase space decomposition.}
The fact that the large-scale zonal velocity $U$ is observed offers a natural way for the phase space decomposition: the low-dimensional subspace contains only $U$ while the remaining high-dimensional subspace involves all the Fourier modes for $\hat{v}_k$ and $\hat{T}_k$. Putting into the general framework in (3), this means:
\begin{equation*}
  \mathbf{X} := U\qquad\mbox{and}\qquad \mathbf{Y} := (\mathbf{v},\mathbf{T}) = (\hat{v}_1, \hat{v}_2,\ldots, \hat{T}_1, \hat{T}_2,\ldots).
\end{equation*}
\paragraph{Step 2. Systematic multiscale statistical closure approximation of the large-scale dynamics in the low-dimensional subspace.}
Since $U$ is coupled with $\hat{v}_k$ and $\hat{T}_k$, a suitable closure equation of $U$ needs to be developed before applying the traditional MC method to forecast $U$ in a closed intrinsic low-dimensional subspace.

Following the general framework in (5) in the main text, the fluctuation modes are decomposed into two subsets of the \emph{resolved scales} for wavenumbers in $\mathcal{I}=\left\{k:\lvert k\rvert\leq M\right\}$ and the long spectrum of the \emph{unresolved scales} in $\mathcal{I}^{c}=\left\{ k:M<\lvert k\rvert\leq N\right\}$, where the modes of $\hat{v}_k$ and $\hat{T}_k$ belonging to $\mathcal{I}$ correspond to the state variable $\mathbf{Y}_\mathbf{1}$ in (5) while the remaining modes are $\mathbf{Y}_\mathbf{2}$. Therefore, the mean flow equation \eqref{eq:topo_model} in the topographic model can be rewritten as
\begin{equation}\label{eq:red_model_U}
\begin{split}
\frac{dU}{dt}= & \sum_{k\in\mathcal{I}}h_{k}^{*}\bar{v}_{k}-d_{0}U+\mathcal{M}_{U}+\mathcal{N}_{U}+\sigma_{0}\dot{W}_{0} \\:=& \sum_{k\in\mathcal{I}}h_{k}^{*}\bar{v}_{k}-d_{0}U+\mathcal{H}_{U}+\sigma_{0}\dot{W}_{0}\\
\mathcal{M}_{U}= & \sum_{k\in\mathcal{I}\cup\mathcal{I}^{c}}h_{k}^{*}\left(\hat{v}_{k}-\bar{v}_{k}\right),\quad\mathcal{N}_{U}=\sum_{k\in\mathcal{I}^{c}}h_{k}^{*}\hat{v}_{k},\qquad \mathcal{H}_{U} = \mathcal{M}_{U} + \mathcal{N}_{U},
\end{split}
\end{equation}
where $\bar{v}_{k}$ is the conditional mean of $v_k$ given the past trajectory of $U$ that will be introduced in the next two steps. Clearly, $\mathcal{H}_{U} = \mathcal{M}_{U} + \mathcal{N}_{U}$ corresponds to $\mathcal{F}_\mathbf{X}$ in the general framework (5). Following (6), $\mathcal{H}_{U}$ is approximated by a RNN that approximates the contribution from both the fluctuation part of the resolved modes and the entire unresolved modes,
\begin{equation*}
  \mathcal{H}_{U}(t+1) =   \mbox{RNN}\left(U(t-m:t),\left\{ \bar{v}_{k}(t-m:t)\right\}_{k\in\mathcal{I}},\mathcal{H}_U(t-m:t)\right).
\end{equation*}
With the governing equation of $\hat{v}_k$ being provided, the intrinsic dimension of the approximated governing equation of $U$ is low. It therefore allows to use a MC simulation with a small number of ensembles $N$ to forecast its PDF up to a given time instant, which is then smoothed using a kernel density estimation,
\begin{equation*}
p(U) = \lim_{J\to\infty}\frac{1}{J}\sum_{j=1}^J \tilde{p}(U^{\{j\}})
\end{equation*}
with $\tilde{p}(U^{\{j\}})$ the $j$-th member from kernel density estimation that is associated with $U^{\{j\}}$.

\paragraph{Step 3. Effective physics-informed conditional Gaussian mixture via data assimilation.}
Each of the ensemble member from the MC simulation in Step 1 provides one trajectory of $U$, denoted by $U^{\{j\}}$. Conditioned on such a trajectory, there is one corresponding distribution of $\mathbf{v}$ and $\mathbf{T}$, namely $p(\mathbf{v},\mathbf{T}\vert U^{\{j\}})$. Note that $p(\mathbf{v},\mathbf{T}\vert U^{\{j\}})$ is a conditional Gaussian distribution for the layered topographic model \eqref{eq:topo_model} since conditioned on $U$ the processes of $v_k$ and $T_k$ are conditional linear. The joint distribution of $U, \mathbf{v}$ and $\mathbf{T}$ is thus given by
\begin{equation}\label{joint_pdf}
  p(U, \mathbf{v},\mathbf{T}) = \lim_{J\to\infty}\frac{1}{J}\sum_{j=1}^J \tilde{p}(U^{j})p(\mathbf{v},\mathbf{T}\vert U^{\{j\}}).
\end{equation}
This corresponds to (9) in the main text of the general PIDD-CG  forecast framework.

\paragraph{Step 4. Time evolution of the conditional statistics in smaller-scale dynamics.}
Following the general framework (10) or (12) in the main text, the time evolutions of the conditional mean and the conditional covariance for the barotropic model with layered topography are given as follows,
\begin{subequations}\label{eq:model_cond}
\begin{align}
\frac{d\bar{v}_{k}}{dt}= & -l_{x}^{2}\hat{h}_{k}U+\left[i\omega_{v,k}\left(U\right)-d_{v,k}\right]\bar{v}_{k}+\sigma_{0}^{-2}\mathcal{F}_{U}\cdot\mathcal{G}_{v,k},\\
\frac{d\bar{T}_{k}}{dt}= & \left[i\omega_{T,k}\left(U\right)-d_{T,k}\right]\bar{T}_{k}-\alpha\bar{v}_{k}+\sigma_{0}^{-2}\mathcal{F}_{U}\cdot\mathcal{G}_{c,k},\label{eq:model_cond_mean}\\
\frac{dr_{v,k}}{dt}= & -2d_{v,k}r_{v,k}+\sigma_{v,k}^{2}-\sigma_{0}^{-2}\lvert\mathcal{G}_{v,k}\rvert^{2},\\
\frac{dr_{T,k}}{dt}= & -2d_{T,k}r_{T,k}-\alpha\left(c_{k}+c_{k}^{*}\right)-\sigma_{0}^{-2}\lvert\mathcal{G}_{c,k}\rvert^{2},\\
\frac{dc_{k}}{dt}= & -\left(d_{v,k}+d_{T,k}\right)c_{k}+i\left[\omega_{T,k}\left(U\right)-\omega_{v,k}\left(U\right)\right]c_{k}\notag\\
&\qquad\qquad\qquad\qquad-\alpha r_{v,k}-\sigma_{0}^{-2}\mathcal{G}_{c,k}\cdot\mathcal{G}_{v,k}^{*},\label{eq:model_cond_var}
\end{align}
\end{subequations}
where $\bar{v}_{k}$ and $\bar{T}_{k}$ are the conditional mean of $\hat{v}_k$ and $\hat{T}_k$ while $r_{v,k}$, $r_{T,k}$ and $c_{k}$ are the conditional variance of $\hat{v}_k$, $\hat{T}_k$ and the cross covariance between $\hat{v}_k$ and $\hat{T}_k$, respectively.
In \eqref{eq:model_cond}, only the leading modes $\lvert k\rvert\leq M$ are resolved explicitly. In addition, the covariance equations have been further simplified by applying a block diagonal approximation, where each block has the size $2\times 2$ including the cross-correlation of the mode with the same wavenumber $c_{k}$. The central (quasi) linear dynamics are explicitly expressed while the complicated nonlinear functions are denoted by $\mathcal{F}_{U}$, $\mathcal{G}_{v,k}$ and $\mathcal{G}_{c,k}$ with
\begin{equation*}
\begin{gathered}
\mathcal{F}_{U}  =\dot{U}-\sum_{m}\hat{h}_{m}^{*}\bar{v}_{m}+d_{0}U,\\
\mathcal{G}_{v,k}  =\sum_{m}\hat{h}_{m}r_{v,km},\qquad
\mathcal{G}_{c,k}  =\sum_{m}\hat{h}_{m}c_{km},
\end{gathered}
\end{equation*}
where $\mathcal{F}_{U}$ corresponds to $\mathcal{F}_\mathbf{X}$ while $\mathcal{G}=(\mathcal{G}_{v,k},\mathcal{G}_{c,k})_{k\in\mathcal{I}}$ correspond to $\mathcal{G}_\mathbf{Y}$ in (11) of the main text.
\paragraph{Step 5. Data-driven modeling of the nonlinear feedbacks in conditional statistics via recurrent neural networks.}
Finally, corresponding to (13), the complicated nonlinear functions are approximated by RNNs,
\begin{equation}\label{eq:rnn_baro}
\begin{split}
\mathcal{F}_{U}(t+1) &= \mbox{RNN}\left( U(t-m:t),\left\{ \bar{v}_{k}(t-m:t)\right\}_{k},\mathcal{F}_{U}(t-m:t)\right),\\
\mathcal{G}(t+1) &= \mbox{RNN}\left( U(t-m:t),\left\{ r_{k}(t-m:t),c_{k}(t-m:t),\mathcal{G}_{k}(t-m:t)\right\} _{k\in\mathcal{I}}\right).
\end{split}
\end{equation}

\subsection{Model parameters}
This section includes the basic numerical setup and model parameters for the barotropic topographic model \eqref{barotropic} as well as the corresponding neural network architecture for the unresolved subscale processes in \eqref{eq:rnn_baro}.

\subsubsection{Parameters in the topographic barotropic model}
The barotropic topographic model displays  distinct dynamical and statistical features with different values of the white noise forcing amplitudes $\sigma_{U}$ and $\sigma_{v,k}$. In particular, the two representative regimes in the main text corresponding to the highly non-Gaussian and the nearly Gaussian regimes,  and the PDFs are equipped with the following parameters:
\begin{itemize}
\item \emph{Strongly non-Gaussian regime:} The zonal mean flow is strongly forced with white noise strength $\sigma_{U}=\frac{1}{\sqrt{2}}$ while only small noises $\sigma_{v,k}=\frac{k^{-1}}{20\sqrt{2}}$ are added to the fluctuation modes.
\item \emph{Near-Gaussian regime:} The fluctuation modes are subject to relatively stronger noise forcing with strength $\sigma_{v,k}=\frac{k^{-1}}{\sqrt{2}}$ compared with the noise strength in the zonal mean flow $\sigma_{U}=\frac{1}{2\sqrt{2}}$.
\end{itemize}
Notice that even in the near-Gaussian regime, nonlinear dynamics takes a dominant role in the multiscale interactions. Therefore, the feedbacks from the fluctuations cannot be neglected.

Next, the topography structure is given by
\[
h = H_{1}\left(\cos x+\sin x\right)+H_{2}\left(\cos2x+\sin2x\right) + \sum_{k=3}^{K}k^{-2}e^{i\theta_0} + c.c.,
\]
with the first two dominant leading modes $H_1=1, H_2=\frac{1}{2}$, and initial phase parameter $\theta_0=-\frac{\pi}{4}$. The `c.c.' denotes the complex conjugate.
This topography can be viewed as an analog to a long north-south ridge \cite{qi2017low}. For the rest parts of the model parameters, a uniform damping is adopted in both the mean and the fluctuation modes $d_U=d_{v,k}\equiv 0.0125$. The damping, diffusion, and mean cross-sweep for tracer field are $d_T=0.1,\kappa_T=0.001,\alpha=1$, respectively, and the rotation parameter is $\beta=1$.
These parameter values are derived from non-dimensionalization of the real physics measurements of the characteristic scales \cite{majda2006nonlinear}.

For the numerical integration in the true model to generate the simulated data, the standard 4th-order Runge-Kutta scheme with time step size $\mathrm{d}t=1\times10^{-3}$ is adopted, which is essential to maintain stability due to the stiffness in the small-scale flow and tracer dynamics, as well as the full conditional statistical equations. In contrast, only a forward Euler scheme with a much larger time step size $\Delta t=0.01$ is utilized in the RNN and thus the numerical cost is further reduced. Notice that this large time step cannot guarantee the numerical stability in the original model.

\subsubsection{Parameters in the neural network}
Recurrent neural networks (RNNs) offer the desirable structure to incorporate temporal processes of sequential data, and keep tracking of hidden processes.
The \emph{long short-time memory} (LSTM) network is a special RNN
that is useful to recover the time-series including very long time
correlations. The LSTM designed to learn the multi-scale temporal
structures overcoming the problem of vanishing gradients. In the computational
cell of the LSTM network, it consists of the basic building cell as
\begin{equation}
\begin{aligned}f_{t}= & \sigma_{g}\left(W_{f}x_{t}+U_{f}h_{t-1}+V_{f}c_{t-1}+b_{f}\right),\\
i_{t}= & \sigma_{g}\left(W_{i}x_{t}+U_{i}h_{t-1}+V_{i}c_{t-1}+b_{i}\right),\\
c_{t}= & f_{t}\otimes c_{t-1}+i_{t}\otimes\tanh\left(W_{c}x_{t}+U_{c}h_{t-1}+b_{c}\right),\\
o_{t}= & \sigma_{g}\left(W_{o}x_{t}+U_{o}h_{t-1}+V_{o}c_{t}+b_{o}\right),\\
h_{t}= & o_{t}\otimes\tanh\left(c_{t}\right).
\end{aligned}
\label{eq:lstm}
\end{equation}
Above, the $\sigma_{g}=\frac{1}{1+e^{-x}}$ is the sigmoid activation
function, and $\otimes$ represents the element-wise product. The
model cell includes forget, input, and output gates $f_{t},i_{t},o_{t}$,
and the cell state $c_{t}$. The hidden process $\left\{ h_{t-m},\cdots,h_{t-1},h_{t}\right\} $
represents the time-series of the unresolved process. The final output
data is given by a final linear layer $y_{t}=W_{x}h_{t}$ applying
on the final state of the hidden process.

The LSTM net is constructed from $m$ LSTM cells $\mathbf{h}_{i+1}=\mathrm{Lc}\left(\mathbf{x}_{i},\mathbf{h}_{i};\mathbf{W}\right)$
with the same structure and parameters $\mathbf{W}$. The cells are
connected by the intermediate hidden state $\mathbf{h}_{i}\in\mathbb{R}^{h}$.
Every LSTM cell takes in the input data $\mathbf{x}_{i}$ at the $i$-th
step and the output $\mathbf{h}_{i}$ from the previous adjacent cell,
and gives out the inner hidden state $\mathbf{h}_{i+1}$ to be used
for prediction of the next state. The full LSTM chain is connected
by $m$ sequential cell structures, that is,
\begin{equation}
\mathbf{h}_{m}=\mathrm{Lc}^{\left(m\right)}\left\{ \mathbf{h}_{0};\mathbf{x}_{t-m},\cdots,\mathbf{x}_{t-i},\cdots,\mathbf{x}_{t-1}\right\} \equiv\mathrm{Lc}\left(\mathbf{x}_{t-1}\right)\circ\cdots\mathrm{Lc}\left(\mathbf{x}_{t-i}\right)\cdots\circ\mathrm{Lc}\left(\mathbf{x}_{t-m}\right)\left(\mathbf{h}_{0}\right).\label{eq:lstm_full-1}
\end{equation}
Above, the data at different time instants $\mathbf{x}_{i}$ is fed
into the corresponding LSTM cell, and $\mathbf{h}_{i}$ is the hidden
state as the output of the previous cell and input for the next cell.
For simplicity, the initial value of the hidden state is often set
as zero, $\mathbf{h}_{0}=0$. The final output $\mathbf{h}_{m}$ from
the last step of the LSTM chain goes through another single layer
fully connected network to give the model approximation of the dynamical
increment for $f$
\begin{equation}
f_{m}^{M}=\sigma\left(\mathbf{W}^{f}\mathbf{h}_{m}+\mathbf{b}^{f}\right),\label{eq:lstm_final}
\end{equation}
where $\mathbf{W}^{f},\mathbf{b}^{f}$ are the model coefficients
in the final layer, and $\sigma$ is a nonlinear activation function
adopting the rectified linear unit (ReLU).

The above standard architecture of the LSTM is applied to estimate the feedback $\mathcal{F}_{U},\mathcal{G}$ to the leading resolved scales in \eqref{eq:red_model_U} and \eqref{eq:model_cond} in \eqref{eq:rnn_baro} to learn the embedded  processes. In addition, a residual structure is adopted in the LSTM neural network to update the correlated time sequence
\[
\theta_{i+1}=\theta_{i}+\mathcal{F},
\] where $\mathcal{F}$ is the LSTM output for the increment.

The major hyperparameters of the neural network used in the tests are as follows. The LSTM chain consists of $m=100$ repeating cells with the same structure, taking a time sequence of time length $t=1$ which is about the decorrelation time of the system state. The dimension of the hidden state is taken as $h_{v}=100$ for the conditional variance equations and $h_{m}=20$ for the conditional mean feedback. The LSTM output is reiterated forward for $n=10$ steps (only a short forward time of $t=0.1$) to account for the integrated error along the time integration steps. The optimization is carried out by stochastic gradient descent using the ADAM scheme for a training batch of size $100$ samples. During the training process, a total of $100$ epochs are repeated starting from the learning rate $\mathrm{lr}=1\times10^{-3}$ which is reduced three times to half of its original values at epoch number 25, 50 and 75. In the training data using the neural network, we pick a larger integration time step $\Delta t=10\mathrm{d}t=0.01$.  The reduced-order model can be integrated using the simple forward Euler scheme with the large time step $\Delta t$ thanks to the autocorrections directly learned during the training process of the LSTM neural network.

\subsection{Forecast results}

Finally, we display detailed test results for the training and prediction
using the PIDD-CG algorithm of the barotropic topographic model with
strong non-Gaussian and intermittent dynamics.

\subsubsection{Transition in true model statistics }

In the numerical test, we consider different statistical regimes with
the inclusion of damping and stochastic forcing effects. Especially,
we would like to check the transition in statistics in the flow and
tracer field solutions with varying model parameters. The standard
model parameters used in the tests are listed in Table \ref{tab:Initial-and-model}
according to the reference values proposed in \cite{qi2016low}. The same damping
amplitude $d_{U}=d_{k}=d_{0}$ is applied to all the spectral modes
and two sets stochastic forcing strength are considered for i) the
near-Gaussian regime $\sigma_{U}=\sigma_{0},\sigma_{k}=2\sigma_{0}$;
and ii) the non-Gaussian regime $\sigma_{U}=2\sigma_{0},\sigma_{k}=0.1\sigma_{0}$.
In the near-Gaussian regime, the zonal flow $U$ is subject to a smaller
white noise forcing compared with the small scale modes forcing $\sigma_{U}<\sigma_{k}$.
Both flow and tracer modes show PDFs close to Gaussian. In the non-Gaussian
regime, a stronger forcing in the zonal flow field $\sigma_{k}<\sigma_{U}$
gives a negatively skewed zonal state $U$, and the small scale modes
become strongly fat tailed. The corresponding passive tracer field
in this case also displays fat tails and also a large skewness in
the leading modes.

\begin{figure}
\subfloat{\includegraphics[scale=0.38]{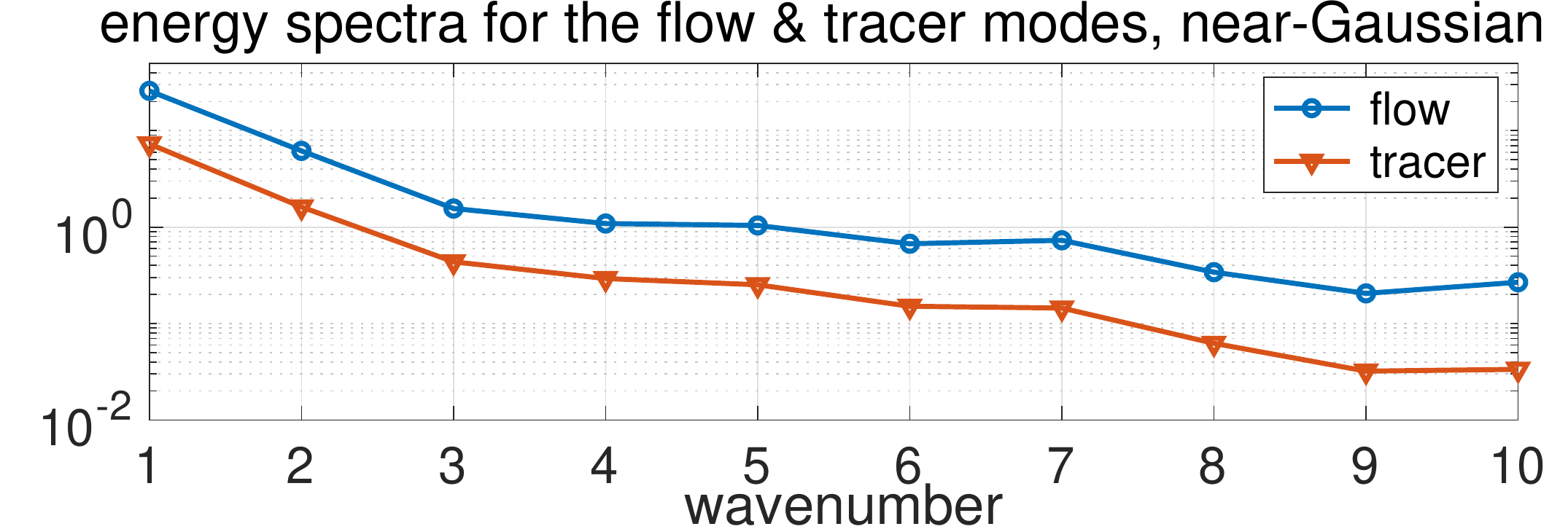}}\subfloat{\includegraphics[scale=0.38]{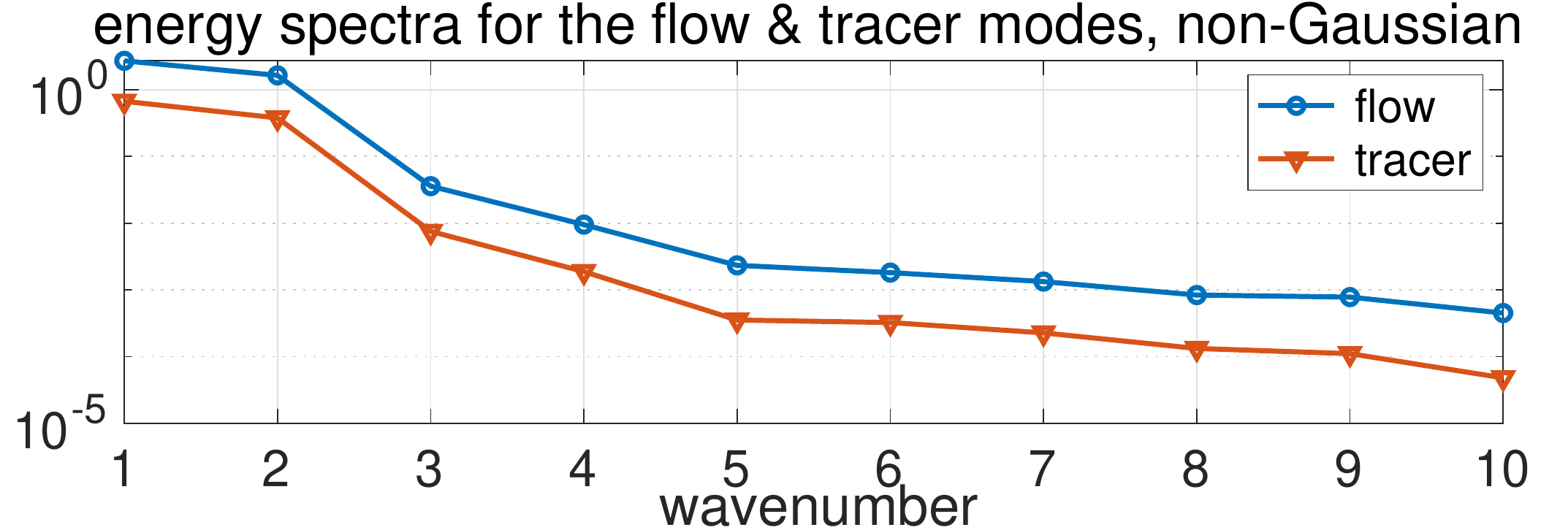}}

\vspace{-1em}

\subfloat{\includegraphics[scale=0.4]{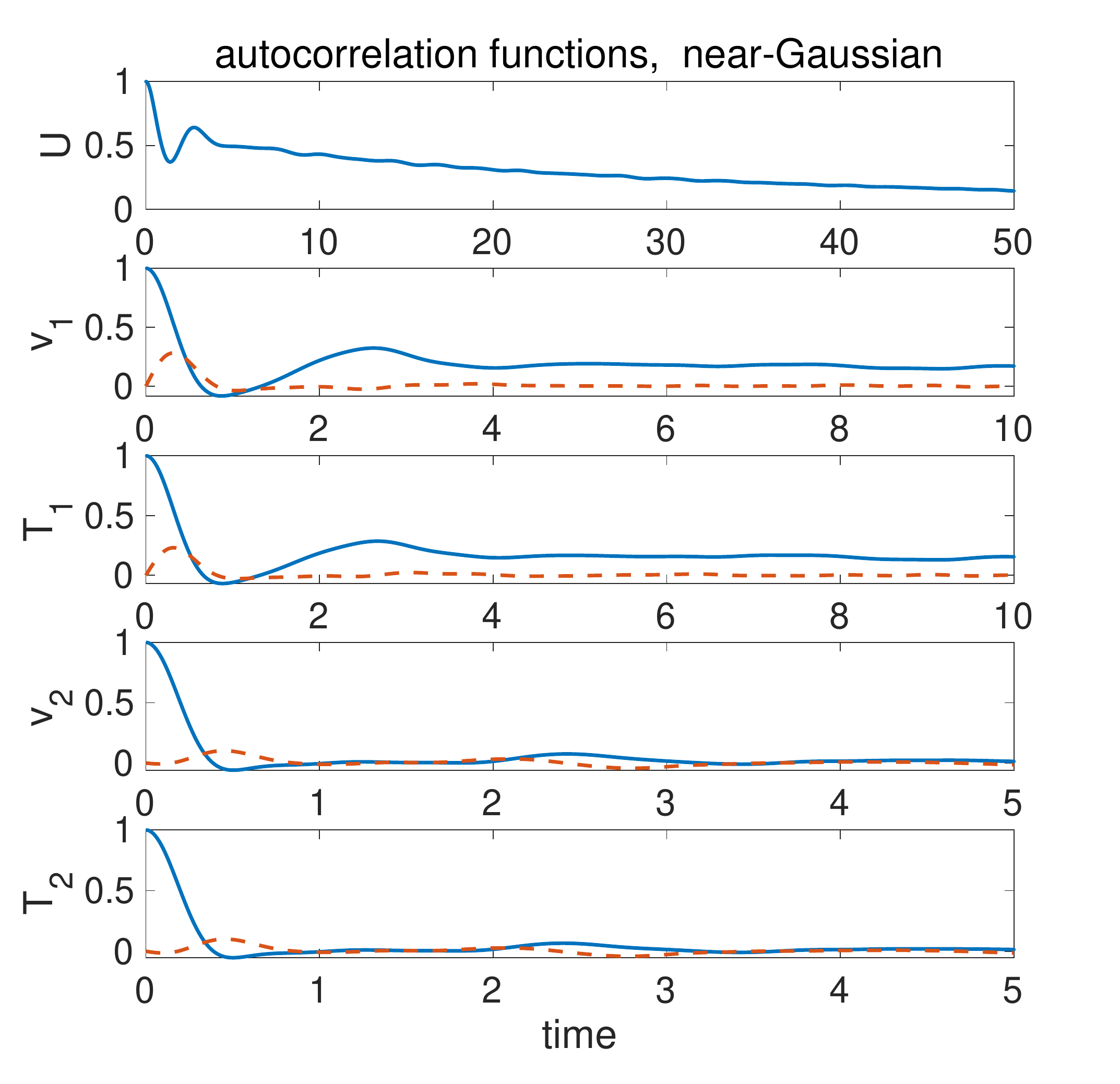}}\hspace{-1.em}\subfloat{\includegraphics[scale=0.4]{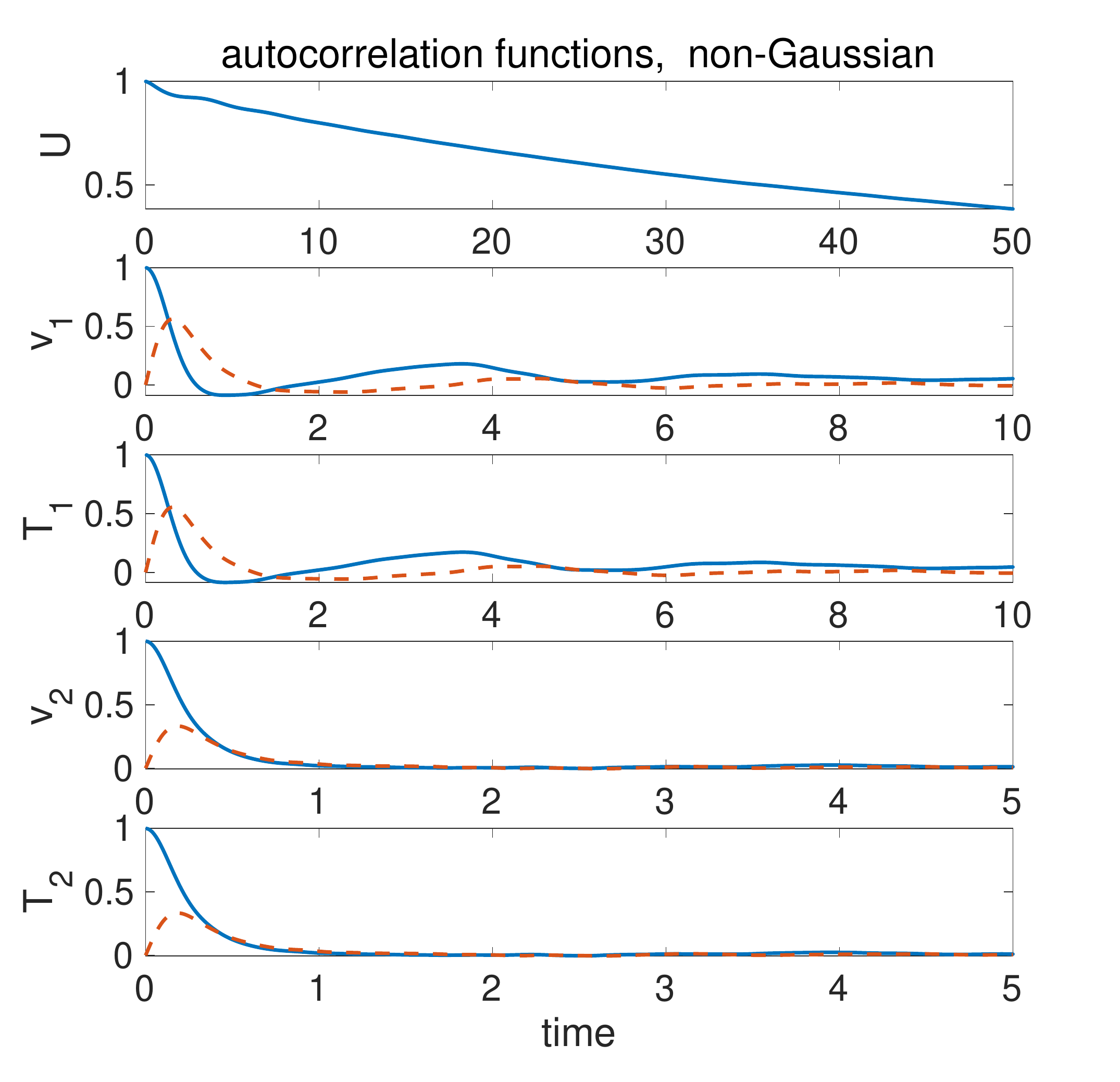}}

\caption{Equilibrium energy spectrum and autocorrelation functions of the barotropic
topographic model in the near-Gaussian (left) and non-Gaussian (right)
regimes.\label{fig:Equilibrium-energy-spectrum}}

\end{figure}

\begin{figure}
\begin{centering}
\subfloat[near-Gaussian regime]{\begin{centering}
\includegraphics[scale=0.3]{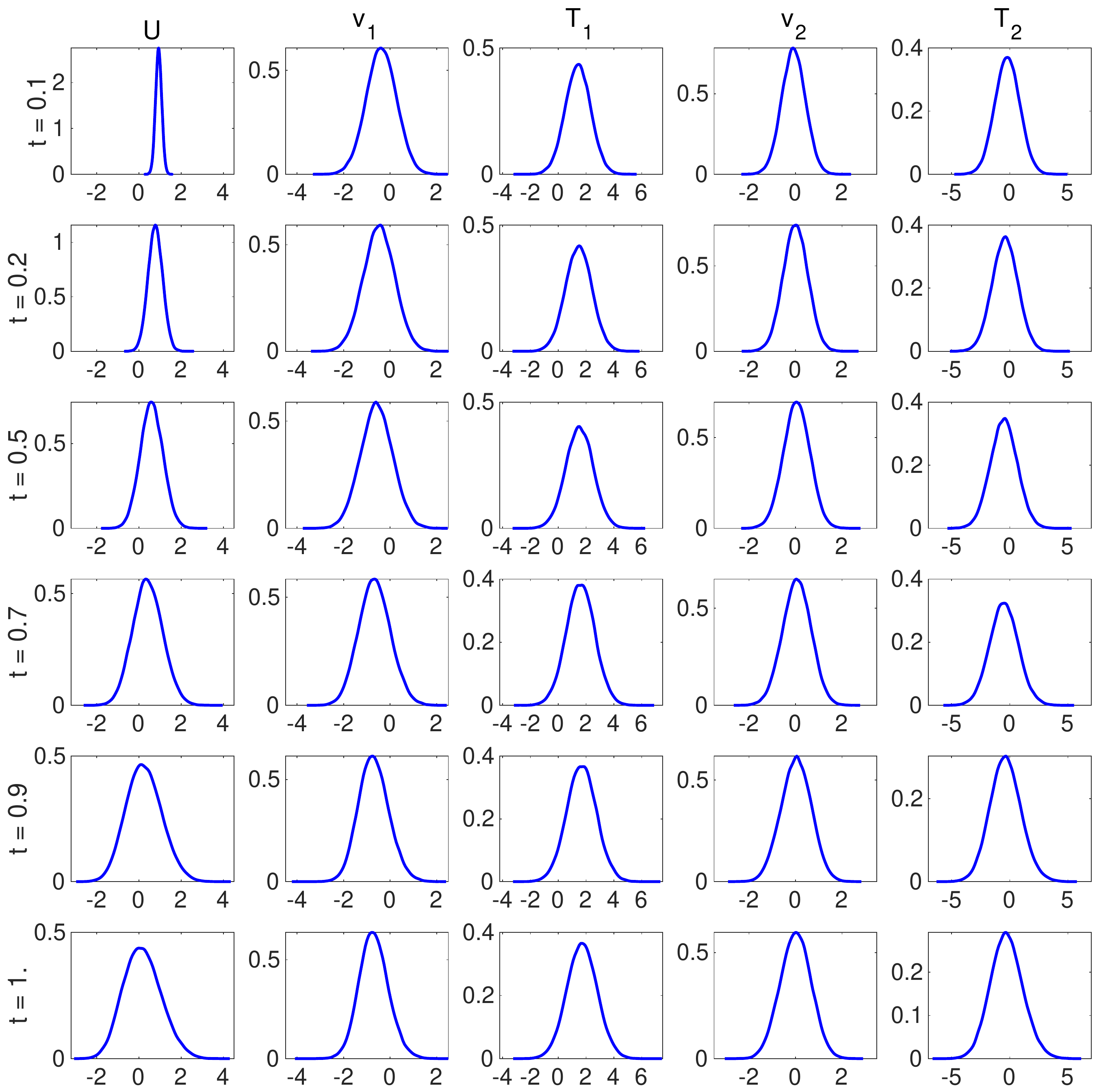}
\par\end{centering}
}\subfloat[non-Gaussian regime]{\begin{centering}
\includegraphics[scale=0.3]{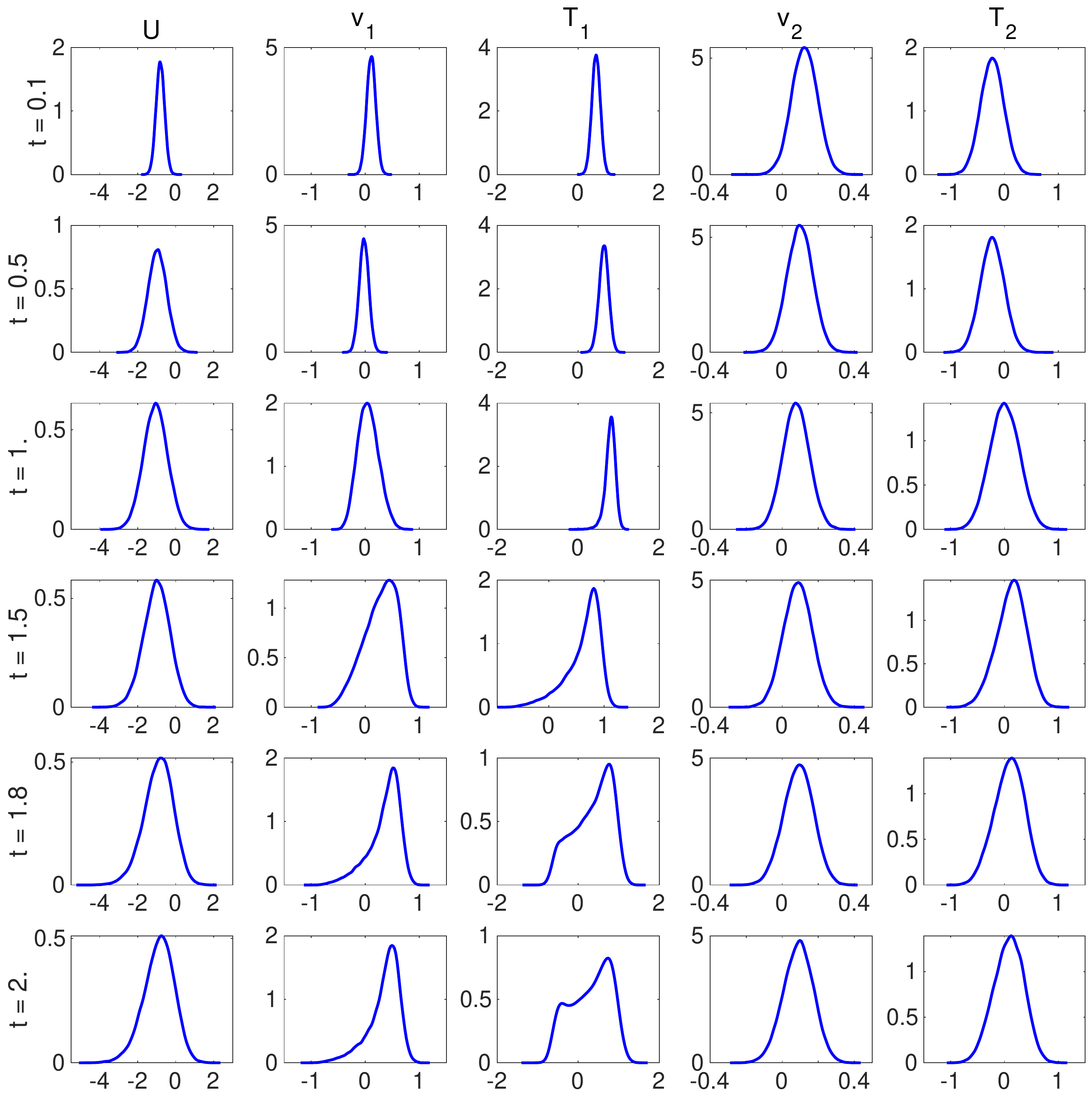}
\par\end{centering}
}
\par\end{centering}
\caption{Marginal PDFs for the zonal flow $U$ and the leading flow and tracer
modes $\hat{v}_{1},\hat{v}_{2}$ and $\hat{T}_{1},\hat{T}_{2}$ at
several time instants until equilibrium state is reached in the near-Gaussian
(left) and non-Gaussian (right) regime.\label{fig:Marginal-PDFs}}
\end{figure}

\begin{table}\centering
\begin{tabular}{cccccccccccc}
\toprule
layered topo. & $N$ & $\beta$ & $H_{1}$ & $H_{2}$ & $\left(l_{x},l_{y}\right)$ & $d_{T}$ & $\kappa_{T}$ & $\alpha$ &  & $d_{0}$ & $\sigma_{0}$\tabularnewline
\midrule
\midrule
$\begin{aligned}H_{1}\left(\cos x+\sin x\right)+\\
H_{2}\left(\cos2x+\sin2x\right)
\end{aligned}
$ & 10 & 1 & 1 & 0.5 & (1, 0) & 0.1 & 0.001 & 1 &  & 0.0125 & $\frac{1}{2\sqrt{2}}$\tabularnewline
\bottomrule
\end{tabular}

\caption{Standard model parameters for the barotropic topographic model simulations.\label{tab:Initial-and-model}}
\end{table}
We first display the key features in the two test regimes with distinct
statistics. The equilibrium energy spectra in the flow and tracer
modes as well as the autocorrelation functions are shown in Figure \ref{fig:Equilibrium-energy-spectrum}.
The time development of the marginal PDFs of the zonal flow $U$ and
the leading fluctuation and tracer modes $\hat{v}_{1},\hat{v}_{2}$ and $\hat{T}_{1},\hat{T}_{2}$
at different time instants until equilibrium are compared in Figure
\ref{fig:Marginal-PDFs} for the near-Gaussian and non-Gaussian regime
respectively. First, in the equilibrium energy spectra it is observed
that even though the first two leading modes are the most energetic, the
other smaller scale modes still contain large amount of energy with
strong feedback to the mean flow thus cannot be directly neglected
in the reduced model approximation. This confirms the crucial role
to consider the conditional Gaussian structure in the PIDD-CG algorithm
to achieve accurate statistical forecast. In addition to the multiscales
in spatial modes, the time series also display multiscale structures
as shown in the autocorrelation functions. The zonal flow $U$ has
a much slower decay in time correlation compared with the rapid decay
in the fluctuation modes. From the comparison of the time-series in
the two cases, the multiscale structure and competition between the
blocked and unblocked regimes is also observed. Especially, we see
the generation of high skewness in the zonal mean flow $U$
is from the strong zonal transport with suppressed fluctuations.
These multiscale features are consistent with the time series in Figure
2 of the main text.

To see the distinct statistical features in the two test regimes,
Figure \ref{fig:Marginal-PDFs} illustrates the
time evolution of the marginal PDFs in time up to the final equilibrium state.
Both regimes start with a Gaussian initial distribution with accurate observation
in $U$ (thus with zero variance) and the fluctuation modes are sampled
from the conditional Gaussian distribution. In the near-Gaussian regime,
the marginal PDFs of the states rapidly develop into the equilibrium
steady state with all near-Gaussian structures. In contrast, in the
non-Gaussian regime, strongly skewed PDFs are gradually developed
in time with highly non-Gaussian features.


\subsubsection{Training and lead time prediction}

The PIDD-CG algorithm includes the data-driven component to learn the
unresolved feedbacks from data. The standard LSTM network is used and thus a training stage is required. In the training
process using the neural network, we pick a larger integration time
step $\Delta t=10\mathrm{d}t=0.01$. The training output is reiterated
recurrently for $n=10$ times to improve the stability of the scheme.

\begin{figure}
\subfloat{\includegraphics[scale=0.42]{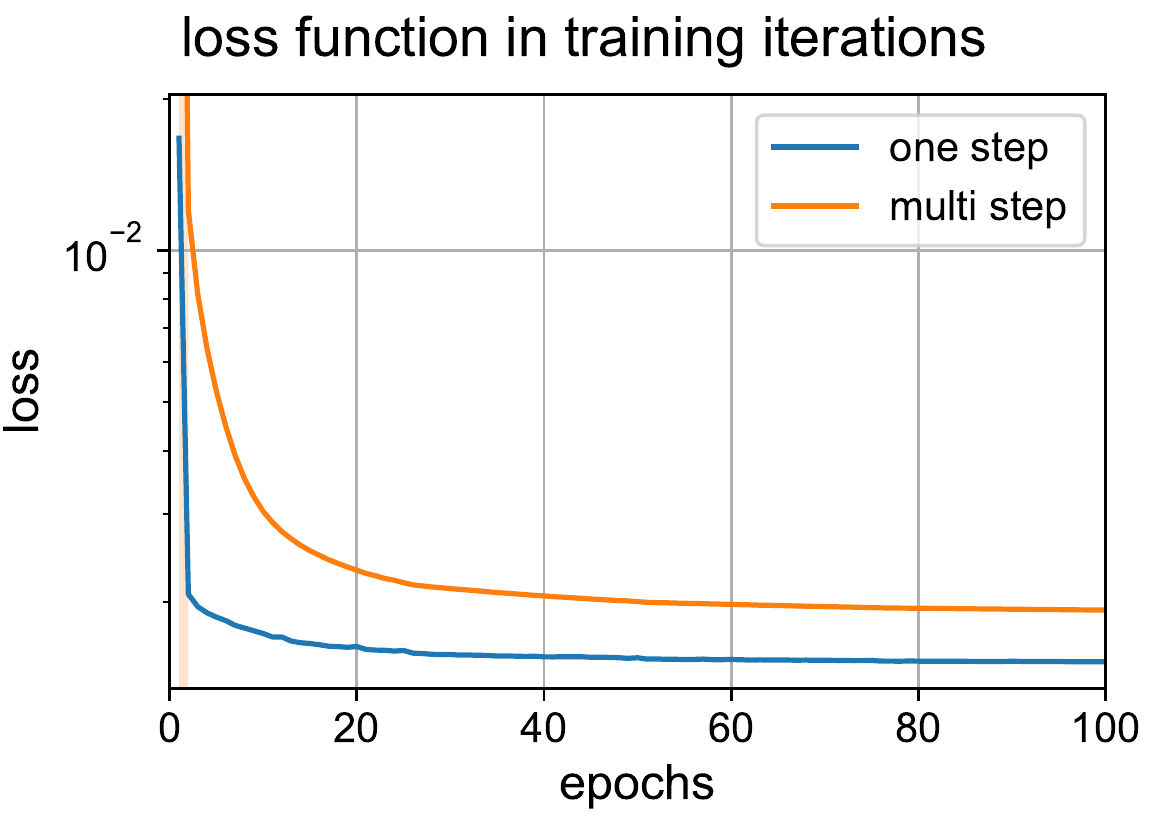}\hspace{-.5em}\includegraphics[scale=0.42]{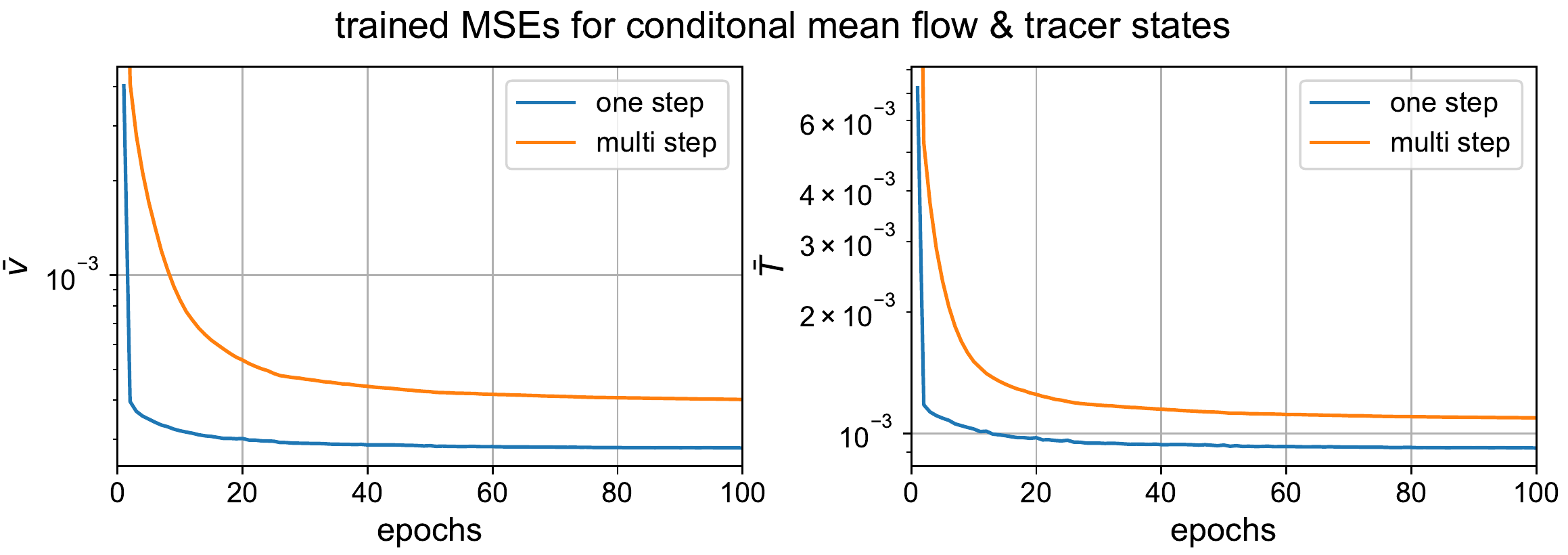}}

\subfloat{\includegraphics[scale=0.42]{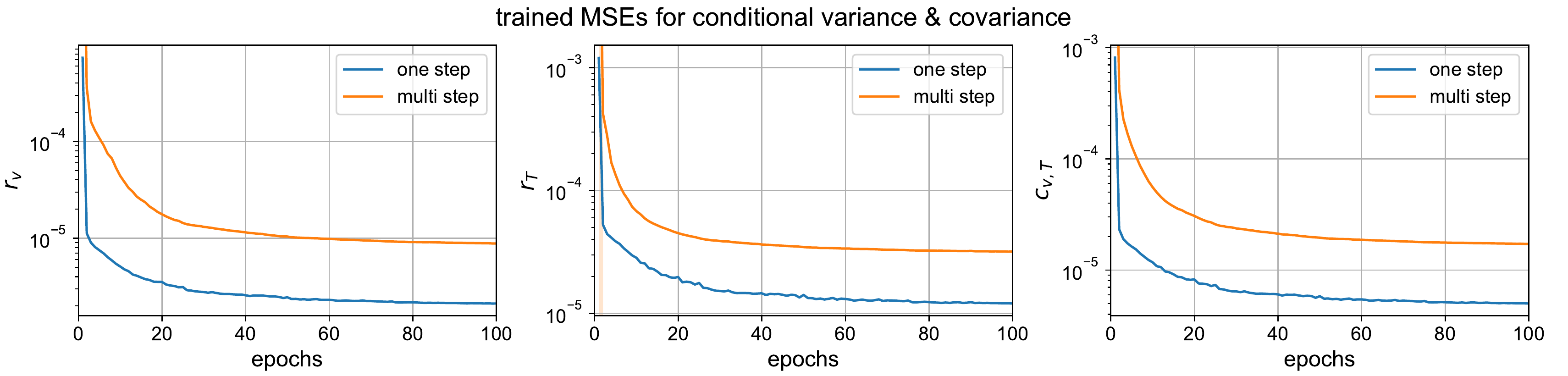}}

\caption{Errors in the loss function and the MSEs in conditional mean, variance,
and covariance during training iterations.\label{fig:Errors-in-the}}
\end{figure}

First, we show the convergence in the training stage. In Figure \ref{fig:Errors-in-the},
the training loss and mean square errors are displayed during the training iterations
with 100 epochs. It can be seen that the training error drops rapidly
during the first few iterations and is quickly saturated at a low level.
The corresponding errors in the conditional mean and covariance can also
be  minimized very quickly. Furthermore, we compare the improvement
with the reiterated multistep forecast $n=10$ compared with single
step update $n=1$ in the training calibration. The multi-step model
achieves a higher accuracy during training and is faster to
converge with a fewer number of iterations.

\begin{figure}
\begin{centering}
\subfloat[prediction error in the mean]{\includegraphics[scale=0.45]{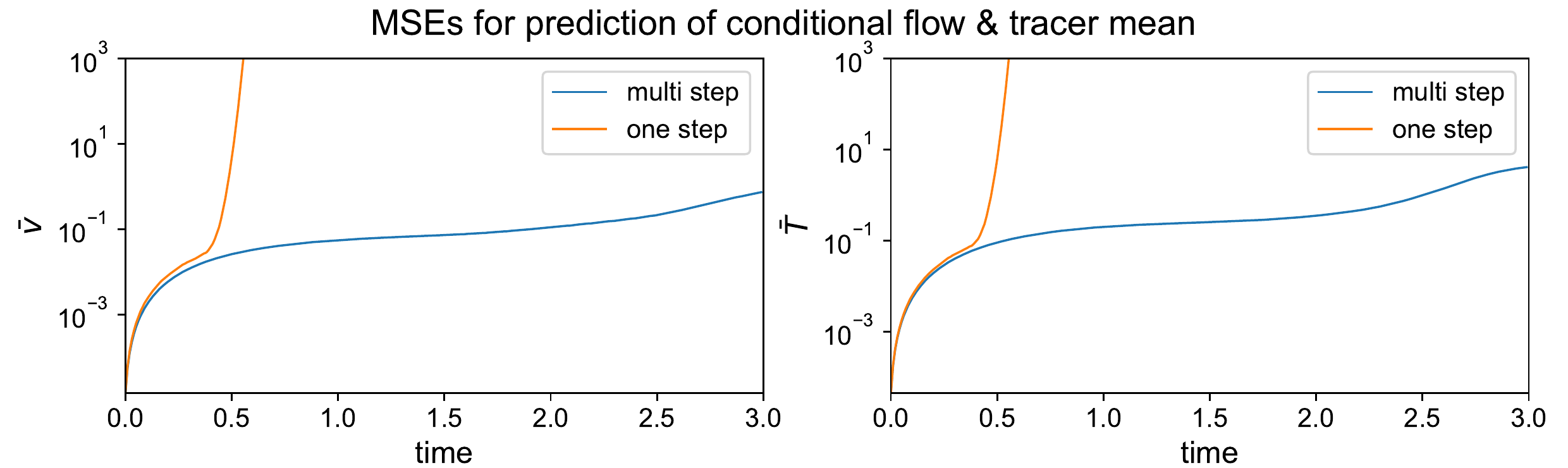}

}
\par\end{centering}
\begin{centering}
\subfloat[prediction error in the variance \& covariance]{\includegraphics[scale=0.45]{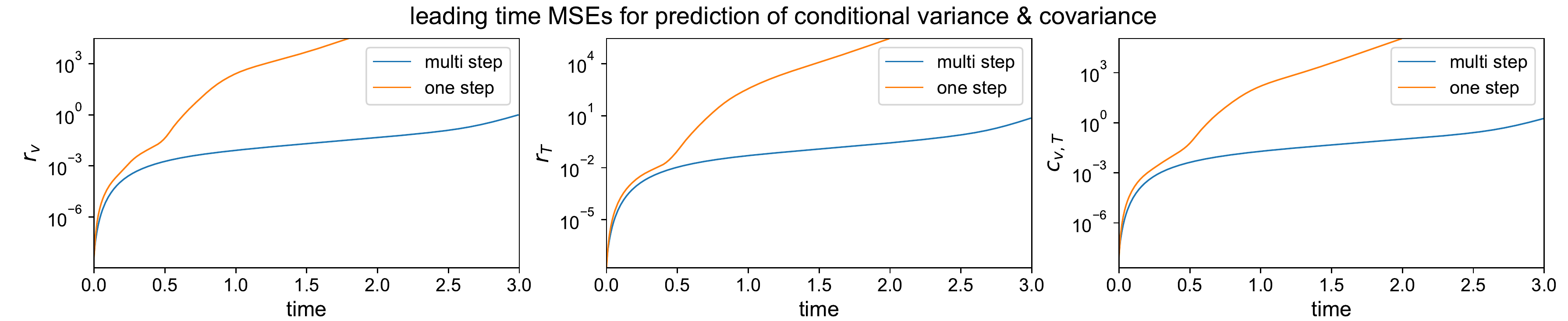}

}
\par\end{centering}
\caption{MSEs with different lead time predictions for errors in the conditional
mean and variance. The two trained models with different iterating
forward steps are compared. \label{fig:MSEs-with-different}}
\end{figure}

Next, we check the lead time prediction using the trained model. The
neural network model prediction is in general challenging with growing
imperfect errors in time due to the high model uncertainty and strong
internal instability. The prediction accuracy gradually grows larger
as the errors accumulate for the prediction in longer lead time shown
in Figure \ref{fig:MSEs-with-different} (also in the trajectory predictions
in Figure 5 of the main text). Again, the multi-step time integration
shows much higher stability during longer time iterations and maintains
high accuracy beyond the decorrealtion time. In comparison, if only
the one-step updating scheme is used in the training stage, the prediction
keeps accurate for short leading time (around $T<0.5$) while it quickly
diverges to much larger errors when the leading time becomes larger.
This is related with the inherent difficulty in the unstable numerical
integration with this large time step $\Delta t$.

\subsubsection{Prediction of the transient PDFs}

\begin{figure}
\subfloat{\hspace{4em}\includegraphics[scale=0.36]{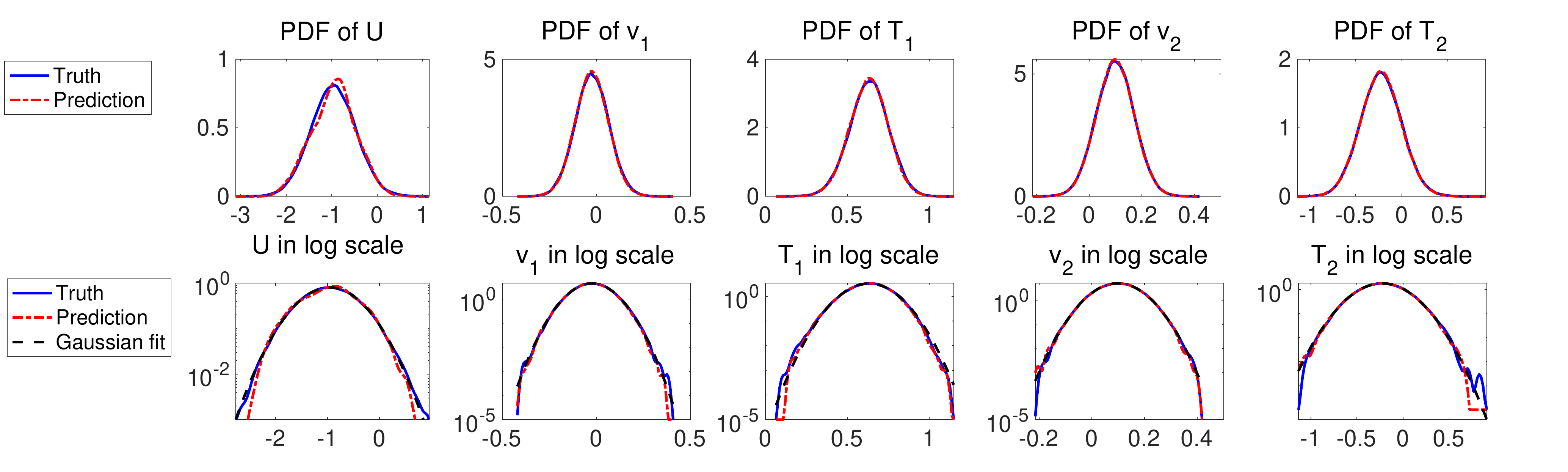}}

\subfloat{\includegraphics[scale=0.4]{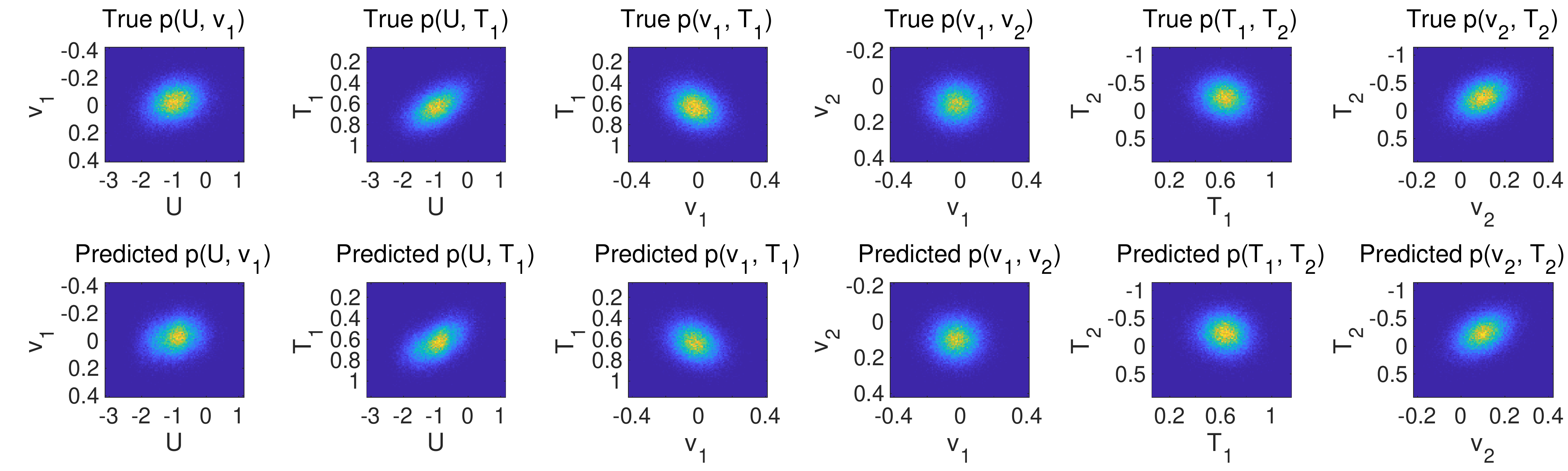}}

\caption{Prediction of the marginal PDFs and joint PDFs in the non-Gaussian
regime at lead time $t=0.5$.\label{fig:Prediction-of-pdf-1}}
\end{figure}

\begin{figure}
\subfloat{\hspace{4em}\includegraphics[scale=0.36]{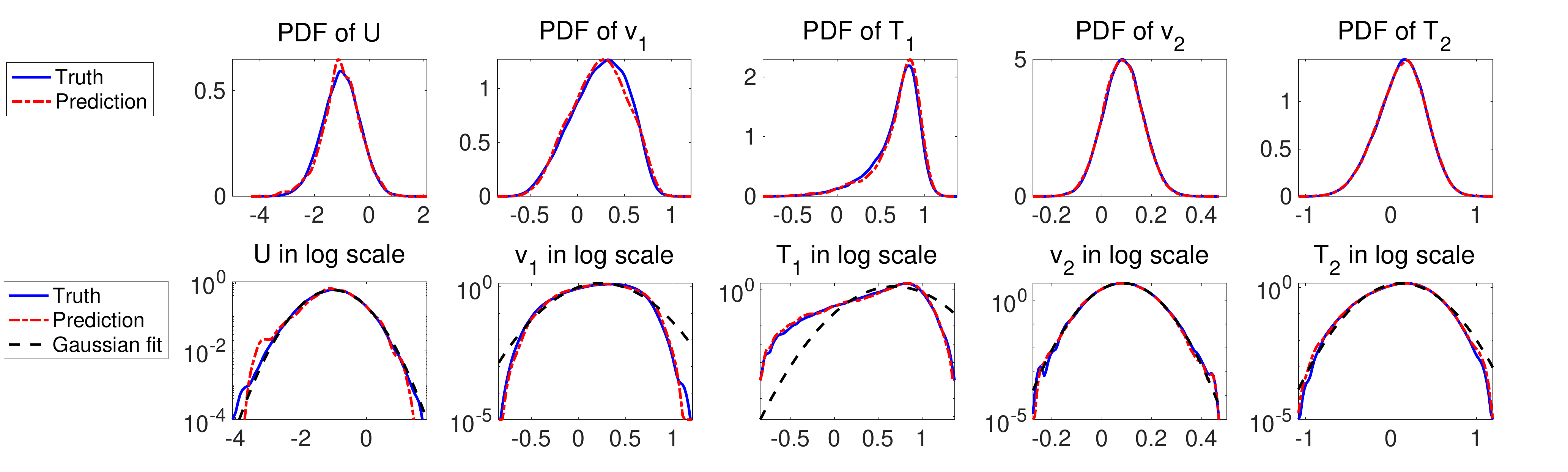}}

\subfloat{\includegraphics[scale=0.4]{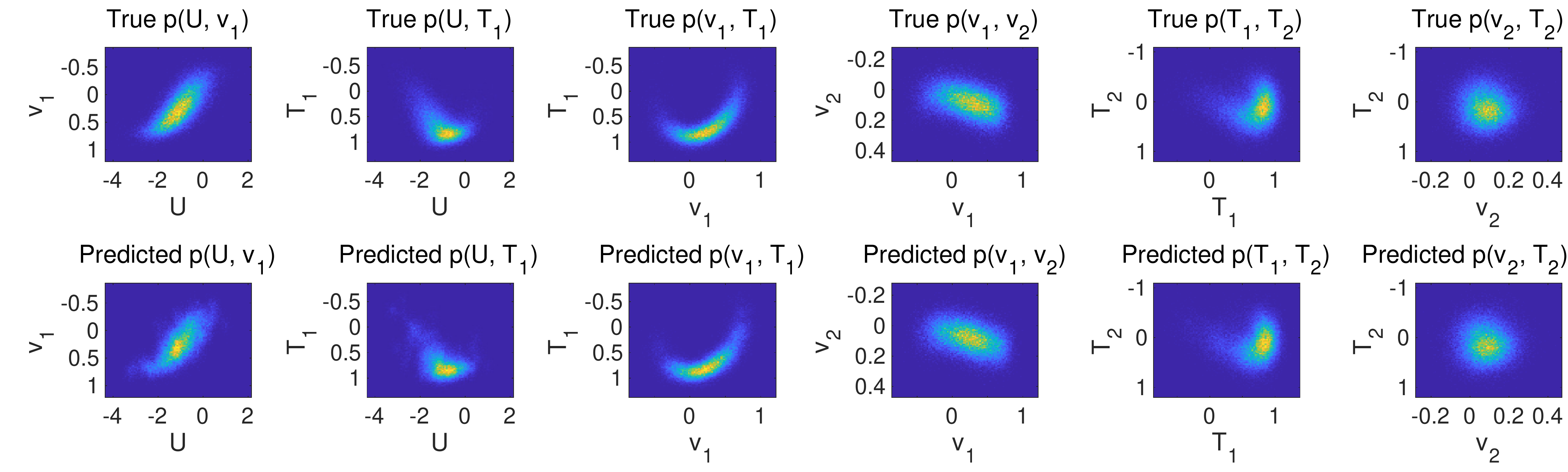}}

\caption{Prediction of the marginal PDFs and joint PDFs in the non-Gaussian
regime at lead time $t=1.5$.\label{fig:Prediction-of-pdf-2}}
\end{figure}

\begin{figure}
\subfloat{\hspace{4em}\includegraphics[scale=0.36]{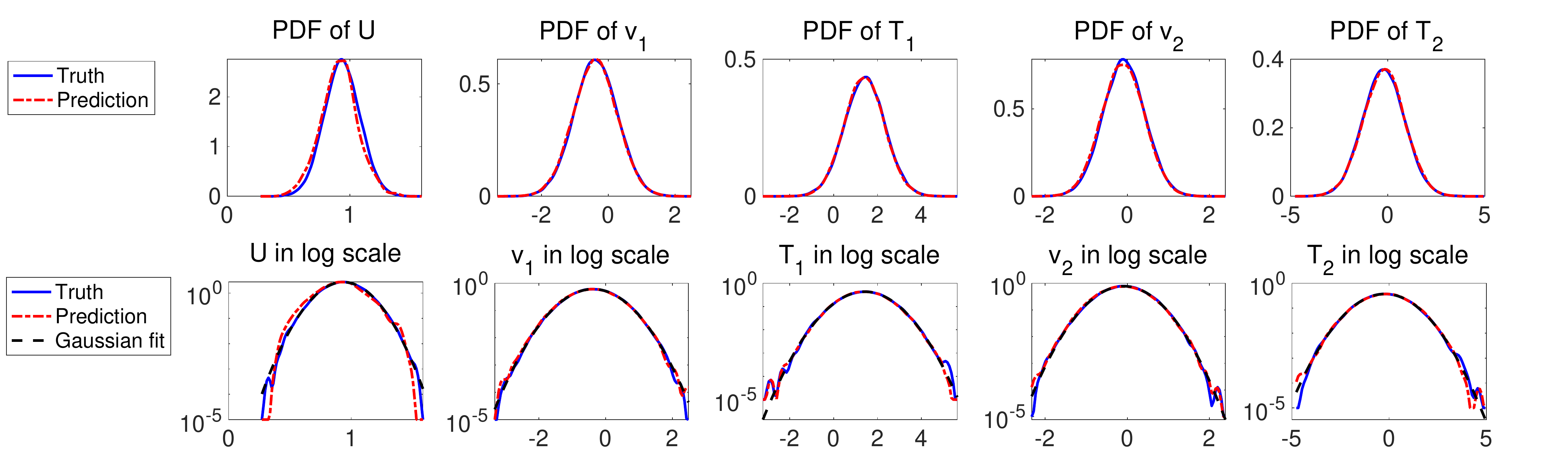}}

\subfloat{\includegraphics[scale=0.4]{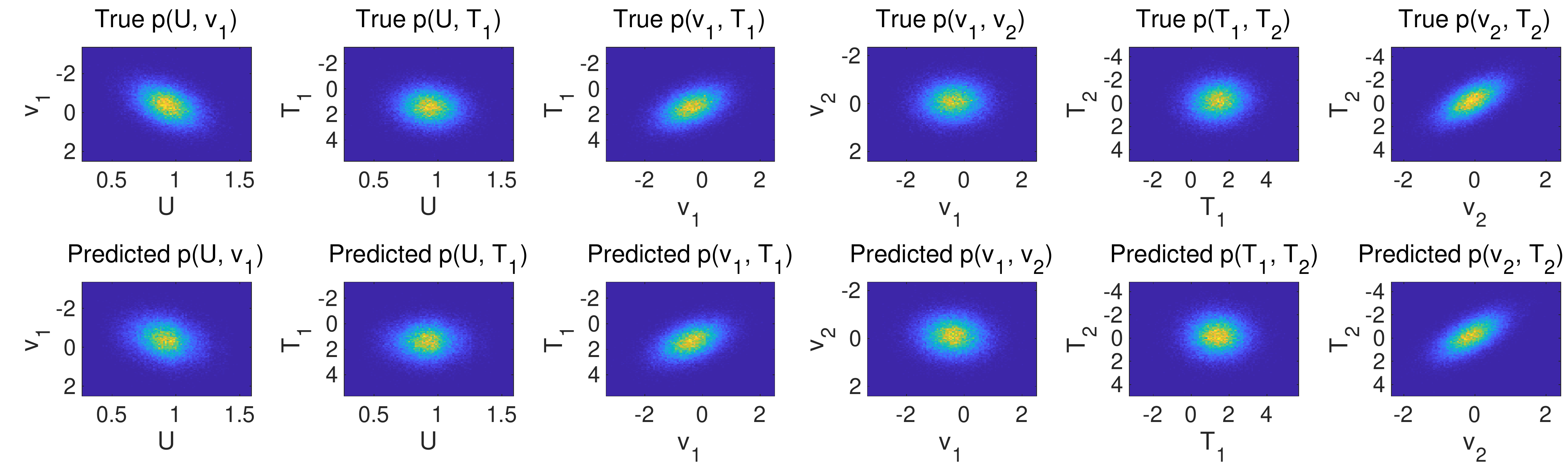}}

\caption{Prediction of the marginal PDFs and joint PDFs in the near-Gaussian
regime at lead time $t=0.1$.\label{fig:Prediction-of-pdf-3}}
\end{figure}

\begin{figure}
\subfloat{\hspace{4em}\includegraphics[scale=0.36]{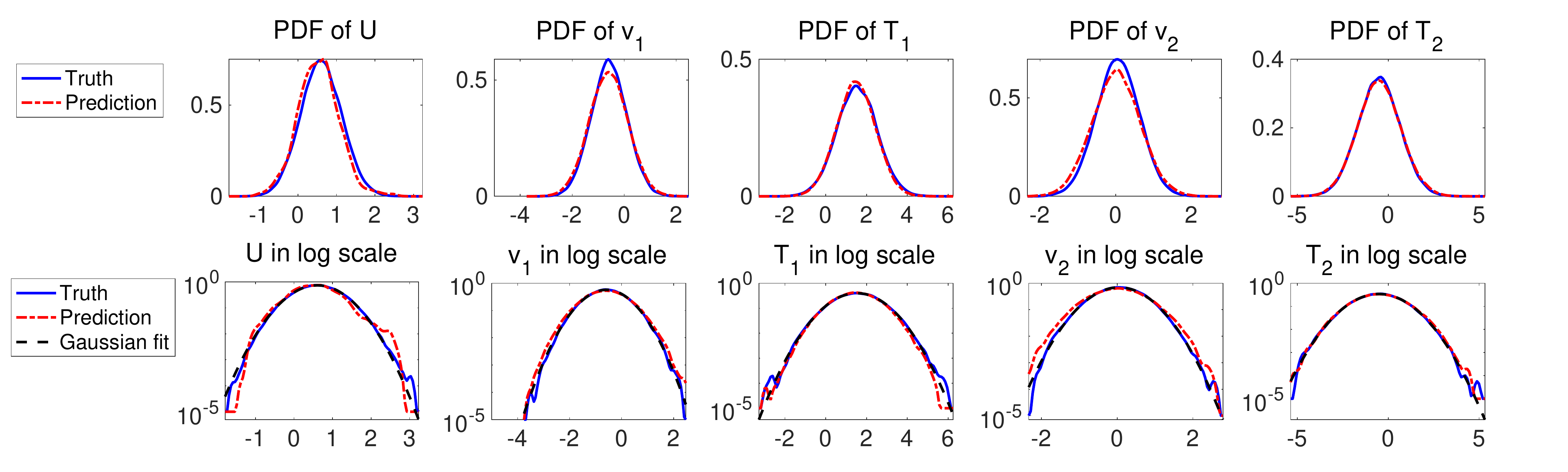}

}

\subfloat{\includegraphics[scale=0.4]{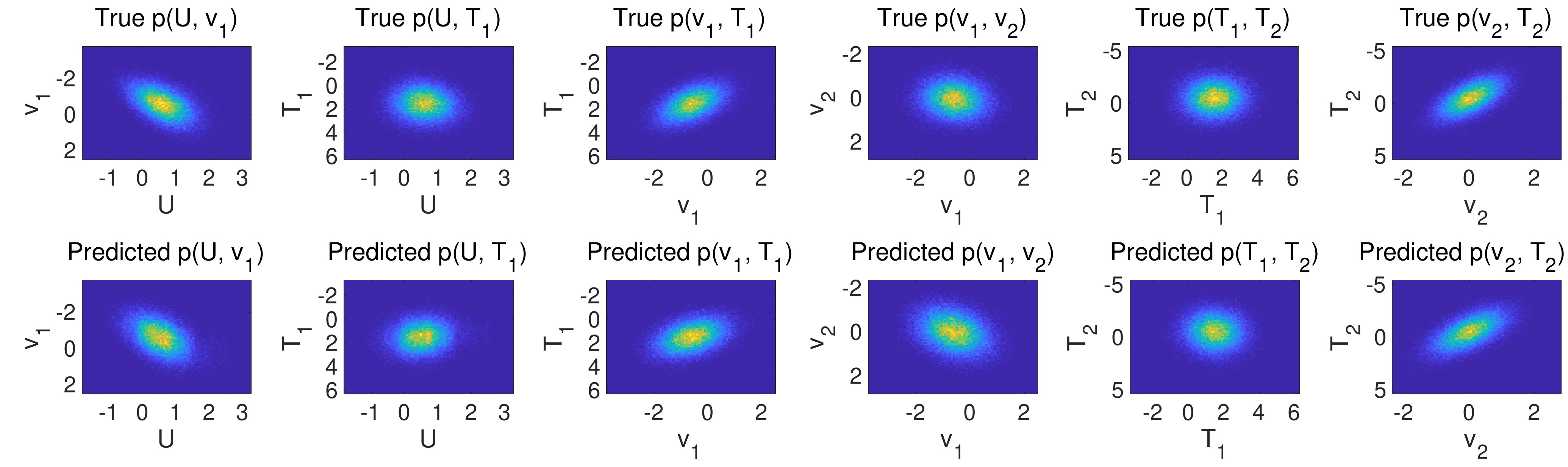}}

\caption{Prediction of the marginal PDFs and joint PDFs in the near-Gaussian
regime at lead time $t=0.5$.\label{fig:Prediction-of-pdf-4}}

\end{figure}

Finally, we display more detailed prediction results of using the
PIDD-CG algorithm to efficiently capture the transient PDFs of key
model states. Using the same initial distribution as in the main text,
Figure \ref{fig:Prediction-of-pdf-1}-\ref{fig:Prediction-of-pdf-4}
display the predicted marginal PDFs of the resolved states and the
joint distributions between the zonal flow $U$ and the first two
leading flow and tracer modes $\hat{v}_{1},\hat{v}_{2}$ and $\hat{T}_{1},\hat{T}_{2}$
in the transient states before equilibrium is reached. Especially,
we observe the development of non-Gaussian features from the initial
mixed Gaussian state. Notice the difference scales in the values of
$U$ and leading modes at different time instants. The direct Monte-Carlo
simulation requires a sample size of 50000 particles to capture the
PDFs in accuracy, while in contrast the PIDD-CG algorithm only needs
$N=100$ samples to achieve comparable accuracy with the truth. The
PIDD-CG algorithm maintains high accuracy in capturing the highly
non-Gaussian statistics regardless of the relatively high full dimension
of the system. On the other hand, in the near-Gaussian regime, the
convergence to equilibrium is faster and is also accurately captured
with accuracy in the much efficient PIDD-CG algorithm.


\end{document}